\documentclass{article}
\usepackage{amsmath, amssymb, amsthm}
\usepackage{thmtools}  
\usepackage{graphicx}  

\usepackage{cleveref}
\usepackage{geometry}
\usepackage{stoch_limit}

\title{The Quantum Stochastic Limit for Non-Markovian Driving Processes}
\author{Tanner J.\ Reese}

\begin{document}
\maketitle

\begin{abstract}
We express the stochastic limit (weak-coupling limit)
of a quantum system interacting with its environment
as the limit of solutions to a parametrized family of quantum stochastic differential equations (QSDEs).
We prove the convergence of such solutions in the case where
the driving processes may be non-Markovian and the system may be infinite-dimensional with unbounded Hamiltonian.
We achieve this by imposing a particular condition on the $ n $-point functions of the driving process
which we verify for annihilation, creation, and gauge operators
on a bosonic Fock space with sufficient mixing.
Additionally, we include not only the usual fast-evolving system Hamiltonian
which is averaged out in the limit,
but also a slow-evolving system Hamiltonian which survives the limit.
We also relax the requirement for the fast-evolving Hamiltonian
to have a purely discrete spectrum allowing accumulation points under some circumstances.
Moreover, our convergence mode for the processes
is convergence of all time-centered $ n $-point functions
which is stronger than the typical semigroup convergence for the stochastic limit.
\end{abstract}

\section{Introduction}
\label{sect:intro}  

Suppose a quantum system is interacting with its environment
and the combined system (which we will call the universe)
is evolving according to unitary dynamics.
Conventionally, one represents the system and environment
by their Hilbert spaces of states, $ \cH_M $ and $ \cH_E $, respectively.
Naturally, the Hamiltonian of the system is a self-adjoint operator $ K $ on $ \cH_M $
and the Hamiltonian of the environment is a self-adjoint operator $ Z $ on $ \cH_E $.
In general, $ K $ and $ Z $ are not necessarily bounded operators.
Then, the states of the universe are $ \cH_E \otimes \cH_M $
and we assume its dynamics are given by the Hamiltonian
\begin{equation} \label{eq:schro-qsde-fast}
\eps{H}_\fast = 1 \otimes K + Z \otimes 1 + \ep \frac{1}{2} \sum_{\alpha=1}^{d_E}
    (\Ea \otimes \ha + (\Ea)^* \otimes (\ha)^*)
\end{equation}
where $ \ep > 0 $ and for each $ \alpha $, $ \Ea $ is an operator on $ \cH_E $
and $ \ha $ is an operator on $ \cH_M $.
The summation describes the interaction between the system and environment
the strength of which is modulated by $ \ep $.
Moreover, depending on the choice of $ K $, $ Z $, $ \ep $, $ \{\Ea\} $, and $ \{\ha\} $,
there may exist a unitary evolution $ \{\eps{U}_t\}_{t \in \bR} $ generated by $ \eps{H}_\fast $
such that $ \frac{\dr}{\dr[t]} \eps{U}_t = -i\eps{H}_\fast \eps{U}_t = -i\eps{U}_t \eps{H}_\fast $.
\bigskip

Next, we select a trace class operator $ \varrho $ on $ \cH_E $
to represent the potentially mixed state of the environment
and use $ \tr_E $ to refer to the partial trace over $ \cH_E $.
In \cite{Dav74}, Davies showed that by scaling the time by $ \ep^{-2} $,
passing to the interaction picture,
and tracing over the environmental degrees of freedom,
the unitary evolution $ \eps{U}_t $ approximates a quantum Markov semigroup (QMSs)
when $ \cH_M $ is finite-dimensional, $ d_E = 1 $,
and $ \cH_E $ is a fermionic Fock space with a free-field Hamiltonian.
One way of stating this rigorously is that
for each $ t \geq 0 $ and any bounded operator $ F $ on $ \cH_M $,
\begin{equation} \label{eq:davies-limit}
\Lambda_t(F) = \lim_{\ep \dto 0} \,\, \tr_E\p{\varrho U^{(0)}_{t / \ep^2}
    \eps{U}_{-t / \ep^2} (1 \otimes F) \eps{U}_{t / \ep^2}
U^{(0)}_{-t / \ep^2} }
\end{equation}
where $ \{\Lambda_t\}_{t \geq 0} $ is a QMS
which is a strongly continuous semigroup
of completely positive $ \sigma $-weakly continuous linear maps
with $ \Lambda_0(X) = X $ and $ \Lambda_t(1) = 1 $
(see \cite{Chan15} for more about QMSs).
In fact, Davies demonstrates the limit exists uniformly for $ t $
on compact intervals away from $ t = 0 $.
\bigskip

Following the terminology of \cite{ALV02},
we refer to any limit in which a weak (i.e. order $ \ep $) fast-evolving (i.e. $ t / \ep^2 $) interaction
converges to a diffusion term as a \textbf{stochastic limit}.
Davies' work made rigorous earlier limits from the physics literature (see \cite{Hove54}).
There have since been many further developments of the stochastic limit.
In \cite{AFL90}, Accardi, Frigerio, and Lu reframe the stochastic limit
as a weak convergence claim about quantum stochastic processes using a language not dissimilar from this work.
In \cite{DdR07} and \cite{DdR08}, Derezi\'nski and De Roeck show the existence of
a stochastic limit in the dilation (i.e. without the partial trace).
Recently, in \cite{LP25}, Lopatin and Pechen demonstrate the limit for an infinite-dimensional system
considering both a bosonic and fermionic Fock space as the environment.
\bigskip

Because the characteristic time scale on which approximately Markovian behavior occurs
is slower than the system and environment dynamics by a factor of $ \ep^2 $,
it is natural to define a universe Hamiltonian with the rescaled time.
Treating $ K $, $ Z $, $ \Ea $, and $ \ha $ as operators on $ \cH_E \otimes \cH_M $,
this Hamiltonian is
\begin{equation} \label{eq:schro-qsde}
\eps{H}_\slow := \frac{1}{\ep^2} \eps{H}_\fast
= \frac{1}{\ep^2} K + \frac{1}{\ep^2} Z + \frac{1}{2 \ep} \sum_{\alpha=1}^{d_E} (\Ea \ha + (\Ea)^* (\ha)^*)
\end{equation}
such that $ \frac{\dr}{\dr[t]} \eps{U}_{t / \ep^2} = -i\eps{H}_\slow \eps{U}_{t / \ep^2} $.
Although the interaction term has order $ \frac{1}{\ep} $,
this is merely a different presentation of the stochastic limit from above.
\bigskip

\subsection{Stochastic Limit as limit of Flows on a $ C^* $-algebra}
\label{subsect:stoch-as-flow}  

In an effort to illustrate the connection between
the dynamics of closed and open quantum systems and classical SDEs,
we will rephrase \cref{eq:schro-qsde} as flows on $ C^* $-algebras and ultimately as quantum SDEs.
Then, the limit \cref{eq:davies-limit} may be viewed as
uniform convergence of flows on compact time intervals
or, alternatively, the weak convergence of solutions to QSDEs.
We represent the system and environment
by the $ C^* $-algebras generated by their observables, respectively, $ \cM $ and $ \cE $.
In the above language, these are $ \cM := B(\cH_M) $ and $ \cE := B(\cH_E) $
where $ B(\cdot) $ is the set of bounded linear operators.
However, in general, one could take $ \cM $ and $ \cE $ to be
proper $ {}^* $-subalgebras of $ B(\cH_M) $ and $ B(\cH_E) $.
Correspondingly, the dynamics of the system are generated by a $ {}^* $-derivation $ \fk $ on $ \cM $
and the dynamics of the environment are generated by a $ {}^* $-derivation $ \fz $ on $ \cE $.
Note that $ \fk $ and $ \fz $ are not necessarily bounded and so may have merely dense domains.
In the above language, these are $ \fk(f) := i[K, f] $ and $ \fz(X) := i[Z, X] $.
However, in general, one could take $ \fk $ and $ \fz $ to be any $ {}^* $-derivations
even those not given by commutators
(i.e. $ \fk $ and $ \fz $ need not be inner derivations).
\medskip

Next, the universe is represented by a $ C^* $-algebra $ \cA $
such that $ \cM, \cE \subseteq \cA $,
the subalgebra $ \cM $ commutes with $ \cE $,
and $ \cA $ is generated by $ \cM \cup \cE $.
In the above language, one could take $ \cA = B(\cH_E \otimes \cH_M) $
so that $ \cA $ is the spatial tensor product of $ \cM $ and $ \cE $
(see Chapter 6 of \cite{Murp90} for a description of tensor products on operator algebras).
Alternatively, one could take $ \cA = \cE \mxtens \cM $
so that $ \cA $ is the maximal tensor product of $ \cM $ and $ \cE $.
This latter choice is the free-est choice of $ C^* $-norm
on the algebraic tensor product $ \cE \otimes \cM $.
So, the inclusions of $ \cM $ and $ \cE $ into any $ \cA $ fulfilling the above conditions
factor through $ \cE \mxtens \cM $.
Then, the (slow) dynamics of the universe are generated by
\[ \edf := \frac{1}{\ep^2} \fk + \frac{1}{\ep^2} \fz + \frac{1}{\ep} \sum_{\alpha=1}^{d_E} i[\Ea \ha, \cdot] \]
where $ \fk $ and $ \fz $ must be lifted to $ \cA $.
Because the sums in \cref{eq:schro-qsde-fast} and \cref{eq:schro-qsde} include Hermitian conjugates,
without loss of generality, one may instead assume $ \Ea $ and $ \ha $ are self-adjoint simplifying the sum.
As above, depending on the choice of $ \fk $, $ \fz $, and other parameters,
there may exist a flow of $ {}^* $-algebra homomorphisms $ \{\exi_t : \cA \to \cA\}_{t \in \bR} $
generated by $ \edf $ such that $ \frac{\dr}{\dr[t]} \exi_t = \exi_t \circ \edf $.
In the above language, if a unitary evolution $ \eps{U}_\bt $ generated by $ \eps{H}_\fast $ exists
then we can write $ \exi_t(X) := \eps{U}_{-t / \ep^2} X \eps{U}_{t / \ep^2} $ for all $ t \in \bR $.
\bigskip

If $ \{\zeta_t : \cE \to \cE\}_{t \in \bR} $ is the flow generated by $ \fz $
then one can correct the flow $ \exi_t $ by $ \zeta_{-t / \ep^2} $ to obtain
\begin{align}
\frac{\dr}{\dr[t]} \exi_t \zeta_{-t / \ep^2}
&= \exi_t \edf \zeta_{-t / \ep^2} - \frac{1}{\ep^2} \exi_t \fz \zeta_{-t / \ep^2}
= \exi_t \zeta_{-t / \ep^2} \circ
    \zeta_{t / \ep^2} \p{\edf - \frac{1}{\ep^2} \fz} \zeta_{-t / \ep^2} \notag \\
\text{where} \quad
\label{eq:stoch-qsde-no-slow}
\edf_t &:= \zeta_{t / \ep^2} \p{\edf - \frac{1}{\ep^2} \fz} \zeta_{-t / \ep^2} 
= \frac{1}{\ep^2} \fk + \frac{1}{\ep} \sum_{\alpha=1}^\infty i[\Ea_{t / \ep^2} \ha, \cdot]
\quad \text{and} \,\, \Ea_{t / \ep^2} := \zeta_{t / \ep^2}(\Ea)
\end{align}
eliminating $ \fz $ at the cost of the time-inhomogeneous $ \Ea_t $.
So, $ \edf_t $ is the (right) generator for $ t \mapsto \exi_t \zeta_{-t / \ep^2} $.
Note that because $ \edf_t $ is time-dependent,
it no longer commutes with $ \exi_t \zeta_{-t / \ep^2} $ in general.
Now, it is possible to take $ \Ea_t $ to be defined in any way
not merely as the flow of an operator by some unitary dynamics.
In particular, one could take $ t \mapsto \Ea_t $ to be an arbitrary
(i.e. potentially non-Markovian) continuous collection of operators.
Additionally, we may include a component of the system Hamiltonian
which evolves at the same speed as the limiting process:
\begin{equation} \label{eq:stoch-gen-qsde}
\edf_t := \fks + \frac{1}{\ep^2} \fkf + \frac{1}{\ep} \sum_{\alpha=1}^{d_E} i[\Ea_{t / \ep^2} \ha, \cdot]
\end{equation}
where $ \fks $ and $ \fkf $ are $ {}^* $-derivations on $ \cM $
while for each $ \alpha $, $ \{\Ea_t\}_{t \in \bR} $ is a
continuous collection of (potentially unbounded) elements of $ \cE $.
The $ \fks $ term lies outside of the traditional stochastic limit regime.
However, it arises naturally in the classical context.
The current author and Wehr proved the weak convergence of solutions
to a classical analogue of \cref{eq:stoch-gen-qsde} as an SDE with $ \fkf = 0 $
where the driving processes $ \Ea $ were functions of an It\^o diffusion \cite{RW26}.
In \cref{subsect:qu-sdes}, the description of a quantum SDE will be introduced.
The flow $ \{\exi_t\} $ may be thought of as a general solution
(described in \cref{subsect:gen-sol-space}) to the QSDE \cref{eq:stoch-limit}.
\medskip

\tableofcontents

\section{Quantum Topological \& Probability Spaces}
\label{sect:qu-top-prob}  

As stated in \cref{subsect:stoch-as-flow},
a central motivation for the language used in this work
is to illustrate the connection between quantum and classical SDEs.
The basis of this connection is the Gelfand representation (or first Gelfand-Naimark theorem)
which establishes a bijective correspondence between
locally compact Hausdorff (LCH) topological spaces and commutative $ C^* $-algebras.
In particular, for any commutative $ C^* $-algebra, $ \cM $, we take $ \hcM $ to be the set of
non-zero $ {}^* $-homomorphisms $ \hat{x} : \cM \to \mathbb{C} $.
Then, $ \hcM $ endowed with the weak* topology is an LCH space.
Conversely, for any LCH space $ X $, we let $ C_0(X) $
be the space of (complex) continuous functions on $ X $ vanishing at infinity.
Then, $ C_0(X) $ endowed with the uniform norm is a $ C^* $-algebra.
In \cite{DB86}, Doran and Belfi provide a history of the Gelfand-Naimark theorems and their wider impact
(the second theorem being the representation of any $ C^* $-algebra
as a norm-closed $ {}^* $-subalgebra of operators on a Hilbert space).
If we allow $ \cM $ to be a non-commutative $ C^* $-algebra
then we obtain a ``quantum" analogue of an LCH space.
\medskip

\begin{defn} \label{defn:qu-top}
If $ \cM $ is a separable $ C^* $-algebra
then we define the corresponding \textbf{quantum topological space (QTS)} $ \hcM $
as the set of isomorphism classes of non-zero topologically irreducible $ {}^* $-representations
(irreps) $ \hat{x} : \cM \to B(\cH_{\hat{x}}) $.
We also define $ \chcM $ as the collection of isomorphism classes
of non-degenerate separable $ {}^* $-representations
so that $ \hcM \subseteq \chcM $.
For any $ S \subseteq \chcM $, we define the \textbf{hull-kernel closure} of $ S $ as
\[
\wbar{S} := \left\{ \hat{y} \in \chcM \,:\,
\bigcap_{\hat{x} \in S} \ker(\hat{x}) \subseteq \ker(\hat{y}) \right\}.
\]
Then, the \textbf{hull-kernel topology} on $ \chcM $
is the topology generated by this closure operation.
We will call $ \hcM $ \textbf{unital} when $ \cM $ is unital.
\end{defn}

It should be remarked that this definition of the hull-kernel topology
is essentially an operator algebraic version of the Zariski topology on the spectra of rings.
Moreover, the hull-kernel topology in \cref{defn:qu-top} has a close relationship
with the weak* topology on the dual space $ \cM^* $ (see \cite{Fell60}).
In this context, a version of the first Gelfand-Naimark theorem exists which
associates $ \cM $ with a certain space of functions on $ \hcM $ (see \cite{Fuji98}).
One may also construct a \textbf{quantum metric space} from a QTS
by choosing an appropriate convex subset of $ \cM $ to be the $ 1 $-Lipschitz functions on $ \hcM $
(see \cite{Rief04} and \cite{Latr16} for more about this theory).
For the remainder of the work, we assume that all quantum topological spaces are unital
so $ \cM $ is unital and separable.
Note that $ \hcM $ being unital implies that it is compact (see Proposition 3.1.8 of \cite{Dixm77}).
\medskip

Next, we consider the morphisms between QTSs.
If $ X $ and $ Y $ are classical LCH spaces
then for any \textsl{proper continuous} function $ \gamma : X \to Y $,
one has the pullback $ \what{\gamma} : C_0(Y) \to C_0(X) $
defined by $ \what{\gamma}(f) := f \circ \gamma $.
One may generalize this approach to the quantum context
to define mappings between QTSs.
\medskip

\begin{defn} \label{defn:qu-map}
Suppose $ \hcM_1 $ and $ \hcM_2 $ are quantum topological spaces.
We say $ \gamma : \hcM_1 \to \chcM_2 $ is a \textbf{quantum map}
when there exists a norm-one $ {}^* $-algebra homomorphism $ \what{\gamma} : \cM_2 \to \cM_1 $
such that $ \wbar{\hat{x} \circ \what{\gamma}} = \gamma(\hat{x}) $
for all $ \hat{x} \in \hcM_1 $
where $ \wbar{\hat{x}} $ indicates the non-degenerate part of the representation.
\end{defn}

Note that when such a $ \what{\gamma} $ exists, it uniquely determines $ \gamma $.
Although one might be inclined to think of the elements of $ \hcM $ as points,
this analogy is not entirely accurate.
For example, if $ \hat{x} \in \hcM $ is an irrep and $ \gamma $ is a quantum map
then $ \gamma(\hat{x}) $ may be reducible.
\Cref{defn:qu-map} may be viewed as establishing a contravariant functor
from the category of $ C^* $-algebras into the category of quantum topological spaces
which extends the contravariant functor given by the Gelfand representation.
(Note: One must restrict the morphisms of the category of LCH spaces
to proper maps for this to work.
However, for compact Hausdorff spaces these are precisely the continuous functions.)
\medskip

A similar correspondence exists for probability spaces.
To wit, for any probability space $ (\Omega, \cF, \bP) $ with expectation $ \bE $,
we take $ L^\infty := L^\infty(\Omega, \cF, \bP) $ to be
the $ \bP $-essentially bounded $ \cF $-measurable complex functions on $ \Omega $.
Then, $ L^\infty $ is a commutative $ W^* $-algebra
with predual $ L^1 \cong L^\infty_* $ (meaning $ (L_1)^* \cong L_\infty $)
so that $ \bE \in L^\infty_* $ namely $ \bE $ corresponds to $ 1 \in L_1 $.
Conversely, let $ \cA $ be a commutative $ W^* $-algebra with
a norm-one positive normal linear functional $ \nu \in \cA_* $.
One may define $ \cF_\cA $ to be the lattice of projections of $ \cA $
and $ \Omega_\cA $ to be the ultrafilters of the lattice $ \cF_\cA $.
Thus, $ \cA \cong L^\infty(\Omega_\cA, \cF_\cA, \nu) $.
\medskip

\begin{defn} \label{defn:qu-prob}
Let $ \cA $ be a $ W^* $-algebra
then we define the corresponding \textbf{quantum measurable space (QMS)} to be $ \hcA $.
Next, let $ \nu \in \cA_* $ (where $ \cA_* $ is the predual of $ \cA $)
be a norm-one positive normal linear functional
then $ (\hcA, \nu) $ is a \textbf{quantum probability space (QPS)}.
When clear from context, we will omit $ \nu $ and use $ \hcA $ to refer
to the quantum probability space.
\end{defn}

An exploration of this approach can be found in \cite{Chan15}.
However there, a quantum probability space is defined as $ (\cA, \nu) $.
Here, we use $ (\hcA, \nu) $ to try to emphasize the association with the classical case.
For any $ W^* $-subalgebra $ \cB $ of $ \cA $,
one can define a conditional expectation with respect to $ \cB $
as a $ \cB $-bimodular norm-one $ {}^* $-linear projection $ A \mapsto \meanWhen{A}{\cB} : \cA \to \cB $
such that $ \mean{\meanWhen{A}{\cB}} = \mean{A} $ for all $ A \in \cA $.
This construction was introduced and proven to exist and be unique
for faithful tracial normal states by Umegaki \cite{Umeg54}.
For faithful non-tracial states,
the conditional expectation may not exist, but will be unique if it does.
The conditions under which existence holds
were investigated by Accardi \& Cecchini among others \cite{AC82}.
For non-faithful non-tracial states, it may not exist and not be unique.
\medskip

Classical random variables valued in a topological space
are merely measurable functions from a probability space into that topological space.
Hence, we define quantum random variables as quantum maps from a QPS to a QTS.
Then, stochastic processes are indexed collections of such maps.
\medskip

\begin{defn} \label{defn:qu-rv-sproc}
Suppose $ \hcM $ is a quantum topological space and $ \hcA $ is a quantum probability space.
Then, a \textbf{$ \hcM $-valued quantum random variable}
is a quantum map $ X : \hcA \to \chcM $ with pullback $ \hX : \cM \to \cA $
which is a norm-one $ {}^* $-algebra homomorphism.
By abuse of notation, we use $ f(X) $ to mean $ \what{X}(f) $ for all $ f \in \cM $.
Naturally, a \textbf{$ \hcM $-valued quantum stochastic process}
is a collection of quantum random variables $ \{X_t : \hcA \to \chcM\}_{t \in T} $
with pullbacks $ \{\hX_t : \cM \to \cA\}_{t \in T} $
for some $ T \subseteq \bR $.
We write $ f(X_t) $ to mean $ \hX_t(f) $.
\end{defn}

If $ X $ is real-valued quantum random variable
then $ \hX : C_0(\bR) \to \cA $ is a $ {}^* $-algebra homomorphism.
So, we may lift $ \hX $ to a map between the enveloping $ W^* $-algebras: $ \hX^{**} : C_0(\bR)^{**} \to \cA^{**} $
(see section 2 of chapter 3\ in \cite{Take02}
for basic properties of the enveloping $ W^* $-algebra).
However, $ \cA $ is a $ W^* $-algebra so $ \hX^{**} : C_0(\bR)^{**} \to \cA $.
By the Riesz-Markov theorem, $ C_0(\bR)^* $ is isomorphic to the space of regular Borel measure.
Then, for every Borel set $ E \subseteq \bR $, the indicator function $ \chi_E $
is a bounded linear functional on the space of regular Borel measures by integration
so $ \chi_E \in C_0(\bR)^{**} $.
Thus, $ E \mapsto \hX^{**}(\chi_E) $ is a projection-valued measure $ P_X $
and we can say $ \wtld{X} := \int_\bR x \dr[P_X(x)] \in \cA $ so that $ \wtld{X} $ is self-adjoint.
Conversely, for any $ f \in C_0(\bR) $, by the spectral theorem, $ f(\wtld{X}) \in \cA $
so we may define $ \hX(f) := f(\wtld{X}) $.
Therefore, one can represent a real-valued quantum random variable as
a (potentially unbounded) self-adjoint element of $ \cA $.
\medskip

\begin{defn} \label{defn:filter-markov}
Suppose $ (\hcA, \bE) $ is a QPS.
If $ \{\cF_t\}_{t \in T} $ (for some $ T \subseteq \bR $)
is a collection of $ W^* $-subalgebras of $ \cA $
such that $ \cF_s \subseteq \cF_t $ for all $ s \leq t $
then we say $ \cF_\bt $ is a \textbf{filtration} on $ \hcA $
and $ (\hcA, \bE, \cF_\bt) $ is a \textbf{filtered QPS}.
\medskip

Now, suppose $ \{X_t\}_{t \in T} $ is an $ \hcM $-valued stochastic process on $ \hcA $.
We say $ X $ is $ \cF_\bt $-adapted if $ f(X_t) \in \cF_t $ for all $ f \in \cM $ and $ t \in T $.
Additionally, an $ \cF_\bt $-adapted process $ X $ is called a \textbf{Markov} process
if for all $ s, t \in T $ with $ s \leq t $
and any $ f \in \cM $, one has $ \meanWhen{f(X_t)}{\cF_s} = \meanWhen{f(X_t)}{\hX_s(\cM)} $
where we assume the conditional expectations exist and are unique.
\end{defn}

If $ \{X_t\}_{t \geq 0} $ is a Markov process and $ 0 \leq s \leq t $
then $ \meanWhen{f(X_t)}{\cF_s} \in \hX_s(\cM) $ 
so there exists $ P_{s \gets t} f \in \cM $
such that $ (P_{s \gets t} f)(X_s) = \meanWhen{f(X_t)}{\cF_s} $.
Then, the generator of $ X $ at time $ s $ is $ \at{\frac{\dr}{\dr t} P_{s \gets t}}_{t = s} $
where the derivative is taken pointwise in $ \cM $.
More about quantum Markov processes may be found in chapter 7 of \cite{Chan15}.
\medskip

\subsection{Quantum SDEs}
\label{subsect:qu-sdes}  

Suppose $ (\hcA, \bE) $ is a quantum probability space
and $ \hcM $ is a quantum topological space.
We will be interested in potentially unbounded random variables.
To this end, we introduce a norm $ p $ on $ \cA $
such that for any $ A \in \cA $, $ |\mean{A}| \leq p(A) $
Additionally, we assume that there exists a norm $ p_\ell $
such that $ p = p_\ell \lor p_\ell^* $ (where $ p_\ell^*(A) := p_\ell(A^*) $)
and $ p_\ell $ is left-modular
meaning $ p_\ell(AB) \leq \|A\| \,\, p_\ell(B) $ for all $ A, B \in \cA $.
We use $ \cA[p] $ to refer to the completion of $ \cA $
with respect to the norm $ p $.
\bigskip

We are interested in quantum analogues of ODEs and SDEs.
Suppose $ E_\bt := \{E_t\}_{t \geq 0} $ is a real quantum stochastic process
represented as a collection of self-adjoint elements of $ \cA[p] $.
Also, assume $ \df $ is a $ {}^* $-derivation on $ \cM $.
Intuitively, we might attempt to construct a QSDE with $ E $ as the driving process
and $ \df $ as the vector field along which it acts:
\[ \begin{split}
\dr[f(X_t)] &= E_t \cdot (\df f)(X_t) \dr[t] \\
\text{which is convenient notation for} \quad
\hX_t(f) &= \int_0^t \dr[s] E_s \cdot \hX_s(\df f)
\quad \text{for all} \;\; t \geq 0
\end{split} \]
where the integral is $ p $-convergent.
However, the noncommutativity of the collection $ E_\bt $
can introduce difficulties to this definition.
To wit, if $ E_\bt $ does not fully commute with itself
then, in general, for $ t > 0 $, $ E_t $ will not commute with $ (\df f)(X_t) $.
Because of this, $ f \mapsto E_t \cdot (\df f)(X_t) $
will not behave as a derivation in general
(specifically, a derivation onto $ \cA $ treated as an $ \cM $-module using $ \hX_t $)
and so $ f \mapsto f(X_t) $ will fail to be a $ {}^* $-algebra homomorphism.
\bigskip

This difficulty arises because, in a noncommutative context,
the act of observing the values of $ E_\bt $ forces them to change.
In physics, this is sometimes referred to as a \textsl{quantum back action}.
To resolve this, we must account for the back action on $ E_\bt $.
However, this can only be done if $ \df $ is an inner derivation.
Otherwise, we need the driving process $ E_\bt $ to be simultaneously measurable
(i.e. commute) with itself and any other coefficients present.
\bigskip

Suppose $ \{\Ea\}_{\alpha \in \bN} $ and $ \{\Jb\}_{\beta \in \bN} $
are collections of real quantum processes
so that $ \Ea_t, \Jb_t \in \cA[p] $ for all $ t \in \bR $.
In this work, we use $ \bN $ to mean $ \{1, 2, \ldots\} $.
We use $ \cE \subseteq \cA $ to denote the $ W^* $-subalgebra
generated by $ \{\Ea_t\}_{\alpha \in \bN, t \geq 0} $
and $ \{\Jb_t\}_{\beta \in \bN, t \geq 0} $
(i.e. the minimal $ W^* $-subalgebras whose $ p $-completion contains the elements).
We require that for all $ t $ and $ \beta $ that $ \Jb_t \in Z(\cE) $
where $ Z(\cdot) $ denotes the center.
This is because $ \{\Jb\} $ will be the coefficients
of potentially outer derivations in $ \cM $ and, as discussed above,
these must commute with the other coefficients.
\bigskip

Additionally, let $ \{\ha\}_{\alpha \in \bN} \subseteq \cM $
be a collection of self-adjoint elements
and $ \{\frb\}_{\beta \in \bN} $ be a collection of
(potentially unbounded) $ {}^* $-derivations on $ \cM $
which generate flows $ \{\rho^\beta_t : \cM \to \cM\}_{\beta=1}^\infty $
(i.e. $ {}^* $-algebra homomorphisms).
Assume $ X_\init $ is a quantum map from $ \hcA $ to $ \hcM $
such that $ X_\init $ is independent of $ \cE $.
That is, for all $ f \in \cM $ and $ A \in \cE $,
\begin{equation} \label{eq:init-indep}
[A, f(X_\init)] = 0
\quad \text{and} \quad
\mean{A f(X_\init)} = \mean{A} \mean{f(X_\init)}.
\end{equation}
Then, we say an $ \hcM $-valued quantum process $ \{X_t\}_{t \geq 0} $ over $ \cA $
is a \textbf{solution} of the following QSDE with initial condition $ X_\init $
(with respect to $ p $) if $ X_0 = X_\init $ and there exists a collection
of $ {}^* $-algebra homomorphisms $ \{\vp_t : \dom(\vp_t) \to \cA[p]\}_{t \geq 0} $
with $ \cA \subseteq \dom(\vp_t) \subseteq \cA[p] $
such that for all $ t \geq 0 $,
\begin{equation} \label{eq:exmp-qsde} \begin{split}
\dr[f(X_t)] &= \sum_{\alpha=1}^\infty \vp_t(\Ea_t) \,\, (i [\ha, f])(X_t) \dr[t]
  + \sum_{\beta=1}^\infty \Jb_t \,\, (\frb f)(X_t) \dr[t]
\quad \text{for all} \;\; f \in \bigcap_\beta \dom(\frb) \\
\text{and} \quad \dr[\vp_t(A)] &= \sum_{\alpha=1}^\infty \vp_t(i [\Ea_t, A]) \,\, \ha(X_t) \dr[t]
\quad \text{for all} \;\; A \in \dom(\vp_t)
\end{split} \end{equation}
where the derivatives exists with respect to $ p $.
As a requirement of the existence of the solution,
we assume that $ \Ea_t \in \dom(\vp_t) $ for all $ \alpha $.
This construction ensures that for all $ f \in \cM $ and $ t \geq 0 $,
$ f(X_t) $ commutes with $ \Jb_s $ and $ \vp_t(\Ea_s) $
for any $ s $, $ \alpha $, and $ \beta $.
\medskip

\subsection{General Solution Space}
\label{subsect:gen-sol-space}  

We are interested in solving QSDEs like \cref{eq:exmp-qsde}
for arbitrary initial conditions and arbitrary quantum probability spaces.
To this end, we introduce a $ C^* $-algebra which allows us to work
with all initial conditions simultaneously.
In fact, this algebra is a maximal tensor product like
the one mentioned in \cref{subsect:stoch-as-flow}.
\medskip

Suppose $ \{\Ea\} $, $ \{\ha\} $, $ \{\Jb\} $, and $ \{\frb\} $
as well as $ \cE $ are defined the same as above.
Then, we use $ \tcA := \cE \mxtens \cM $
to denote the \textbf{general solution space} for \cref{eq:exmp-qsde}.
For each $ \beta $, we may extend the flow $ \rhob_\bt : \cM \to \cM $ to $ \tcA $
by saying $ \trhob_t(A \otimes f) := A \otimes \rhob_t(f) $
for all $ A \in \cE $ and $ f \in \cM $
and extending by linearity and closure.
Naturally, we use $ \tfrb $ to denote the generator of $ \trhob_\bt $
so that it extends $ \frb $.
\medskip

Let $ \{p_t\}_{t \geq 0} $ be an increasing collection
of (potentially unbounded) norms on $ \cE $
such that $ p_0 = p $ and $ p_t(E^*) = p_t(E) $ for all $ E \in \cE $.
Then, we say $ \tp_t := p_t \otimes \|\!\cdot\!\| $.
We use $ \tcA[\tp_t] $ to denote the completion
of $ \{A \in \tcA \,:\, \tp_t(A) < \infty\} $ with respect to $ \tp_t $.
Suppose $ \{\xi_{s \gets t} : \tcA[\tp_t] \to \tcA[\tp_s]\}_{0 \leq s \leq t} $
are bounded $ {}^* $-linear maps and
for all $ U \in \tcA[\tp_t] $, $ U \in \bigcap_{\tau \in [0, t]} \dom(\df_\tau) $
with $ \tau \mapsto \df_\tau(U) $ being $ \tp_t $-continuous for $ \tau \in [0, t] $ and
\begin{equation} \label{eq:defn-sgrpd} \begin{split}
\xi_{s \gets t}(U) &= U + \int_s^t \dr[\tau] \xi_{s \gets \tau}(\df_\tau(U)) \\
\text{where} \quad
\df_t(U) &:= \sum_{\alpha=1}^\infty i [\Ea_t \otimes \ha, U]
  + \sum_{\beta=1}^\infty (\Jb_t \otimes 1) \,\, \tfrb(U)
\end{split} \end{equation}
and the integral converges in $ \tp_s $.
Further, we assume $ \xi_{s \gets t}(UV) = \xi_{s \gets t}(U) \xi_{s \gets t}(V) $
when the product $ UV \in \tcA[\tp_t] $ exists.
Note that this is a consequence of $ \df_t $ being a $ {}^* $-derivation.
Because of \cref{prop:gen-sol-to-part} below,
we call $ \{\xi_{s \gets t}\} $ the \textbf{general solution} to \cref{eq:exmp-qsde}.
If a collection $ \{\xi_{s \gets t}\}_{0 \leq s \leq t \leq \bar{t}} $
fulfilling \cref{eq:defn-sgrpd} exists
then we say \cref{eq:exmp-qsde} is \textbf{generally solvable}.
\medskip

\begin{prop} \label{prop:gen-sol-to-part}
Suppose $ X_\init $ is a $ \hcM $-valued quantum random variable over $ \cA $
that commutes with $ \cE \subseteq \cA $
and $ \{\xi_{s \gets t} : \tcA[\tp_t] \to \tcA[\tp_s]\}_{0 \leq s \leq t} $
fulfills \cref{eq:defn-sgrpd}.
We say $ D_t := \{A \in \cE[p] \,:\, \tp_t(A \otimes 1) < \infty\} $ for all $ t \geq 0 $.
Then, we define the $ \hcM $-valued process $ \{X_t\}_{t \geq 0} $
and homomorphisms $ \{\vp_t : D_t \to \cA[p]\}_{t \geq 0} $ as
\begin{equation} \label{eq:defn-part-sol} \begin{split}
f(X_t) &:= \wbar{X}_\init(\xi_{0 \gets t}(1 \otimes f))
\quad \text{for all} \;\; f \in \cM \\
\text{and} \quad
\vp_t(A) &:= \wbar{X}_\init(\xi_{0 \gets t}(A \otimes 1))
\quad \text{for all} \;\; A \in D_t
\end{split} \end{equation}
where $ \wbar{X}_\init : \tcA \to \cA $
such that $ \wbar{X}_\init(A \otimes f) = A \hX_\init(f) $
for all $ A \in \cE $ and $ f \in \cM $.
Then, $ X_\bt $ and $ \vp_\bt $ are a solution to \cref{eq:exmp-qsde}.
\end{prop}
\begin{proof}
Since $ p_\ell $ is left-modular, we have
\[
p_\ell(\wbar{X}_\init(A \otimes f)) = p_\ell(f(X_0) A)
\leq \|f(X_0)\| \cdot p_\ell(A) = p_\ell(A) \|f\|
\]
for any $ A \in \cE $ and $ f \in \cM $
so $ p_\ell \circ \wbar{X}_\init \leq p_\ell \otimes \nrM{\cdot} $.
Since $ p_\ell^* $ is right-modular, a similar argument holds
so $ p \circ \wbar{X}_\init \leq p \otimes \nrM{\cdot} = \tp_0 $.
Thus, $ \wbar{X}_\init : \tcA[\tp_0] \to \cA[p] $ is a bounded operator.
So, for any $ f \in \bigcap_\beta \dom(\frb) $, we can write
\begin{align*}
\frac{\dr}{\dr[t]} f(X_t) &= \wbar{X}_\init\p{
    \frac{\dr}{\dr[t]} \xi_{0 \gets t}(1 \otimes f)
} \\
&= \wbar{X}_\init\p{\xi_{0 \gets t}\p{
    \sum_{\alpha=1}^\infty i [\Ea_t \ha, 1 \otimes f]
    + \sum_{\beta=1}^\infty \Jb_t \,\, \tfrb(1 \otimes f)
}} \\
&= \wbar{X}_\init\p{\xi_t\p{
    \sum_{\alpha=1}^\infty \Ea_t \,\, i[\ha, f]
    + \sum_{\beta=1}^\infty \Jb_t \,\, \frb(f)
}} \\
&= \sum_{\alpha=1}^\infty \wbar{X}_\init(\xi_{0 \gets t}(\Ea_t))
    \,\, \wbar{X}_\init\p{ \xi_{0 \gets t}(i [\ha, f]) }
+ \sum_{\beta=1}^\infty \wbar{X}_\init(\xi_{0 \gets t}(\Jb_t))
    \,\, \wbar{X}_\init(\xi_{0 \gets t}(\frb f)) \\
&= \sum_{\alpha=1}^\infty \vp_t(\Ea_t) \,\, i[\ha, f](X_t)
    + \sum_{\beta=1}^\infty \Jb_t \,\, (\frb f)(X_t).
\end{align*}
Additionally, $ f(X_0) = \wbar{X}_\init(\xi_{0 \gets 0}(1 \otimes f)) = \wbar{X}_\init(f) = f(X_\init) $.
Note that $ \xi_{0 \gets t}(\Jb_t) = \Jb_t $ since $ \Jb_t $
commutes with $ \{\Ea\} $ and is in the kernels of $ \{\tfrb\} $.
Similarly, for any $ A \in D_t $, $ A \otimes 1 \in \tcA[\tp_t] $ and
since $ \ha $ commutes with $ A $ and $ \tfrb(A \otimes 1) = 0 $,
\begin{align*}
\frac{\dr}{\dr[t]} \vp_t(A) &= \wbar{X}_\init\p{
    \frac{\dr}{\dr[t]} \xi_{0 \gets t}(A \otimes 1)
} \\
&= \wbar{X}_\init\p{\xi_{0 \gets t}\p{
    \sum_{\alpha=1}^\infty i [\Ea_t \ha, A \otimes 1]
    + \sum_{\beta=1}^\infty \Jb_t \,\, \tfrb(A \otimes 1)
}} \\
&= \wbar{X}_\init\p{\xi_{0 \gets t}\p{
    \sum_{\alpha=1}^\infty i [\Ea_t, A] \ha
}}
= \sum_{\alpha=1}^\infty \vp_t(i[\Ea_t, A]) \ha(X_t).
\end{align*}
Also, $ \vp_0(A) = \wbar{X}_\init(\xi_{0 \gets 0}(A \otimes 1)) = \wbar{X}_\init(A) = A $.
Therefore, $ X_\bt $ and $ \vp_\bt $ fulfill \cref{eq:exmp-qsde} with $ X_0 = X_\init $.
\end{proof}

\section{The Stochastic Limit as a limit of Quantum SDEs}
\label{sect:stoch-as-qsdes}  

We now formulate the stochastic limit as a limit of the solutions to QSDEs.
Suppose $ \fks $ and $ \fkf $ are $ {}^* $-derivations on $ \cM $,
$ \{\Ea\}_{\alpha \in \bN} $ are real quantum stochastic processes
represented by self-adjoint elements of $ \cA[p] $,
and $ \{\ha\}_{\alpha \in \bN} $ are (bounded) self-adjoint elements of $ \cM $.
We assume $ \fks $ and $ \fkf $ generate flows $ \kps_\bt : \cM \to \cM $
and $ \kpf_\bt : \cM \to \cM $, respectively.
For any $ \ep > 0 $ and initial condition $ X_\init $,
let $ \eX $ be the $ \hcM $-valued process
such that $ \eX_0 = X_\init $ and for all $ t \geq 0 $,
\begin{equation} \label{eq:stoch-limit} \begin{split}
\dr[f(\eX_t)] &= (\fks f)(\eX_t) \dr[t]
  + \frac{1}{\ep^2} (\fkf f)(\eX_t) \dr[t]
  + \frac{1}{\ep} \sum_{\alpha=1}^\infty
    i [\ha, f](\eX_t) \;\; \evp_t(\Ea_{t / \ep^2}) \dr[t] \\
&\text{for all} \;\; f \in \dom(\fks) \cap \dom(\fkf) \\
\text{and} \quad \dr[\evp_t(U)] &= \frac{1}{\ep} \sum_{\alpha=1}^\infty
\ha(\eX_t) \evp_t(i[\Ea_{t / \ep^2}, U]) \dr[t]
\quad \text{for all} \;\; U \in \dom(\evp_t).
\end{split} \end{equation}
Because we care about all initial conditions,
we employ the general solution space described in \cref{subsect:gen-sol-space}.
Our main interest is in the stochastic limit from \cref{eq:stoch-limit}.
For the remainder of the text, for all sufficiently small $ \ep > 0 $,
we will assume that \cref{eq:stoch-limit} is generally solvable.
We use $ \{\exi_{s \gets t}\}_{0 \leq s \leq t} $ to refer to its general solution
and $ \{\etp_t\}_{0 \leq t} $ to the controlling norms.
So, we may write $ \exi_t $ from \cref{subsect:stoch-as-flow} as $ \exi_t := \exi_{0 \gets t} $.
For convenience, we can treat $ \exi_{0 \gets t} $ as a collection of quantum maps.
That is, we say $ \etX_t $ is a quantum map from $ \htcA $ to $ \hcM $
given by $ f(\etX_t) := \exi_{0 \gets t}(1 \otimes f) $ for all $ f \in \cM $.
Then, $ f(\eX_t) = \wbar{X}_\init(f(\etX_t)) $.
\medskip

We are interested in the convergence in law of $ \eX $ as $ \ep \dto 0 $.
In principle, one may include $ \{\Jb\} $ terms in this limit as well,
but to limit the complexity of the calculation, we will omit them here.
In order for such a limit to exist, we must impose the following conditions.
\medskip

\begin{cond} \label{cond:Ea-cov-stat}
The processes $ \{\Ea\} $ are \textbf{covariance stationary}.
That is, for all $ \alpha, \beta \in \bN $ and $ 0 \leq s \leq t $, one has
\begin{equation} \label{eq:cov-stat}
\mean{\Ea_s \Eb_t} = \mean{\Ea_0 \Eb_{t - s}}.
\end{equation}
\end{cond}
\bigskip

We would like $ \{\ha\} $ to be eigenvectors of $ \fkf $.
However, $ \{\ha\} $ are self-adjoint
(to ensure that $ i[\Ea \ha, \cdot] $ are derivations)
while $ \fkf $ is the generator of a unital $ {}^* $-algebra homomorphism
which means $ \fkf $ has an imaginary spectrum.
Hence, $ \{\ha\} $ cannot be eigenvectors.
Instead, we may require that they be the real parts of eigenvectors.
\medskip

\begin{cond} \label{cond:ha-eigen}
The operators $ \fks $ and $ \fkf $ commute.
Also, there exist $ \{\hoa\}_{\alpha=1}^\infty \subseteq \cM $
and $ \{\wa\}_{\alpha=1}^\infty \subseteq \bR $
such that for each $ \alpha $, $ \ha = \Re(\hoa) $
and $ \hoa $ is an eigenvector of $ \fkf $
with eigenvalue $ i\wa $ so $ \fkf \hoa = i\wa \hoa $.
Without loss of generality, we may assume $ \wa \geq 0 $ for all $ \alpha $.
Moreover, abusing notation, there exist $ \|\ho\|_1, \|\fks \ho\|_1 \in \bRp $ such that
\[
\sum_{\alpha=1}^\infty \|\hoa\| \leq \|\ho\|_1
\quad \text{and} \quad
\sum_{\alpha=1}^\infty \|\fks \hoa\| \leq \|\fks \ho\|_1.
\]
\end{cond}
\medskip

The requirement that $ \fks $ and $ \fkf $ commute is not critical,
but significantly simplifies the calculations
and there are no notable situations of interest that require them not to commute.
In order for the fast dynamics of $ \fkf $ to be averaged out in the limit,
it is necessary to impose some separation conditions on
the frequencies $ \{\wa\} $ in \cref{cond:ha-eigen}.
However, these conditions need not be as stringent as discreteness of the spectrum.
\medskip

\begin{cond} \label{cond:ha-freq-sep}
For each $ \alpha \in \bN $, we define the separation of $ \wa $
from other frequencies as
\[ \Delta\wa := \inf\{|\wa - \wb| \,:\, \beta \in \bN, \wb \neq \wa\}. \]
Abusing notation, we require that there exists $ \|\ho\|_{\Delta,1} \in \bRp $ such that
\[ \sum_{\alpha=1}^\infty \frac{\|\hoa\|}{\Delta\wa} \leq \|\ho\|_{\Delta,1}. \]
\end{cond}
\medskip

Note that if the collection $ \{\wa\} $ is discrete then $ \Delta := \min_\alpha \Delta\wa > 0 $
meaning that one could take $ \|\ho\|_{\Delta,1} \leq \Delta^{-1} \|\ho\|_1 $
so that \cref{cond:ha-freq-sep} would be a consequence of \cref{cond:ha-eigen}.
\medskip

\subsection{Graph-based Control}
\label{subsect:graph-ctl}  

In order to control the growth of $ \exi_\bt $,
we must bound the correlations of $ \{\Ea\} $.
We will do this using a graph-based method
which is not unlike the cluster expansions used elsewhere in mathematical physics.
In the sequel, we take $ \cG $ to be a collection of (finite nonempty) simple graphs
such that for any $ G \in \cG $, its vertex set is $ V(G) := [1, |G|] $ and every vertex has an edge.
Additionally, we let $ \{\vartheta_G\}_{G \in \cG} $ be positive real coefficients.
\medskip

\begin{cond} \label{cond:graph-ctl}
As above, let $ p $ be a mean-bounding norm on $ \cE $
and $ \{\Ea\}_{\alpha=1}^\infty $
be a collection of real quantum processes on $ \cE $
with $ \Ea_t \in \cE[p] $ for all $ t \in \bR $.
For any times $ t_1, \ldots, t_n \geq 0 $
and any $ \alpha_1, \ldots, \alpha_n \in \bN $,
we enforce that
\begin{equation} \label{eq:graph-mean-ctl}
\left|\mean{E^{\alpha_1}_{t_1} \cdots E^{\alpha_n}_{t_n}}\right|
\leq \Theta_\cG(t_1, \ldots, t_n)
\quad \text{where} \quad
\Theta_\cG(t_1, \ldots, t_n) := \sum_{\substack{G \in \cG \\ |G| = n}}
  \vartheta_G \prod_{i \sim_G j} e^{-\eta |t_i - t_j|}
\end{equation}
with $ \sim_G $ being the adjacency relation on $ G $
and each edge appearing once in the product.
Further, to control the behavior of the covariance,
we must have that for any $ j \in [1, n) $,
if $ t_k \leq t_j \land t_{j+1} $ for all $ k \in [1, n] \setminus \{j, j+1\} $ then
\begin{equation} \label{eq:graph-cov-ctl} \begin{split}
&\ab{\mean{E^{\alpha_1}_{t_1} \cdots E^{\alpha_j}_{t_j} \p{
    E^{\alpha_j}_{t_j} E^{\alpha_{j+1}}_{t_{j+1}} - \mean{E^{\alpha_j}_{t_j} E^{\alpha_{j+1}}_{t_{j+1}}}
  } E^{\alpha_{j+2}}_{t_{j+2}} \cdots E^{\alpha_n}_{t_n}}} \\
&\qquad \leq e^{-\eta |t_{j+1} - t_j|} \cdot
    \Theta_\cG(t_1, \ldots, t_{j-1}, t_j \land t_{j+1}, t_{j+2}, \ldots t_n).
\end{split} \end{equation}
We also require that
\begin{equation} \label{eq:graph-inf-rad-conv}
\lim_{n \to \infty} \frac{1}{n} \sqrt[n]{\sum_{\substack{G \in \cG \\ |G| = n}} \vartheta_G} = 0
\end{equation}
where $ |G| $ and $ c(G) $ are, respectively,
the number of vertices and connected components in $ G $.
\end{cond}
\bigskip

To effectively make use of \cref{eq:graph-mean-ctl} and \cref{eq:graph-cov-ctl}
we introduce the power series $ (\ep, t) \mapsto \vartheta(\ep; t) $ given by
\begin{equation} \label{eq:vartheta-defn}
\vartheta(\ep; t) := \sum_{G \in \cG} \vartheta_G \frac{1}{(|G| - 1)!}
    2^{3|G| - c(G) - 2} \|\ho\|_1^{|G| - 1} \,\, \eta^{c(G) - |G|}
    \cdot t^{c(G) - 1} \ep^{|G| - 2c(G)}.
\end{equation}
The form of this power series is motivated by the calculations
in \cref{eq:vartheta-form0} and \cref{eq:vartheta-form1}.
For any graph $ G $ on $ [1, n] $, we also define
\begin{align} \label{eq:ThetaG-defn}
\Theta_G(t_1, \ldots, t_n) &:= \prod_{i \sim_G j} e^{-\eta |t_i - t_j|}
\quad \text{for all} \,\, t_1, \ldots, t_n \in \bR \\
\notag \text{such that} \quad \Theta_\cG(t_1, \ldots, t_n)
&= \sum_{\substack{G \in \cG \\ |G| = n}} \vartheta_G \Theta_G(t_1, \ldots, t_n).
\end{align}
Bounds involving $ \vartheta(\ep; t) $ and $ \Theta_G $
can be found in \cref{subsect:graph-ctl-props}.
\medskip

\subsection{Seminorms to Control Mean}
\label{subsect:mu-snorm}

We are interested in bounding the dependence between increments
of the solutions to \cref{eq:stoch-limit}.
To do this for all possible initial conditions,
we work with general solution space $ \tcA $
and we define the conditional expectation $ \meanM{\cdot} : \tcA \to \cM $
as $ \meanM{E \otimes f} = \mean{E} f \in \cM $
for all $ E \in \cE $ and $ f \in \cM $
then extend by linearity and closure.
We now introduce an appropriate seminorm on $ \tcA $
in order to reason about the correlations between the processes $ \{\Ea\} $.
\medskip

\begin{defn} \label{defn:mu-snorm}
For any sequence $ 0 \leq t_1 \leq \ldots \leq t_n $,
we define the following seminorm on $ \cE $:
\[ \begin{split}
\mu_n(t_1, \ldots, t_n; X) &:= \sup \Big\{
  \ab{\mean{
    E^{\alpha_1}_{t_{\sigma(1)}} \cdots E^{\alpha_j}_{t_{\sigma(j)}}
    X E^{\alpha_{j+1}}_{t_{\sigma(j+1)}} \cdots E^{\alpha_n}_{t_{\sigma(n)}}
  }} \\
  &\qquad \,\Big|\, \alpha_1, \ldots, \alpha_n \in [1, d_E], \,
  \sigma \text{ is a permutation on $ [1, n] $} \\
  &\qquad \text{such that} \,
  t_{\sigma(1)} \leq \ldots \leq t_{\sigma(j)} \,\, \text{ and } \,\,
  t_{\sigma(j+1)} \geq \ldots \geq t_{\sigma(n)}
\Big\}.
\end{split} \]
Now, for any time $ t \geq 0 $ and parameter $ \ep > 0 $,
we can define the seminorm $ \emu_t : \cE \to \ebRp $ as
\[
\emu_t(X) := \sum_{n=0}^\infty (2\ep \|\ho\|_1)^n
  \int_0^{t / \ep^2} \dr{t_n} \cdots \int_0^{t_2} \dr{t_1}
  \mu_n(t_1, \ldots, t_n; X).
\]
We also define $ \etmu_t := \emu_t \otimes \nrM{\cdot} $.
\end{defn}

To ensure that $ \etmu_t $ is finite for a sufficiently large set,
we bound it by $ \etp_t $.
The first of the following assumptions may be thought of
as a generalization of the requirement
that $ \mean{A} \leq p(A) $ for all $ A \in \cA $.
\medskip

\begin{cond} \label{cond:etmu-etp-bdd}
For any $ t \geq 0 $, there exists real $ C_t > 0 $
such that $ \etmu_t \leq C_t \etp_t $.
\end{cond}

\subsection{Interaction Picture}
\label{subsect:int-pic}  

To facilitate computation, we will shift \cref{eq:stoch-limit} to an interaction picture.
From \cref{cond:ha-eigen}, we know $ \fks $ and $ \fkf $ commute.
So, we can define $ \ekp_t := \kps_t \circ \kpf_{t / \ep^2} $ for all $ t \in \bR $
such that $ \ekp_\bt $ is the flow for $ \fks + \frac{1}{\ep^2} \fkf $.
Thus, for all $ f \in \dom(\fks) \cap \dom(\fkf) $,
\[
\dr[\ekp_t(f)] = \ekp_t((\fks + \ep^{-2} \fkf) f) \dr[t]
= (\fks + \ep^{-2} \fkf) \ekp_t(f) \dr[t].
\]
For every $ t \geq 0 $, we define $ \egX_t $ as
\begin{equation} \label{eq:int-proc-defn} f(\egX_t) = (\ekp_{-t} f)(\eX_t). \end{equation}
\medskip

From \cref{cond:ha-eigen}, for each $ \alpha $,
we know $ \hoa $ is an eigenvector of $ \fkf $
so we define $ \ghoa_t := \kps_t(\hoa) $ such that
\[
\ekp_t(\ha) = \kps_t(\kpf_{t / \ep^2}(\Re(\hoa)))
= \Re(\kps_t(e^{it\wa / \ep^2} \hoa))
= \Re(e^{it\wa / \ep^2} \ghoa_t).
\]
Then, \cref{eq:stoch-limit} becomes
\begin{align}
\dr[f(\egX_t)] &= \dr[(\ekp_{-t} f)(\eX_t)] \notag \\
&= -(\ekp_{-t} (\fks + \ep^{-2} \fkf) f)(\eX_t) \dr[t]
  + \Bigg( \big.
    ((\fks + \ep^{-2} \fkf) \ekp_{-t} f)(\eX_t) \dr[t] \notag \\
    &\qquad + \frac{1}{\ep} \sum_{\alpha=1}^\infty
      \evp_t(\Ea_{t / \ep^2}) \;\; (i [\ha, \ekp_{-t}(f)])(\eX_t) \dr[t]
  \Bigg) \notag \\
&= \frac{1}{\ep} \sum_{\alpha=1}^\infty
  \evp_t(\Ea_{t / \ep^2}) \;\; (i \ekp_{-t} [\ekp_t(\ha), f])(\eX_t) \dr[t] \notag \\
\label{eq:int-stoch-limit}
&= \frac{1}{\ep} \sum_{\alpha=1}^\infty
  \evp_t(\Ea_{t / \ep^2}) \;\; (i [\Re(e^{it\wa / \ep^2} \ghoa_t), f])(\egX_t) \dr[t]
\end{align}
effectively canceling the drift term.
Note that \cref{eq:int-proc-defn} is not quite the standard interaction picture used in quantum mechanics
which would instead apply $ \ekp_{-t} $ on the outside.
The only effect of this choice on the interaction picture generator is
to reverse the direction in which $ \ha $ is transformed
(so for the typical interaction picture,
one would have $ \ekp_{-t}(\ha) $ instead of $ \ekp_t(\ha) $).
\medskip

We can perform a similar modification to the general solution.
Since $ \ekp_\bt : \cM \to \cM $ is a bounded flow on $ \cM $,
we may extend it to a flow $ \etkp_\bt : \tcA \to \tcA $ on $ \tcA $
by defining $ \etkp_t(E \otimes f) = E \otimes \ekp_t(f) $.
We perform the analogous procedure on $ \kps_\bt $ and $ \kpf_\bt $
to obtain $ \tkps_\bt $ and $ \tkpf_\bt $, respectively.
Naturally, $ \tfks $ and $ \tfkf $ are the generators of $ \tkps_\bt $ and $ \tkpf_\bt $.
Because $ \tp_t = p_t \otimes \nrM{\cdot} $ and $ \nrM{\cdot} $ is invariant under $ \etkp_s $,
we have $ \tp_t \circ \etkp_s = \tp_t $.
Then, for every $ 0 \leq s \leq t $,
we define $ \egxi_{s \gets t} : \tcA[\tp_t] \to \tcA[\tp_s] $ as
\[ \egxi_{s \gets t}(U) := \etkp_s \xi_{s \gets t}(\etkp_{-t} U) \]
for all $ U \in \tcA[\tp_t] $.
Then, treating $ s $ as fixed, \cref{eq:defn-sgrpd} becomes
\begin{equation} \label{eq:int-stoch-limit-sgrpd} \begin{split}
\dr[\egxi_{s \gets t}(U)] &= \dr[\etkp_s \exi_{s \gets t}(\etkp_{-t} U)] \\
&= - \etkp_s \exi_{s \gets t}(\etkp_{-t} (\tfks + \ep^{-2} \tfkf) U) \dr[t]
  + \etkp_s \exi_{s \gets t}\p{
    (\tfks + \ep^{-2} \tfkf) \etkp_{-t} U
    + \frac{1}{\ep} \sum_{\alpha=1}^\infty
      i [\Ea_{t / \ep^2} \ha, \etkp_{-t} U]
  } \dr[t] \\
&= \etkp_s \exi_{s \gets t}\p{
    \frac{1}{\ep} \sum_{\alpha=1}^\infty
    i \etkp_{-t} [\Ea_{t / \ep^2} \ekp_t(\ha), U]
  } \dr[t] \\
&= \etkp_s \exi_{s \gets t}\p{\etkp_{-t}\p{
  \frac{1}{\ep} \sum_{\alpha=1}^\infty
    i [\Ea_{t / \ep^2} \Re(e^{it\wa / \ep^2} \ghoa_t)
, U]
}} \dr[t] \\
&= \egxi_{s \gets t}(\egdf_t U) \dr[t]
\end{split} \end{equation}
where, for sake of convenience, we define
\[
\egdf_t U := \frac{1}{\ep} \sum_{\alpha=1}^\infty
  i [\Ea_{t / \ep^2} \Re(e^{it\wa / \ep^2} \ghoa_t), U].
\]
As above, for each $ t \geq 0 $,
we introduce the quantum maps $ \egtX_t $ from $ \what{\tcA} $ to $ \hcM $
by saying $ f(\egtX_t) = \egxi_{0 \gets t}(1 \otimes f) $ for all $ f \in \cM $.
Then, $ f(\egX_t) = \wbar{X}_\init(f(\egtX_t)) $.
\medskip

\section{Results}
\label{sect:results}  

Before stating the convergence of the stochastic limit,
we derive the generator of the Markov semigroup that arises in the limit.
To do this, we introduce the well-known Lindblad operators \cite{Lind76}.
For any $ a, b \in \cM $,
we define the \textbf{Lindblad operator} $ \gksl_{a,b} $ (on observables)
associated with $ a $ and $ b $ as
\begin{equation} \label{eq:gksl-defn}
\gksl_{a,b}(f) := - a^* b f + 2a^* f b - f a^* b
\quad \text{for all} \;\; f \in \cM.
\end{equation}
Note that $ (a, b) \mapsto \gksl_{a,b} $ is sesquilinear
(i.e. conjugate linear in the first argument and linear in the second).
Moreover, it is conjugate symmetric in $ a $ and $ b $ so
\begin{equation} \label{eq:gksl-conj-symm}
\gksl_{a,b}(f)^* = (-a^* b f + 2a^* f b - f a^* b)^*
= -f^* b^* a + 2 b^* f^* a - b^* a f^* = \gksl_{b,a}(f^*)
\quad \text{for all} \;\; f \in \cM.
\end{equation}
It will also be useful to define the following:
\begin{equation} \label{eq:gksl-parts-defn}
\gksl^+_{a,b}(f) := \frac{\gksl_{a,b}(f) + \gksl_{b^*,a^*}(f)}{2}
\quad \text{and} \quad
\gksl^-_{a,b}(f) := \frac{\gksl_{a,b}(f) - \gksl_{b^*,a^*}(f)}{2}
\quad \text{for all} \;\; f \in \cM
\end{equation}
so that $ \gksl_{a,b} = \gksl^+_{a,b} + \gksl^-_{a,b} $.
Since $ \gksl $ is conjugate symmetric, $ \gksl^+ $ and $ \gksl^- $ are as well
so for all $ f \in \cM $,
\begin{equation} \label{eq:gksl-parts-conj-symm} \begin{split}
\gksl^+_{a,b}(f)^* &= \frac{\gksl_{a,b}(f)^* + \gksl_{b^*,a^*}(f)^*}{2}
= \frac{\gksl_{b,a}(f^*) + \gksl_{a^*,b^*}(f^*)}{2} = \gksl^+_{b,a}(f^*) \\
\gksl^-_{a,b}(f) &= \frac{\gksl_{a,b}(f)^* - \gksl_{b^*,a^*}(f)^*}{2}
= \frac{\gksl_{b,a}(f^*) - \gksl_{a^*,b^*}(f^*)}{2}
= \gksl^-_{b,a}(f^*).
\end{split} \end{equation}
\medskip

Now, we calculate what the coefficients in the generator will be.
For each $ \alpha, \beta \in \bN $ and $ \w \in \bR $, we define the functions
\[ \cab(\w) := \int_0^\infty \dr[\dl] e^{-i\dl\w} \mean{\Ea_0 \Eb_\dl}. \]
Next, we use $ \cab^R $ and $ \cab^I $
to refer to the Fourier transforms of the real and imaginary parts
of $ \mean{\Ea_0 \Eb_\dl} $:
\begin{align}
\label{eq:cab-re-defn}
&\begin{split}
\cab^R(\w) &:= \int_0^\infty \dr[\dl] e^{-i\dl\w} \Re\p{\mean{\Ea_0 \Eb_\dl}} \\
&= \frac{1}{2} \int_0^\infty \dr[\dl] e^{-i\dl\w} \p{\mean{\Ea_0 \Eb_\dl} + \mean{(\Eb_\dl)^* (\Ea_0)^*}} \\
&= \frac{1}{2} \int_0^\infty \dr[\dl] e^{-i\dl\w} \mean{\{\Ea_0, \Eb_\dl\}} \\
\end{split} \\
\label{eq:cab-im-defn}
&\begin{split}
\cab^I(\w) &:= \int_0^\infty \dr[\dl] e^{-i\dl\w} \Im\p{\mean{\Ea_0 \Eb_\dl}} \\
&= \frac{1}{2i} \int_0^\infty \dr[\dl] e^{-i\dl\w} \p{\mean{\Ea_0 \Eb_\dl} - \mean{(\Eb_\dl)^* (\Ea_0)^*}} \\
&= \frac{1}{2i} \int_0^\infty \dr[\dl] e^{-i\dl\w} \mean{[\Ea_0, \Eb_\dl]}
\end{split}
\end{align}
since $ \Ea $ and $ \Eb $ are self-adjoint.
Thus, $ \cab^R $ and $ \cab^I $ are both Hermitian functions of $ \w $
and $ \cab = \cab^R + i\cab^I $.
For each $ \w $, we further separate
the matrices $ (\cab^R(\w))_{\alpha,\beta} $ and $ (\cab^I(\w))_{\alpha,\beta} $
into their Hermitian and anti-Hermitian components:
\begin{align}
\label{eq:cab-re-parts-defn}
\cab^{R+}(\w) := \frac{\cab^R(\w) + (\cba^R(\w))^*}{2}
\quad &\text{and} \quad
\cab^{R-}(\w) := \frac{\cab^R(\w) - (\cba^R(\w))^*}{2i} \\
\label{eq:cab-im-parts-defn}
\cab^{I+}(\w) := \frac{\cab^I(\w) + (\cba^I(\w))^*}{2}
\quad &\text{and} \quad
\cab^{I-}(\w) := \frac{\cab^I(\w) - (\cba^I(\w))^*}{2i}.
\end{align}
For each time $ t \geq 0 $, we define the generator
\begin{align}
\label{eq:limit-int-gener-defn}
\gL_t &:= i \ad(\kps_t(H)) + \frac{1}{2} \sum_{\substack{\alpha, \beta \in \bN \\ \wa = \wb}} \p{
    (\cab^{R+}(\wa) - \cab^{I-}(\wa)) \gksl_{\ghoa_t,\ghob_t}
    + (\cab^{R+}(\wa) + \cab^{I-}(\wa)) \gksl_{(\ghob_t)^*,(\ghoa_t)^*}
} \\
\text{where} \quad H &:= -\sum_{\substack{\alpha, \beta \in \bN \\ \wa = \wb}} \p{
    \frac{\cab^{I+}(\wa) + \cab^{R-}(\wa)}{2} (\hoa)^* \hob
    + \frac{\cab^{I+}(\wa) - \cab^{R-}(\wa)}{2} \hob (\hoa)^*
}. \notag
\end{align}
Here, the dependence on $ t $ is an artifact of the interaction picture.
So, we also define
\begin{equation} \label{eq:limit-gener-defn} \begin{split}
L := \fks + i\ad(H) + \frac{1}{2} \sum_{\substack{\alpha, \beta \in \bN \\ \wa = \wb}} \p{
    (\cab^{R+}(\wa) - \cab^{I-}(\wa)) \gksl_{\hoa,\hob}
    + (\cab^{R+}(\wa) + \cab^{I-}(\wa)) \gksl_{(\hob)^*,(\hoa)^*}
}
\end{split} \end{equation}
so that $ \gL_t = \kps_t(L - \fks) $.
Note that while $ \gL_t $ is bounded, $ L $ is unbounded when $ \fks $ is.
\smallskip

\begin{lem} \label{lem:gener-alt-form}
We may rewrite the generator $ L $ and drift term $ H $ as
\begin{align}
\label{eq:gener-alt-form}
L &= \fks + i\ad(H) + \sum_{\substack{\alpha, \beta \in \bN \\ \wa = \wb}} \p{
    \cab^{R+}(\wa) \gksl^+_{\hoa,\hob} - \cab^{I-}(\wa) \gksl^-_{\hoa,\hob}
} \\
\label{eq:drift-alt-form}
H &= -\frac{1}{2} \sum_{\substack{\alpha, \beta \in \bN \\ \wa = \wb}} \p{
    \cab^{I+}(\wa) \{(\hoa)^*, \hob\} + \cab^{R-}(\wa) [(\hoa)^*, \hob].
}
\end{align}
Consequently, for any time $ t \geq 0 $, the interaction picture
generator $ \gL_t $ may be written as
\begin{equation} \label{eq:int-gener-alt-form}
\gL_t = i\ad(\kps_t(H)) + \frac{1}{2} \sum_{\substack{\alpha, \beta \in \bN \\ \wa = \wb}} \p{
    \cab^{R+}(\wa) \gksl^+_{\ghoa_t,\ghob_t} - \cab^{I-}(\wa) \gksl^-_{(\ghob_t)^*,(\ghoa_t)^*}
}.
\end{equation}
\end{lem}
\begin{proof}  
First, we show \cref{eq:drift-alt-form}:
\begin{align*}
H &= -\sum_{\substack{\alpha, \beta \in \bN \\ \wa = \wb}} \p{
    \frac{\cab^{I+}(\wa) + \cab^{R-}(\wa)}{2} (\hoa)^* \hob
    + \frac{\cab^{I+}(\wa) - \cab^{R-}(\wa)}{2} \hob (\hoa)^*
} \\
&= -\frac{1}{2} \sum_{\substack{\alpha, \beta \in \bN \\ \wa = \wb}} \p{
    \cab^{I+}(\wa) ((\hoa)^* \hob + \hob (\hoa)^*)
    + \cab^{R-}(\wa) ((\hoa)^* \hob - \hob (\hoa)^*)
} \\
&= -\frac{1}{2} \sum_{\substack{\alpha, \beta \in \bN \\ \wa = \wb}} \p{
    \cab^{I+}(\wa) \{(\hoa)^*, \hob\} + \cab^{R-}(\wa) [(\hoa)^*, \hob]
}.
\end{align*}
Second, we show \cref{eq:gener-alt-form}:
\begin{align*}
L - \fks - i\ad(H) &= \frac{1}{2} \sum_{\substack{\alpha, \beta \in \bN \\ \wa = \wb}} \p{
    (\cab^{R+}(\wa) - \cab^{I-}(\wa)) \gksl_{\hoa,\hob}
    + (\cab^{R+}(\wa) + \cab^{I-}(\wa)) \gksl_{(\hob)^*,(\hoa)^*}
} \\
&= \sum_{\substack{\alpha, \beta \in \bN \\ \wa = \wb}} \p{
    \cab^{R+}(\wa) \frac{\gksl_{\hoa,\hob} + \gksl_{(\hob)^*,(\hoa)^*}}{2}
    - \cab^{I-}(\wa) \frac{\gksl_{\hoa,\hob} - \gksl_{(\hob)^*,(\hoa)^*}}{2}
} \\
&= \sum_{\substack{\alpha, \beta \in \bN \\ \wa = \wb}} \p{
    \cab^{R+}(\wa) \gksl^+_{\hoa,\hob} - \cab^{I-}(\wa) \gksl^-_{\hoa,\hob}
}.
\end{align*}
Then, \cref{eq:int-gener-alt-form} follows from \cref{eq:gener-alt-form}
by $ \gL_t = \kps_t \circ (L - \fks) \circ \kps_{-t} $.
\end{proof}

\subsection{Time-Centered Weak Convergence}
\label{subsect:tc-weak-conv}  

We will now state the weak convergence results
for the solutions of \cref{eq:stoch-limit}.
Unfortunately, the methods used in this work do not provide
convergence for \textsl{all} moments ($ n $-point functions) of $ \eX $.
Instead, we will show convergence for moments
in which the times fulfill a certain ordering.
Despite not being as strong as full weak convergence,
this convergence mode is stronger than the semigroup convergence
typically found in the literature for the stochastic limit.
To wit, for any collection of times $ t_1, \ldots, t_n \geq 0 $,
we say $ t_1, \ldots, t_n $ are \textbf{time-centered} (or, simply, \textbf{centered})
if there exists an index $ k \in [1, n] $
such that $ t_1 \leq \ldots \leq t_k $ and $ t_k \geq t_{k+1} \geq \ldots \geq t_n $.
Note that if $ t_1, \ldots, t_n $ are ordered
(either ascending or descending) then they are centered.
Suppose $ \{X_t\}_{t \geq 0} $ is a $ \hcM $-valued stochastic process
on a quantum probability space $ (\hcA, \bE) $.
Then, for any $ t_1, \ldots, t_n \geq 0 $
and $ f_1, \ldots, f_n \in \cM $,
we say $ \mean{f_1(X_{t_1}) \cdots f_n(X_{t_n})} $
is a \textbf{time-centered moment} of $ X_\bt $
if $ t_1, \ldots, t_n $ are centered.
\medskip

\begin{rmk} \label{rmk:tc-justify}
Such moments arise naturally when considering sequential measurements of a quantum system.
Imagine $ X_t $ represents a quantum system at time $ t $
so that for each self-adjoint $ f \in \cM $, the self-adjoint element $ f(X_t) \in \cA $
corresponds to an observable quantity of the system at time $ t $.
Imagine we measure some sequence of quantities, $ f_1, \ldots, f_n \in \cM $,
about the system at times $ t_1 \leq \ldots \leq t_n $.
For sake of simplicity, we assume all of the measurements
take values in a finite set $ \Lambda \subseteq \bR $
so that $ \sigma(f_k) \subseteq \Lambda $ for all $ k \in [1, n] $
Then, we use $ \chi_\lambda(f_k) \in \cA $ to denote the spectral projection
of $ f_k $ onto $ \lambda \in \Lambda $.
We also assume that the expectation is given by
a pure state: $ \mean{A} := \ang{\psi, A \psi} $ for all $ A \in \cA $
(where we have implicitly fixed a faithful representation for the algebra $ \cA $).
\medskip

Now, imagine that the first $ (k - 1) $ measurements
produce values $ \lambda_1, \ldots, \lambda_k \in \Lambda $
and for each $ i \in [1, k) $, we use $ p_i \in \cM $ to denote
the spectral projection of $ f_i $ onto $ \lambda_i $.
Thus, by the Born rule for quantum measurements,
the (unnormalized) state of the system after the first $ (k - 1) $ measurements would be
\[ \psi' = p_{k-1}(X_{t_{k-1}}) \cdots p_1(X_{t_1}) \psi. \]
So, the expected value of the $ k $-th measurement would be
\begin{align*}
\frac{\ang{\psi', f_k(X_{t_k}) \psi'}}{\ang{\psi', \psi'}}
&= \frac{\ang{p_{k-1}(X_{t_{k-1}}) \cdots p_1(X_{t_1}) \psi,
    f_k(X_{t_k}) p_{k-1}(X_{t_{k-1}}) \cdots p_1(X_{t_1}) \psi}}{
    \|p_{k-1}(X_{t_{k-1}}) \cdots p_1(X_{t_1}) \psi\|^2} \\
&= \frac{\mean{p_1(X_{t_1}) \cdots p_{k-1}(X_{t_{k-1}}) f_k(X_{t_k}) p_{k-1}(X_{t_{k-1}}) \cdots p_1(X_{t_1})}}
    {\mean{p_1(X_{t_1}) \cdots p_{k-1}^2(X_{t_{k-1}}) \cdots p_1(X_{t_1})}}.
\end{align*}
Hence, the expected value is always a time-centered moment of the process.
\end{rmk}
\bigskip

Naturally, we say a collection of processes converges time-centered weakly
if we have convergence for the time-centered moments
which is uniform in $ f_1, \ldots, f_n $.
\medskip

\begin{defn} \label{defn:tc-weak-conv}
Let $ (\cA, \mathbb{E}) $ be a quantum probability space, $ b \in \ebRp $, and $ \dl > 0 $.
Suppose $ \{X_t\}_{t \in [0, b]} $ is a $ \hcM $-valued quantum stochastic process
while $ \{\{\eX_t\}_{t \in [0, b]} \,:\, \ep \in (0, \dl)\} $
is a collection of $ \hcM $-valued quantum stochastic processes.
Then, we say the processes $ \eX_\bt $ \textbf{converge time-centered weakly}
to $ X_\bt $ as $ \ep \dto 0 $ if for all centered $ t_1, \ldots, t_n \in [0, b] $,
\[
0 = \lim_{\epsilon \dto 0} \sup_{\substack{
  f_1, \ldots, f_n \in \cM \\ \|f_1\| = \ldots = \|f_n\| = 1
}}
\ab{\big.
  \mean{f_1(\eX_{t_1}) \cdots f_n(\eX_{t_n})}
  - \mean{\Big. f_1(X_{t_1}) \cdots f_n(X_{t_n})}
}.
\]
\end{defn}
\medskip

Suppose $ (\cB, \mathbb{E}_B, \{\cF_t\}_{t \in [0, \infty)}) $
is a filtered quantum probability space
and $ \{\gX_t\}_{t \in [0, b]} $ is a $ \hcM $-valued quantum Markov process
on $ \cB $ adapted to $ \cF_\bt $
whose generator at time $ t $ is $ \gL_t $.
That is, for all $ f \in \cM $,
\begin{equation} \label{eq:Y-markov-gener}
(\gL_t f)(\gX_t) = \lim_{\tau \dto t}
    \frac{\meanWhen[B]{f(\gX_\tau)}{\cF_t} - f(\gX_t)}{\tau - t}.
\end{equation}

\begin{thm} \label{thm:int-weak-conv}
If $ \gX_0 = \egX_0 $ for all $ \ep > 0 $,
$ \{\ha\} $ fulfill \cref{cond:ha-eigen} and \cref{cond:ha-freq-sep},
$ \{\Ea\} $ fulfill \cref{cond:Ea-cov-stat} and \cref{cond:graph-ctl},
and \cref{cond:etmu-etp-bdd} holds
then $ \{\egX_t\}_{t \in [0, \bar{t}]} \to \{\gX_t\}_{t \in [0, \bar{t}]} $
time-centered weakly for any $ \bar{t} \geq 0 $ as $ \ep \dto 0 $.
\end{thm}

We may also remove the interaction picture to obtain
the (time-centered) weak limit of the unmodified system.
Suppose $ \{X_t\}_{t \in [0, \bar{t}]} $ is a $ \hcM $-valued quantum Markov process
on $ \cB $ adapted to $ \cF_\bt $ whose generator at all times is $ L $.
\smallskip

\begin{coro} \label{coro:weak-conv}
If $ X_0 = \eX_0 $ for all $ \ep > 0 $
and the other conditions of \cref{thm:int-weak-conv} hold
then $ \{\eX_t\}_{t \in [0, \bar{t}]} \to \{X_t\}_{t \in [0, \bar{t}]} $
time-centered weakly for any $ \bar{t} \geq 0 $ as $ \ep \dto 0 $.
\end{coro}

\subsection{Examples of Sufficiently Decorrelated Coefficients}
\label{subsect:decor-exmp}  

We now verify that \cref{cond:graph-ctl} is not a vacuous condition.
In particular, we will show that for Gaussian systems that decorrelate sufficiently quickly
the requirements of \cref{cond:graph-ctl} are fulfilled.
Moreover, we show that differential Weyl operators of a bosonic Fock space
(and so annihilation, creation, and gauge operators)
fulfill \cref{cond:graph-ctl} as long as they belong to subspaces of
the single-particle space which are asymptotically orthogonal.
\medskip

\subsubsection{Gaussian System}
\label{subsubsect:gauss-sys}  

As one might anticipate, appropriate conditions on Gaussian systems
will achieve the desired bounds.
\smallskip

\begin{defn} \label{defn:joint-gauss}
We say that a collection of quantum random
variables $ \{B_\iota\}_{\iota \in I} \subseteq \mathcal{A} $
is \textbf{jointly Gaussian} if for all $ \iota_1, \ldots, \iota_n \in I $,
\begin{equation} \label{eq:isserlis-form} \begin{split}
\mean{B_{\iota_1} \cdots B_{\iota_n}} &= 0
\quad \text{when $ n $ is odd} \\
\mean{B_{\iota_1} \cdots B_{\iota_n}}
&= \sum_P \prod_{\{x, y\} \in P} \mean{B_{\iota_x} B_{\iota_y}}
\quad \text{when $ n $ is even}
\end{split} \end{equation}
where $ P $ ranges over pairings on $ [1, n] $.
In the context of quantum field theory,
one would say $ \mathbb{E} $ is a \textsl{quasi-free state}.
\end{defn}

Note that it is possible to derive Isserlis' formula \cref{eq:isserlis-form},
in the quantum context, from more fundamental conditions on the collection $ \{B_\iota\} $
(e.g. having Gaussian linear combinations).
However, here we will leave \cref{eq:isserlis-form} as definitional
as is typically done for the concept of \textsl{quasi-free state}.
\smallskip

\begin{thm} \label{thm:gauss-graph-ctl}
Suppose the collection of processes $ \{\Ea\}_{\alpha=1}^\infty $
is jointly Gaussian and there exist $ C, \eta > 0 $
such that for any $ s, t \geq 0 $
\[ \ab{\mean{\Ea_s \Eb_t}} \leq C e^{-\eta |t - s|}. \]
Then, $ \{\Ea\} $ fulfill \cref{cond:graph-ctl}.
\end{thm}

\subsubsection{Differential Weyl Operators}
\label{subsubsect:diff-weyl-ops}  

Next, we show that \cref{cond:graph-ctl} holds for
annihilation, creation, and gauge operators on a bosonic Fock space.
We will follow \cite{Part92} in much of our terminology for the (bosonic) Fock space.
Suppose $ \cH $ is a Hilbert space and $ \BosFock(\cH) $ is the bosonic Fock space over $ \cH $.
For any $ v \in \cH $, we define the \textbf{coherent vector} for $ v $ as
\[ e(v) := \bigoplus_{n=0}^\infty \frac{v^{\otimes_+ n}}{\sqrt{n!}} \in \BosFock(\cH) \]
so that $ \ang{e(u), e(v)} = e^{\ang{u, v}} $ for all $ u, v \in \cH $.
We also define the \textbf{coherent states} for $ v $ as $ \vp(v) := e^{-\frac{1}{2} \|v\|^2} e(v) $
so that $ \|\vp(v)\| = 1 $.
It is well known that the linear hull of $ \{e(v) \,:\, v \in \cH\} $
(and so also $ \{\vp(v) \,:\, v \in \cH\} $) is dense in $ \BosFock(\cH) $.
\medskip

Additionally, we use $ \ISO(\cH) := \cH \rtimes \UNI(\cH) $ to refer to the Euclidean group of $ \cH $
which is the semidirect product of $ \cH $ treated as translations
and the unitary operators $ \UNI(\cH) $ on $ \cH $.
Then, we use $ \iso(\cH) := \cH \rtimes \uni(\cH) $ for the Lie algebra associated with $ \ISO(\cH) $
where $ \uni(\cH) $ are the bounded skew-Hermitian operators on $ \cH $.
For any $ (x, T) \in \ISO(\cH) $ and $ v \in \cH $, we have the Weyl representation
\begin{align}
W(x, T) e(v) &:= e^{-\frac{1}{2} \|x\|^2 - \ang{x, T v}} e(x + T v) \notag \\
\label{eq:weyl-coh-st}
\text{or for the coherent states} \quad
W(x, T) \vp(v) &= e^{-i\Im\ang{x, T v}} \vp(x + T v).
\end{align}
By a slight abuse of notation, we define the pushforward $ \dW $ of $ W $ such that
for any $ v \in \cH $ and (bounded) Hermitian operator $ A $ on $ \cH $,
since $ (v, -iA) \in \iso(\cH) $, one has $ W(e^{t (v, -iA)}) = e^{-it \dW(v, A)} $.
We will call $ \dW $ the \textbf{differential Weyl representation}.
We may recover the annihilation, creation, and gauge operators from $ \dW $ as follows:
\begin{align*}
a(v) = \frac{-\dW(iv, 1) + i\dW(v, 1)}{2}, &\quad
a^*(v) = \frac{-\dW(iv, 1) - i\dW(v, 1)}{2}, \\
\text{and} \qquad
\lambda(A) &= \dW\p{0, \frac{A + A^*}{2}} + i\dW\p{0, \frac{A - A^*}{2i}}
\end{align*}
for any $ v \in \cH $ and $ A \in B(\cH) $ (see equations 20.29 and 20.30\ in \cite{Part92}).
\medskip

Now, we define our quantum probability space $ (\hcA, \bE) $.
We take $ \cA := \bdd(\BosFock(\cH)) $ and $ \mean{X} = \ang{\vp(0), X \vp(0)} $ for all $ X \in \cA $.
Suppose for every $ t \in \bR $ that $ \cF_t \subseteq \cH $ is a closed subspace
with $ \eta > 0 $ such that for all $ s, t \geq 0 $, $ u \in \cF_s $, and $ v \in \cF_t $,
\[ \ab{\ang{u, v}} \leq e^{-\eta |t - s|} \|u\| \,\, \|v\|. \]
Next, for each $ t \geq 0 $, let $ \{v^\alpha_t\}_{\alpha \in \bN} \subseteq \cF_t $ be vectors
and $ \{A^\alpha_t\}_{\alpha \in \bN} \in i\uni(\cF_t) $ be (bounded) self-adjoint operators.
We assume there exists $ C > 0 $
such that $ \|v^\alpha_t\|, \|A^\alpha_t\| \leq C $ for all $ \alpha $ and $ t $.
\medskip

\begin{thm} \label{thm:gauge-graph-ctl}
If for each $ \alpha \in \bN $ and $ t \geq 0 $,
we define $ \Ea_t := \dW(v^\alpha_t, A^\alpha_t) $
then $ \{\Ea\} $ fulfill \cref{cond:graph-ctl}.
\end{thm}

Note that \cref{thm:gauss-graph-ctl} may be seen as a corollary of \cref{thm:gauge-graph-ctl}
when the Gaussian system is given by the differential Weyl operators $ \dW(v^\alpha, 1) $
for some collection $ \{v^\alpha\} \subseteq \cH $.
\medskip

\section{Proofs}
\label{sect:proofs}

We now provide the proofs for \cref{thm:int-weak-conv} and \cref{coro:weak-conv}
as well as \cref{thm:gauss-graph-ctl} and \cref{thm:gauge-graph-ctl}.
\subsection{Properties of Lindblad operators}

In the following subsection, we derive several useful identities involving
the Lindblad operators $ \gksl^+_{a,b} $ and $ \gksl^-_{a,b} $.
\medskip

\begin{lem} \label{lem:dbl-drv-idents}
For any $ a, b, f \in \cM $,
\begin{align*}
[a, [b, f]] - [b, [a, f]] &= [[a, b], f] \\
[a, [b, f]] + [b, [a, f]] &= -2\gksl^+_{a^*,b}(f) \\
\{a, [b, f]\} + \{b, [a, f]\} &= [\{a, b\}, f] \\
\{a, [b, f]\} - \{b, [a, f]\} &= -2\gksl^-_{a^*,b}(f).
\end{align*}
\end{lem}
\begin{proof}  
The first identity follows from the Jacobi identity.
For the second identity,
\begin{align*}
[a, [b, f]] + [b, [a, f]]
&=  (a b f - a f b - b f a + f b a)
  + (b a f - b f a - a f b + f a b) \\
&=  (a b f - 2 a f b + f a b)
  + (b a f - 2 b f a + f b a) \\
&=  -\gksl_{a^*,b}(f) - \gksl_{b^*,a}(f)
= -2\gksl^+_{a^*,b}(f).
\end{align*}
For the third identity,
\begin{align*}
\{a, [b, f]\}) + \{b, [a, f]\}
&= (a b f - a f b + b f a - f b a)
+ (b a f - b f a + a f b - f a b) \\
&= (ab + ba) f - f (ab + ba)
= [\{a, b\}, f].
\end{align*}
For the fourth identity,
\begin{align*}
\{a, [b, f]\} - \{b, [a, f]\}
&=  (a b f - a f b + b f a - f b a)
  - (b a f - b f a + a f b - f a b) \\
&=  (a b f - 2 a f b + f a b)
  - (b a f - 2 b f a + f b a) \\
&=  -\gksl_{a^*,b}(f) + \gksl_{b^*,a}(f)
= -2\gksl^+_{a^*,b}(f).
\end{align*}
\end{proof}

\begin{prop} \label{prop:sum-dbl-drv-to-gksl}
Suppose $ f \in \cM $ with $ f^* = f $, $ \{z_{ab}\}_{a,b=1}^\infty \subseteq \bC $,
and $ \{g_a\}_{a=1}^\infty \subseteq \cM $.
\begin{align}
\text{If} \quad \sup_{a,b} |z_{ab}| < \infty
\quad &\text{and} \quad
\sum_{a=1}^\infty \nr{g_a} < \infty
\quad \text{then} \notag \\
\label{eq:sum-re-dbl-drv-to-gksl}
\sum_{a,b=1}^\infty \Re(z_{ab} [g_a^*, [g_b, f]])
&= \sum_{a,b=1}^\infty \p{\big. -2z_{ab}^+ \gksl^+_{g_a,g_b}(f) + iz_{ab}^- [[g_a^*, g_b], f]} \\
\text{and} \quad
\label{eq:sum-im-dbl-drv-to-gksl}
\sum_{a,b=1}^\infty \Im(z_{ab} \{g_a^*, [g_b, f]\})
&= \sum_{a,b=1}^\infty \p{\big. -2z_{ab}^- \gksl^-_{g_a,g_b}(f) - iz_{ab}^+ [\{g_a^*, g_b\}, f]}
\end{align}
where $ z_{ab}^+ $ and $ z_{ab}^- $
are the Hermitian and anti-Hermitian components of $ z_{ab} $.
That is, for each $ a, b \in \bN $,
\[
z_{ab}^+ := \frac{z_{ab} + z_{ba}^*}{2}
\quad \text{and} \quad
z_{ab}^- := \frac{z_{ab} - z_{ba}^*}{2i}.
\]
\end{prop}
\begin{proof}  
First, we prove \cref{eq:sum-re-dbl-drv-to-gksl}.
By symmetrizing the sum with respect to $ a $ and $ b $, we get
\begin{equation} \label{eq:symm-sum-dbl-drv}
\sum_{a,b=1}^\infty \Re(z_{ab} [g_a^*, [g_b, f]])
= \frac{1}{2} \sum_{a,b=1}^\infty \p{\big.
    \Re(z_{ab} [g_a^*, [g_b, f]])
    + \Re(z_{ba} [g_b^*, [g_a, f]])
}.
\end{equation}
Note that for any $ a, b \in \bN $,
one has $ z_{ab} = z_{ab}^+ + iz_{ab}^- $ and $ z_{ba}^* = z_{ab}^+ - iz_{ab}^- $.
Then, using the self-adjointness of $ f $,
the fact that $ z_{ab}^+ $ and $ z_{ab}^- $ are Hermitian,
and \cref{lem:dbl-drv-idents}, we obtain
\begin{align*}
\Re(z_{ab} [g_a^*, [g_b, f]]) + \Re(z_{ba} [g_b^*, [g_a, f]])
&= \Re(z_{ab} [g_a^*, [g_b, f]]) + \Re(z_{ba}^* [g_b, [g_a^*, f]]) \\
&= \Re\p{\big.
      (z_{ab}^+ + iz_{ab}^-) [g_a^*, [g_b, f]]
    + (z_{ab}^+ - iz_{ab}^-) [g_b, [g_a^*, f]]
} \\
&= -2\Re(z_{ab}^+ \gksl^+_{g_a,g_b}(f)) + \Re(iz_{ab}^- [[g_a^*, g_b], f]).
\end{align*}
where we use $ [g_b^*, [g_a, f]]^* = [g_b, [g_a^*, f]] $ in the first equality.
Now, because $ z_{ab}^+ $ is Hermitian
and $ \gksl^+_{g_a,g_b} $ is conjugate symmetric from \cref{eq:gksl-parts-conj-symm},
by unsymmetrizing the following sum, we obtain
\begin{align}
\sum_{a,b=1}^\infty \Re(z_{ab}^+ \gksl^+_{g_a,g_b}(f))
&= \frac{1}{2} \sum_{a,b=1}^\infty \p{
    z_{ab}^+ \gksl^+_{g_a,g_b}(f) + (z_{ab}^+)^* (\gksl^+_{g_a,g_b}(f))^*
} \notag \\
&= \frac{1}{2} \sum_{a,b=1}^\infty \p{
   z_{ab}^+ \gksl^+_{g_a,g_b}(f) + z_{ba}^+ \gksl^+_{g_b,g_a}(f)
} \notag \\
\label{eq:sum-re-dbl-drv-symm}
&= \sum_{a,b=1}^\infty z_{ab}^+ \gksl^+_{g_a,g_b}(f).
\end{align}
Next, because $ z_{ab}^- $ is Hermitian, we also obtain
\begin{align}
\sum_{a,b=1}^\infty \Re(iz_{ab}^- [[g_a^*, g_b], f])
&= \frac{1}{2} \sum_{a,b=1}^\infty \p{
    iz_{ab}^- [[g_a^*, g_b], f] + (i z_{ab}^-)^* [[g_a^*, g_b], f]^*
} \notag \\
&= \frac{1}{2} \sum_{a,b=1}^\infty \p{
    iz_{ab}^- [[g_a^*, g_b], f] - i (z_{ab}^-)^* [[g_a, g_b^*], f]
} \notag \\
&= \frac{1}{2} \sum_{a,b=1}^\infty \p{
    iz_{ab}^- [[g_a^*, g_b], f] + iz_{ba}^- [[g_b^*, g_a], f]
} \notag \\
\label{eq:sum-re-dbl-drv-anti}
&= \sum_{a,b=1}^\infty iz_{ab}^- [[g_a^*, g_b], f].
\end{align}
Combining \cref{eq:sum-re-dbl-drv-symm} and \cref{eq:sum-re-dbl-drv-anti}, we get
\begin{align*}
\sum_{a,b=1}^\infty \Re(z_{ab} [g_a^*, [g_b, f]])
&= \frac{1}{2} \sum_{a,b=1}^\infty \p{
    -2\Re(z_{ab}^+ \gksl^+_{g_a,g_b}(f))
    + \Re(iz_{ab}^- [[g_a^*, g_b], f])
} \\
&= \frac{1}{2} \sum_{a,b=1}^\infty \p{
    -2z_{ab}^+ \gksl^+_{g_a,g_b}(f) + iz_{ab}^- [[g_a^*, g_b], f]
}
\end{align*}
proving \cref{eq:sum-re-dbl-drv-to-gksl}.
\medskip

Second, we prove \cref{eq:sum-im-dbl-drv-to-gksl}.
By symmetrizing the sum with respect to $ a $ and $ b $, we get
\[
\sum_{a,b=1}^\infty \Im(z_{ab} \{g_a^*, [g_b, f]\})
= \frac{1}{2} \sum_{a,b=1}^\infty \p{\big.
    \Im(z_{ab} \{g_a^*, [g_b, f]\})
    + \Im(z_{ba} \{g_b^*, [g_a, f]\})
}
\]
Because $ f $ is self-adjoint while $ z_{ab}^+ $ and $ z_{ab}^- $ are Hermitian,
from \cref{lem:dbl-drv-idents}, we obtain
\begin{align}
\Im(z_{ab} \{g_a^*, [g_b, f]\}) + \Im(z_{ba} \{g_b^*, [g_a, f]\})
&= \Im(z_{ab} \{g_a^*, [g_b, f]\}) + \Im(z_{ba}^* \{g_b, [g_a^*, f]\}) \notag \\
&= \Im\p{\big.
      (z_{ab}^+ + iz_{ab}^-) \{g_a^*, [g_b, f]\}
    + (z_{ab}^+ - iz_{ab}^-) \{g_b, [g_a^*, f]\}
} \notag \\
\label{eq:im-dbl-drv-to-gksl}
&= \Im(z_{ab}^+ [\{g_a^*, g_b\}, f]) - 2\Im(iz_{ab}^- \gksl^-_{g_a,g_b}(f))
\end{align}
where we use $ \{g_b^*, [g_a, f]\}^* = -\{g_b, [g_a^*, f]\} $ in the first equality.
Now, because $ z_{ab}^+ $ is Hermitian, by unsymmetrizing the following sum, we obtain
\begin{align}
\sum_{a,b=1}^\infty \Im(z_{ab}^+ [\{g_a^*, g_b\}, f])
&= \frac{1}{2i} \sum_{a,b=1}^\infty \p{
    z_{ab}^+ [\{g_a^*, g_b\}, f] - (z_{ab}^+)^* [\{g_a^*, g_b\}, f]^*
} \notag \\
&= \frac{1}{2i} \sum_{a,b=1}^\infty \p{
    z_{ab}^+ [[g_a^*, g_b\}, f] + z_{ba}^+ [\{g_b^*, g_a\}, f]
} \notag \\
\label{eq:sum-im-dbl-drv-symm}
&= -\sum_{a,b=1}^\infty iz_{ab}^+ [\{g_a^*, g_b\}, f].
\end{align}
Next, because $ z_{ab}^- $ is Hermitian and $ \gksl^-_{g_a,g_b} $ is conjugate symmetric
from \cref{eq:gksl-parts-conj-symm}, we obtain
\begin{align}
\sum_{a,b=1}^\infty \Im(iz_{ab}^- \gksl^-_{g_a,g_b}(f))
&= \frac{1}{2i} \sum_{a,b=1}^\infty \p{
    iz_{ab}^- \gksl^-_{g_a,g_b}(f) - (iz_{ab}^-)^* (\gksl^-_{g_a,g_b}(f))^*
} \notag \\
&= \frac{1}{2i} \sum_{a,b=1}^\infty \p{
    iz_{ab}^- \gksl^-_{g_a,g_b}(f) + iz_{ba}^- \gksl^-_{g_b,g_a}(f)
} \notag \\
\label{eq:sum-im-dbl-drv-anti}
&= \sum_{a,b=1}^\infty z_{ab}^- \gksl^-_{g_a,g_b}(f).
\end{align}
Combining \cref{eq:sum-im-dbl-drv-symm} and \cref{eq:sum-im-dbl-drv-anti}, we get
\begin{align*}
\sum_{a,b=1}^\infty \Im(z_{ab} \{g_a^*, [g_b, f]\})
&= \frac{1}{2} \sum_{a,b=1}^\infty \p{
    \Im(z_{ab}^+ [\{g_a^*, g_b\}, f])
    - 2\Im(iz_{ab}^- \gksl^-_{g_a,g_b}(f))
} \\
&= \frac{1}{2} \sum_{a,b=1}^\infty \p{
    -iz_{ab}^+ [\{g_a^*, g_b\}, f] - 2z_{ab}^- \gksl^-_{g_a,g_b}(f)
}
\end{align*}
proving \cref{eq:sum-im-dbl-drv-to-gksl}.
\end{proof}

\begin{lem} \label{lem:dbl-drv-parts}
For any $ E_1, E_2 \in \cE $ and $ h_1, h_2, f \in \cM $,
treating $ \cE $ and $ \cM $ as subsets of $ \cE \mxtens \cM $,
one has
\begin{equation} \label{eq:double-deriv}
[E_1 h_1, [E_2 h_2, f]] =
    \frac{1}{4} [E_1, E_2] \{h_1, [h_2, f]\}
  + \frac{1}{4} \{E_1, E_2\} [h_1, [h_2, f]].
\end{equation}
\end{lem}
\begin{proof}  
We will use $ \ad(\cdot) $ for the adjoint representation on $ \cM $
so that $ \ad(g) f = [g, f] $ for any $ f, g \in \cM $.
Since $ E_1 $ commutes with $ h_1 $
and $ E_2 $ commutes with $ h_2 $, we can write
\[
[E_1 h_1, [E_2 h_2, f]]
= \frac{1}{4} [\{E_1, h_1\}, [\{E_2, h_2\}, f]]
\]
where $ \{\cdot, \cdot\} $ is the anticommutator.
Then, since $ E_2 $ commutes with $ f $,
\begin{equation} \label{eq:dbl-drv-split}
[\{E_1, h_1\}, [\{E_2, h_2\}, f]]
= [\{E_1, h_1\}, \{E_2, [h_2, f]\}]
= \{[E_1, \{E_2, [h_2, f]\}], h_1\}
+ \{E_1, [h_1, \{E_2, [h_2, f]\}]\}.
\end{equation}
For the first term in \cref{eq:dbl-drv-split},
since $ E_1 $ commutes with $ [h_2, f] $
and $ [E_1, E_2] $ commutes with $ h_1 $, we get
\begin{equation} \label{eq:im-dbl-drv}
\{[E_1, \{E_2, [h_2, f]\}], h_1\}
= \{\{[E_1, E_2], [h_2, f]\}, h_1\}
= \{[E_1, E_2] [h_2, f], h_1\}
= [E_1, E_2] \{h_1, [h_2, f]\}
\end{equation}
For the second term in \cref{eq:dbl-drv-split},
since $ h_1 $ commutes with $ E_2 $
and $ [h_1, [h_2, f]] $ commutes with $ E_1 $ and $ E_2 $,
we get
\begin{equation} \label{eq:re-dbl-drv}
\{E_1, [h_1, \{E_2, [h_2, f]\}]\}
= \{E_1, \{E_2, [h_1, [h_2, f]]\}\}
= \{E_1, E_2 [h_1, [h_2, f]]\}
= \{E_1, E_2\} [h_1, [h_2, f]]
\end{equation}
Combining the expressions in \cref{eq:im-dbl-drv} and \cref{eq:re-dbl-drv},
we get \cref{eq:double-deriv}.
\end{proof}

\subsection{Properties of $ \etmu_t $}
\label{subsect:ctl-corr}  

In this subsection, we prove several useful properties of $ \etmu_\bt $
as defined in \cref{defn:mu-snorm}.
\medskip

\begin{lem} \label{lem:mean-bnd-etmu0}
For any $ \ep > 0 $ and $ U \in \tcA $,
\[ \nrM{\meanM{U}} \leq \etmu_0(U). \]
\end{lem}
\begin{proof}  
From \cref{defn:mu-snorm}, for any $ A \in \cE $,
one has $ \emu_0(A) = \mu_0(; A) = |\mean{A}| $.
Then, for any $ f \in \cM $, we have
\[
\nrM{\meanM{A \otimes f}} = \nrM{\mean{A} f}
= \ab{\mean{A}} \cdot \nrM{f}
\leq \emu_0(A) \cdot \nrM{f}.
\]
Because this holds for all $ A $ and $ f $, for all $ U \in \tcA $,
we have $ \|\meanM{U}\|_\cM \leq (\emu_0 \otimes \|\cdot\|_\cM)(U) = \etmu_0(U) $.
\end{proof}

\begin{lem} \label{lem:etmu-bimod}
Let $ \ep > 0 $ and $ t \geq 0 $
then for any $ g_1, g_2 \in \cM $ and any $ U \in \tcA $,
\begin{equation} \label{eq:etmu-bimod}
\etmu_t(g_1 U g_2) \leq \nrM{g_1} \cdot \nrM{g_2} \cdot \etmu_t(U).
\end{equation}
\end{lem}
\begin{proof}  
For any $ A \in \cE $ and $ f \in \cM $,
using submultiplicativity of $ \nrM{\cdot} $,
\begin{align*}
(\emu_t \otimes \nr{\cdot})(g_1 (A \otimes f) g_2)
&= (\emu_t \otimes \nr{\cdot})(A \otimes (g_1 f g_2)) \\
&\leq \emu_t(A) \cdot \nr{g_1 f g_2} \\
&\leq \|g_1\| \cdot \|g_2\| \cdot
  \p{\big. \emu_t(A) \cdot \nrM{f}}.
\end{align*}
Because this holds for all $ A \in \cE $ and $ f \in \cM $,
by the definition of the product seminorm, we obtain \cref{eq:etmu-bimod}.
\end{proof}

\begin{lem} \label{lem:etmu-drv-bnd-Ea}
Let $ \ep > 0 $ be a parameter, $ t \geq 0 $ a time,
and $ \alpha \in \bN $.
Then, for any $ A \in \cE $,
\begin{equation} \label{eq:emu-drv-bnd-Ea}
\etmu_t(\Ea_{t / \ep^2} A), \emu_t(A \Ea_{t / \ep^2})
\leq \frac{\ep}{2\|\ho\|_1} \edmu_t(A).
\end{equation}
\end{lem}
\begin{proof}  
Let $ n \geq 0 $ be an integer and $ t_1 \leq \ldots \leq t_n $ be times
such that $ t_1, \ldots, t_n \leq t / \ep^2 $.
Then, for any $ \alpha_1, \ldots, \alpha_n \in \bN $ and permutation $ \sigma $
so that $ t_{\sigma(1)} \leq \ldots \leq t_{\sigma(j)} $
and $ t_{\sigma(j+1)} \geq \ldots \geq t_{\sigma(n)} $, we have
\[
\ab{\mean{E^{\alpha_1}_{t_{\sigma(1)}} \cdots E^{\alpha_j}_{t_{\sigma(j)}}
  (\Ea_{t / \ep^2} A) E^{\alpha_{j+1}}_{t_{\sigma(j+1)}} \cdots E^{\alpha_n}_{t_{\sigma(n)}}
}}
\leq \mu_{n+1}(t_1, \ldots, t_n, t / \ep^2; A).
\]
Since this holds for all $ \alpha_\bt $ and $ \sigma $, we have
\[
\mu_n(t_1, \ldots, t_n; \Ea_{t / \ep^2} A)
\leq \mu_{n+1}(t_1, \ldots, t_n, t / \ep^2; A)
\]
and the same bound applies to $ \mu_n(t_1, \ldots, t_n; A \Ea_{t / \ep^2}) $.
Thus,
\begin{align*}
\emu_t(\Ea_{t / \ep^2} A) &= \sum_{n=0}^\infty (2\ep \|\ho\|_1)^n
    \int_0^{t / \ep^2} \dr[t_n] \cdots \int_0^{t_2} \dr[t_1]
    \mu_n(t_1, \ldots, t_n; \Ea_{t / \ep^2} A) \\
&\leq \sum_{n=0}^\infty (2\ep \|\ho\|_1)^n
    \int_0^{t / \ep^2} \dr[t_n] \cdots \int_0^{t_2} \dr[t_1]
    \mu_{n+1}(t_1, \ldots, t_n, t / \ep^2; A) \\
&= \sum_{n=1}^\infty (2\ep \|\ho\|_1)^{n-1}
    \int_0^{t / \ep^2} \dr[t_{n-1}] \cdots \int_0^{t_2} \dr[t_1]
    \mu_n(t_1, \ldots, t_{n-1}, t / \ep^2; A) \\
&= \frac{1}{2\ep \|\ho\|_1} \sum_{n=1}^\infty (2\ep \|\ho\|_1)^n
    \cdot \ep^2 \frac{\dr}{\dr[t]}
    \int_0^{t / \ep^2} \dr[t_n] \cdots \int_0^{t_2} \dr[t_1]
    \mu_n(t_1, \ldots, t_{n-1}, t_n; A)
= \frac{\ep}{2 \|\ho\|_1} \edmu_t(A)
\end{align*}
and the same bound applies to $ \emu_t(A \Ea_{t / \ep^2}) $.
\end{proof}

\begin{lem} \label{lem:etmu-ctl-egxi}
As long as $ \etmu_\bt $ fulfills \cref{cond:etmu-etp-bdd},
for any $ \ep > 0 $, times $ 0 \leq s \leq t $, and $ U \in \tcA $,
\begin{equation} \label{eq:etmu-ctl-xi}
\etmu_s(\egxi_{s \gets t}(U)) \leq \etmu_t(U).
\end{equation}
\end{lem}
\begin{proof}
Let $ A \in \cE $ and $ f \in \cM $
then, using \cref{lem:etmu-drv-bnd-Ea},
\begin{align*}
\etmu_t\p{[\Ea_{t / \ep^2} \Re(e^{it\wa / \ep^2} \ghoa_t), A \otimes f]}
&= (\emu_t \otimes \nrM{\cdot})\p{
  \Ea_{t / \ep^2} A \cdot \Re(e^{it\wa / \ep^2} \ghoa_t) f
  - A \Ea_{t / \ep^2} \cdot f \Re(e^{it\wa / \ep^2} \ghoa_t)
} \\
&\leq \emu_t(\Ea_{t / \ep^2} A) \cdot \nrM{\Re(e^{it\wa / \ep^2} \ghoa_t) f}
  + \emu_t(A \Ea_{t / \ep^2}) \cdot \nrM{f \Re(e^{it\wa / \ep^2} \ghoa_t)} \\
&\leq \frac{2 \ep}{2 \|\ho\|_1} \nrM{\hoa} \cdot \p{\edmu_t(A) \cdot \nrM{f}}.
\end{align*}
Then, we sum over $ \alpha $ to obtain
\begin{align*}
\etmu_t(\egdf_t (A \otimes f))
&\leq \frac{1}{\ep} \sum_{\alpha=1}^\infty \etmu_t\p{
  [\Ea_{t / \ep^2} \Re(e^{it\wa / \ep^2} \ghoa_t), A \otimes f]
} \\
&\leq \frac{1}{\ep} \sum_{\alpha=1}^\infty
  \frac{\ep}{\|\ho\|_1} \|\hoa\| \cdot \p{\edmu_t(A) \cdot \|f\|} \\
&= \p{\edmu_t(A) \cdot \|f\|} \cdot \frac{1}{\|\ho\|_1} \sum_{\alpha=1}^\infty \|\hoa\|
= \edmu_t(A) \cdot \|f\|.
\end{align*}
Because this holds for all $ A $ and $ f $, one has
\begin{align}
\etmu_s(U) + \int_s^t \dr[\tau] \etmu_\tau(\egdf_\tau U)
&\leq \etmu_s(U) + \int_s^t \dr[\tau] (\edmu_\tau \otimes \nrM{\cdot})(U) \notag \\
\label{eq:etmu-diff-bnd-intGen}
&= (\emu_s \otimes \nrM{\cdot})(U) + ((\emu_t - \emu_s) \otimes \nrM{\cdot})(U)
= \edtmu_t(U)
\end{align}
for all $ U \in \tcA $.
Because of \cref{cond:etmu-etp-bdd}, the topologies generated by $ \etmu_\bt $
are coarser than those generated by $ \tp_\bt $
so $ \tp_t $-continuity of $ \tau \mapsto \egdf_\tau U $ implies $ \etmu_\bt $-continuity.
Thus, using the condition \cref{eq:pt-ctl-integ},
\cref{eq:etmu-diff-bnd-intGen} implies
that $ \etmu_\bt $ controls $ \egdf_\bt $.
Then, by \cref{prop:mu-bnd-sgrpd}, for any $ s \leq t $,
we obtain \cref{eq:etmu-ctl-xi}.
\end{proof}

\subsection{Properties of $ \vartheta(\ep; t) $ and $ \Theta_G $}
\label{subsect:graph-ctl-props}

We now demonstrate how $ \vartheta(\ep; t) $ and $ \Theta_G $
can be used to bound $ \etmu_t $.
\bigskip

\begin{lem} \label{lem:vartheta-conv}
\Cref{eq:graph-inf-rad-conv} implies that for all $ \ep, t \in \bR $,
the power series $ \vartheta(\ep; t) $ is absolutely convergent.
Moreover, for any $ \ep_1, \ep_2, t_1, t_2 \geq 0 $,
if $ \ep_1 \leq \ep_2 $ and $ t_1 \leq t_2 $
then $ \vartheta(\ep_1; t_1) \leq \vartheta(\ep_2; t_2) $.
\end{lem}
\begin{proof}  
First, for any $ G \in \cG $, by the definition of $ \cG $,
every vertex of $ G $ has an edge so every connected component has at least two vertices.
Thus, $ |G| \geq 2c(G) $ and so $ |G| - 2c(G) \geq 0 $ for all $ G $.
Also, $ G $ is nonempty so $ c(G) \geq 1 $ and $ c(G) - 1 \geq 0 $.
So, $ \vartheta $ is increasing in $ \ep $ and $ t $ for non-negative arguments
proving the second part of the lemma.
Then, for any $ \ep, t \geq 0 $, we can bound the sum by
\begin{align*}
\vartheta(\ep; t) &= \sum_{G \in \cG} \vartheta_G \frac{1}{(|G| - 1)!}
    2^{2|G| - c(G) - 1} \|\ho\|_1^{|G| - 1} \,\, (\eta^{-1})^{|G| - c(G)}
    \cdot t^{c(G) - 1} \ep^{|G| - 2c(G)} \\
&\leq \sum_{G \in \cG} \vartheta_G \frac{1}{(|G| - 1)!}
    2^{2|G|} (\|\ho\|_1 \lor 1)^{|G|} \,\, (\eta^{-1} \lor 1)^{|G|}
    \cdot (z \lor t)^{c(G) - 1} (z \lor t)^{|G| - 2c(G)} \\
&\leq \sum_{G \in \cG} \vartheta_G \frac{1}{(|G| - 1)!}
    (4 (\|\ho\|_1 \lor 1)(\eta^{-1} \lor 1))^{|G|} \cdot (z \lor t)^{|G|}.
\end{align*}
For sake of brevity, we take $ x := 4 (\|\ho\|_1 \lor 1)(\eta^{-1} \lor 1) \cdot (z \lor t) $
and define
\[ q_n := \sum_{\substack{G \in \cG \\ |G| = n}} \vartheta_G. \]
Now, observe that we can separate the sum along values of $ |G| $:
\begin{align*}
\sum_{G \in \cG} \vartheta_G \frac{1}{(|G| - 1)!}
    (4 (\|\ho\|_1 \lor 1) (\eta^{-1} \lor 1) (z \lor t))^{|G|}
&= \sum_{n=1}^\infty \frac{x^n}{(n - 1)!}
    \sum_{\substack{G \in \cG \\ |G| = n}} \vartheta_G \\
&= \sum_{n=1}^\infty q_n \frac{x^n}{(n - 1)!}.
\end{align*}
Applying Cauchy's radical test to this series,
by \cref{eq:graph-inf-rad-conv}, we see that
\[
\lim_{n \to \infty} \sqrt[n]{q_n \frac{x^n}{(n - 1)!}}
= \lim_{n \to \infty} x \sqrt[n]{q_n \frac{1}{n!}}
= x \lim_{n \to \infty} \frac{e}{n} \sqrt[n]{q_n} = 0
\]
showing that $ \vartheta(\ep; t) $ is absolutely convergent for any $ z $ and $ t $.
\end{proof}

\begin{lem} \label{lem:ThetaG0-bnd}
Suppose $ s \geq 0 $ and $ S_1, \ldots, S_n $ are independent random variables
uniformly distributed on $ [0, s] $ then for any graph $ G $ on $ [1, n] $,
\[ \mean{\Theta_G(S_1, \ldots, S_n)} \leq \p{\frac{2}{s \eta}}^{|G| - c(G)}. \]
\end{lem}
\begin{proof}  
For any graph $ G $, if $ H \subseteq G $ is a subgraph
then, by the definition of \cref{eq:ThetaG-defn},
because $ e^{-\eta |t_i - t_j|} \geq 0 $ for any $ t_i, t_j \geq 0 $,
we have $ \Theta_G(S_1, \ldots, S_n) \leq \Theta_H(S_1, \ldots, S_n) $.
Thus, it suffices to prove the bound for a maximal spanning forest $ H \subseteq G $
(i.e. $ H $ is the disjoint union of spanning trees for each component)
so that $ |H| = |G| $ and $ c(H) = c(G) $.
So, without loss of generality, we assume $ G $ is a forest.
\medskip

We prove the inequality for forests by induction on $ m(G) $.
When $ m(G) = 0 $, there are no edges in $ G $ meaning $ \Theta_G(\ldots) = 1 $
and the inequality holds.
Now, we consider when $ m(G) \geq 1 $.
Since $ G $ has edges, it has a leaf $ v \in V(G) = [1, n] $.
Let $ x \in V(G) $ be the unique vertex adjacent to $ v $.
Since $ S_v $ is uniformly distributed on $ [0, s] $,
we have $ |S_v - s/2| \preceq |S_v - c| $ for all $ c \in \bR $
where $ \preceq $ is stochastic ordering.
Note that $ |S_v - s/2| $ is uniformly distributed on $ [0, s/2] $.
Then, because $ e^{-\eta s} $ is decreasing,
using independence of $ S_1, \ldots, S_n $,
\begin{equation} \label{eq:eta-Tv-mean-bnd}
\meanWhen{e^{-\eta |S_v - S_x|}}{\{S_y\}_{y \neq v}}
\leq \meanWhen{e^{-\eta |S_v - s/2|}}{\{S_y\}_{y \neq v}}
= \frac{2}{s} \int_0^{s/2} \dr[\dl] e^{-\eta \dl}
\leq \frac{2}{s} \cdot \frac{1}{\eta}.
\end{equation}
Therefore, using the induction hypothesis,
since $ m(G \setminus v) = m(G) - 1 $,
we can bound the mean of $ \Theta_G(\ldots) $ by
\begin{align*}
\mean{\Theta_G(S_1, \ldots, S_n)}
&= \mean{\meanWhen{e^{-\eta |S_v - S_x|}}{\{S_y\}_{y \neq v}}
    \Theta_{G \setminus v}(S_1, \ldots, S_{v-1}, S_{v+1}, \ldots, S_n)} \\
&\leq \frac{2}{t \eta} \mean{
    \Theta_{G \setminus v}(S_1, \ldots, S_{v-1}, S_{v+1}, \ldots, S_n)} \\
& \leq \frac{2}{s \eta} \cdot \p{\frac{2}{s \eta}}^{m(G) - 1}
= \p{\frac{2}{s \eta}}^{m(G)} = \p{\frac{2}{s \eta}}^{|G| - c(G)}
\end{align*}
since $ |G| = m(G) + c(G) $ for forests.
\end{proof}

We will also need a slightly more sophisticated
version of \cref{lem:ThetaG0-bnd}.

\begin{lem} \label{lem:ThetaG1-bnd}
Suppose $ 0 \leq s \leq t $ and $ S_1, \ldots, S_n $ are independent random variables
uniformly distributed on $ [0, s] $ then for any graph $ G $ on $ [1, n] $
where the vertex $ j $ has an edge in $ G $,
\[
\mean{\Theta_G(S_1, \ldots, S_{j-1}, t, S_{j+1}, \ldots, S_n)}
\leq e^{-\eta |t - s|} \p{\frac{2}{s \eta}}^{|G| - c(G)}.
\]
\end{lem}
\begin{proof}  
As in the proof of \cref{lem:ThetaG0-bnd},
it suffices to prove the bound for a maximal spanning forest of $ G $.
So, without loss of generality, we assume $ G $ is a forest.
\medskip

We prove the inequality for forests by induction on $ m(G) $ the stronger claim that
\[
\mean{\Theta_G(S_1, \ldots, S_{j-1}, t, S_{j+1}, \ldots, S_n)}
\leq \p{\frac{2}{s \eta}}^{m(G)} \exp\p{-\eta|t - s| \cdot \bbb{d_G(j) > 0}}
\]
where $ \bbb{\cdot} $ is an indicator function
and $ d_G(j) $ is the degree of the vertex $ j $ in $ G $.
When $ m(G) = 0 $, $ G $ has no edges
so $ \Theta_G(\ldots) = 1 $ and the result holds.
Otherwise, $ m(G) \geq 1 $ and so $ G $ has edges.
Thus, it must have at least two distinct leaves $ u, v \in V(G) = [1, n] $.
At least one of the leaves is not $ j $
so $ u \neq j $ or $ v \neq j $.
Without loss of generality, we assume $ v \neq j $
and let $ x \in V(G) $ be the unique vertex such that $ v \sim_G x $.
As in the proof of \cref{lem:ThetaG0-bnd},
we have $ |S_v - s/2| \preceq |S_v - c| $ for all $ c \in \bR $.
Now, either $ x \neq j $ or $ x = j $.
If $ x \neq j $ then, as in \cref{eq:eta-Tv-mean-bnd}, one has
\[
\meanWhen{e^{-\eta |S_v - S_x|}}{\{S_y\}_{y \neq v}}
\leq \frac{2}{s \eta}.
\]
Note that because $ x \neq j $, the vertex $ j $ has edges in $ G $
if and only if $ j $ has edges in $ G \setminus v $.
Thus, $ \bbb{d_{G \setminus v}(j) > 0} = \bbb{d_G(j) > 0} $.
Therefore, by the induction hypothesis,
using $ m(G \setminus v) = m(G) - 1 $, we get
\begin{align*}
\mean{\Theta_G(S_1, \ldots, S_{j-1}, t, S_{j+1}, \ldots, S_n)}
&\leq \frac{2}{s \eta} \mean{\Theta_{G \setminus v}(
    S_1, \ldots, S_{v-1}, S_{v+1}, \ldots, S_{j-1}, t, S_{j+1}, \ldots, T_n
)} \\
&\leq \frac{2}{s \eta} \cdot \p{\frac{2}{s \eta}}^{m(G) - 1}
    \exp\p{-\eta |t - s| \cdot \bbb{d_{G \setminus v}(j) > 0}} \\
&\leq \p{\frac{2}{s \eta}}^{m(G)} \exp{-\eta |t - s| \cdot \bbb{d_G(j) > 0}}.
\end{align*}
\medskip

Otherwise, $ x = j $ and $ S_v \leq s \leq t $
so $ |S_v - t| = |S_v - s| + |s - t| $ implying
\[
\meanWhen{e^{-\eta |S_v - t|}}{\{S_y\}_{y \neq v}}
= e^{-\eta |t - s|} \meanWhen{\big. |S_v - s|}{\{S_y\}_{y \neq v}}
\leq e^{-\eta |t - s|} \frac{2}{s \eta}.
\]
Note that because $ x = j $, we know $ d_G(j) > 0 $
so $ \bbb{d_G(j) > 0} = 1 $.
Therefore, by the induction hypothesis,
using $ m(G \setminus v) = m(G) - 1 $,
regardless of whether $ d_{G \setminus v}(j) > 0 $, we obtain
\begin{align*}
\mean{\Theta_G(S_1, \ldots, S_{j-1}, t, S_{j+1}, \ldots, S_n)}
&\leq e^{-\eta |t - s|} \frac{2}{s \eta} \mean{\Theta_{G \setminus v}(
    S_1, \ldots, S_{v-1}, S_{v+1}, \ldots, S_{j-1}, t, S_j, \ldots, S_n
)} \\
&\leq e^{-\eta |t - s|} \frac{2}{s \eta} \cdot \p{\frac{2}{s \eta} \|\eta\|_1}^{m(G) - 1}
= \p{\frac{2}{s \eta}}^{|G| - c(G)} \exp\p{-\eta |t - s| \cdot \bbb{d_G(j) > 0}}
\end{align*}
since $ |G| = m(G) + c(G) $ for forests.
As a particular case of the induction claim,
when $ d_G(j) > 0 $, we get the desired inequality.
\end{proof}

\begin{prop} \label{prop:etmu-vartheta-bnd}
Given that $ \{\Ea\} $ fulfill \cref{cond:graph-ctl},
for any times $ 0 \leq s \leq t \leq \tau $ and $ \alpha, \beta \in \bN $,
\begin{align*}
\emu_s(1) \leq 2s\|\ho\|_1 \vartheta(\ep; s), \qquad
 \emu_s(\Ea_{t / \ep^2}) &\leq \ep e^{-\eta |t - s| / \ep^2} \vartheta(\ep; s), \\
\text{and} \quad
\emu_s\p{\Ea_{t / \ep^2} \Eb_{\tau / \ep^2} - \mean{\Ea_{t / \ep^2} \Eb_{\tau / \ep^2}}}
&\leq \ep e^{-\eta |\tau - s| / \ep^2} \vartheta(\ep; s).
\end{align*}
Moreover, $ \etmu_s(A) \leq \emu_s(A) $ for any $ A \in \cE $
so the above bounds apply to $ \etmu_s $ as well.
\end{prop}
\begin{proof}  
First, suppose $ 0 \leq s_1 \leq \ldots \leq s_n \leq s \leq t $
and $ \alpha, \beta \in \bN $.
Now, let $ \alpha_1, \ldots, \alpha_n \in \bN $, $ j \in [1, n] $ be an index,
and $ \sigma $ be a permutation on $ [1, n] $
such that $ s_{\sigma(1)} \leq \ldots \leq s_{\sigma(j)} $
and $ s_{\sigma(j+1)} \geq \ldots \geq s_{\sigma(n)} $.
Then, by \cref{eq:graph-mean-ctl} in \cref{cond:graph-ctl},
\begin{align}
\label{eq:mun-term0-mean}
\ab{\mean{E^{\alpha_1}_{s_{\sigma(1)}} \cdots E^{\alpha_n}_{s_{\sigma(n)}}}}
&\leq \Theta_\cG(s_{\sigma(1)}, \ldots, s_{\sigma(j)}) \\
\label{eq:mun-term1-mean}
\ab{\mean{E^{\alpha_1}_{s_{\sigma(1)}} \cdots E^{\alpha_j}_{s_{\sigma(j)}}
    \Ea_t E^{\alpha_{j+1}}_{s_{\sigma(j+1)}} \cdots E^{\alpha_n}_{s_{\sigma(n)}}}}
&\leq \Theta_\cG(s_{\sigma(1)}, \ldots, s_{\sigma(j)},
    t, s_{\sigma(j+1)}, \ldots s_{\sigma(n)}).
\end{align}
Since $ \mu_n(s_1, \ldots, s_n; 1) $ and $ \mu_n(s_1, \ldots, s_n; \Ea_t) $
are the suprema over the means in \cref{eq:mun-term0-mean} and \cref{eq:mun-term1-mean},
we may write
\begin{align}
\label{eq:mun0-sum-bnd}
\mu_n(s_1, \ldots, s_n; 1) &\leq \sum_{j, \sigma}
    \Theta_\cG(s_{\sigma(1)}, \ldots, s_{\sigma(n)}) \\
\label{eq:mun1-sum-bnd}
\mu_n(s_1, \ldots, s_n; \Ea_s) &\leq \sum_{j, \sigma}
    \Theta_\cG(s_{\sigma(1)}, \ldots, s_{\sigma(j)},
   t, s_{\sigma(j+1)}, \ldots, s_{\sigma(n)})
\end{align}
where $ j $ and $ \sigma $ range over
valid choices of the index and permutation.
Similarly, by \cref{eq:graph-cov-ctl}, we obtain
\begin{align}
&\ab{\mean{E^{\alpha_1}_{s_{\sigma(1)}} \cdots E^{\alpha_j}_{s_{\sigma(j)}}
    \p{\Ea_t \Eb_\tau - \mean{\Ea_t \Eb_\tau}}
    E^{\alpha_{j+1}}_{s_{\sigma(j+1)}} \cdots E^{\alpha_n}_{s_{\sigma(n)}}}}
\notag \\
&\qquad \leq e^{-\eta |\tau - t|} \Theta_\cG(s_{\sigma(1)}, \ldots, s_{\sigma(j)},
    t, s_{\sigma(j+1)}, \ldots, s_{\sigma(n)}) \notag \\
\text{and thus} \quad
&\mu_n\p{s_1, \ldots, s_n; \Ea_t \Eb_\tau - \mean{\Ea_t \Eb_\tau}} \notag \\
\label{eq:mun2-sum-bnd}
&\qquad \leq e^{-\eta |\tau - t|} \sum_{j, \sigma} \Theta_\cG(s_{\sigma(1)}, \ldots, s_{\sigma(j)},
    t, s_{\sigma(j+1)}, \ldots, s_{\sigma(n)}).
\end{align}
Next, let $ S_1, S_2, \ldots $ be a sequence of independent random variables
uniformly distributed on $ [0, s] $.
For each $ j \in [1, n] $,
because $ s_{\sigma(1)} \leq \ldots \leq s_{\sigma(j)} $
and $ s_{\sigma(j+1)} \geq \ldots \geq s_{\sigma(n)} $,
we know $ \sigma(1) \leq \ldots \leq \sigma(j) $
and $ \sigma(j+1) \geq \ldots \geq \sigma(n) $.
Thus, the valid permutations $ \sigma $ for $ j $
are in bijection with size $ j $ subsets of $ n $.
So, there are $ \binom{n}{j} $ choices for $ \sigma $
and thus there are $ 2^n $ choices for $ j $ and $ \sigma $.
Therefore, after substituting $ S_1, \ldots, S_n $ for $ s_1, \ldots, s_n $
we use \cref{lem:ThetaG0-bnd} to bound the sum \cref{eq:mun0-sum-bnd} by
\begin{align}
\mean{\mu_n(S_1, \ldots, S_n; 1)}
&\leq \sum_{j, \sigma} \mean{\Theta_\cG(S_{\sigma(1)}, \ldots, S_{\sigma(n)})} \notag \\
&= \sum_{j,\sigma} \sum_{\substack{G \in \cG \\ |G| = n}} \vartheta_G
    \mean{\Theta_G(S_{\sigma(1)}, \ldots, S_{\sigma(n)})} \notag \\
&\leq \sum_{j,\sigma} \sum_{\substack{G \in \cG \\ |G| = n}} \vartheta_G
    \p{\frac{2}{s \eta}}^{|G| - c(G)}
\label{eq:mean-mun0-bnd}
= \sum_{\substack{G \in \cG \\ |G| = n}} \vartheta_G
    2^n \p{\frac{2}{s \eta}}^{|G| - c(G)}.
\end{align}
Similarly, using \cref{lem:ThetaG1-bnd},
we bound \cref{eq:mun1-sum-bnd} and \cref{eq:mun2-sum-bnd} by
\begin{align}
\label{eq:mean-mun1-bnd}
\mean{\mu_n(S_1, \ldots, S_n; \Ea_t)}
&\leq \sum_{\substack{G \in \cG \\ |G| = n + 1}} \vartheta_G
    2^n e^{-\eta |t - s|} \p{\frac{2}{s \eta}}^{|G| - c(G)} \\
\label{eq:mean-mun2-bnd}
\mean{\mu_n\p{S_1, \ldots, S_n; \Ea_t \Eb_\tau - \mean{\Ea_t \Eb_\tau}}}
&\leq e^{-\eta |\tau - t|} \sum_{\substack{G \in \cG \\ |G| = n + 1}} \vartheta_G
    2^n e^{-\eta |t - s|} \p{\frac{2}{s \eta}}^{|G| - c(G)}.
\end{align}
Now, using \cref{eq:mean-mun0-bnd}, we can bound $ \emu_s(1) $ by
\begin{align}
\emu_s(1) &= \sum_{n=0}^\infty (2\ep \|\ho\|_1)^n
    \int_0^{s / \ep^2} \dr[s_n] \cdots \int_0^{s_2} \dr[s_1]
    \mu_n(s_1, \ldots, s_n; 1) \notag \\
&= \sum_{n=0}^\infty (2\ep \|\ho\|_1)^n \frac{(s / \ep^2)^n}{n!}
    \mean{\mu_n(S_1 / \ep^2, \ldots, S_n / \ep^2; 1)} \notag \\
&\leq \sum_{n=0}^\infty \frac{1}{n!} \p{\frac{2s}{\ep} \|\ho\|_1}^n
    \sum_{\substack{G \in \cG \\ |G| = n}} \vartheta_G
    2^n \p{\frac{2 \ep^2}{s \eta}}^{|G| - c(G)} \notag \\
&= \sum_{G \in \cG} \vartheta_G \frac{1}{|G|!}
    \p{4 \frac{s}{\ep} \|\ho\|_1}^{|G|} \p{\frac{2 \ep^2}{s \eta}}^{|G| - c(G)} \notag \\
&= \sum_{G \in \cG} \vartheta_G \frac{1}{|G|!}
    2^{2|G| + |G| - c(G)} \|\ho\|_1^{|G|} \,\, \eta^{-(|G| - c(G))}
    \cdot s^{|G| - (|G| - c(G))} \ep^{-|G| + 2|G| - 2c(G)} \notag \\
\label{eq:vartheta-form0}
&\leq 4s\|\ho\|_1 \cdot \sum_{\substack{G \in \cG}} \vartheta_G \frac{1}{(|G| - 1)!}
    2^{3|G| - c(G) - 2} \|\ho\|_1^{|G| - 1} \,\, \eta^{c(G) - |G|}
    \cdot s^{c(G) - 1} \ep^{|G| - 2c(G)}
= 4s\|\ho\|_1 \vartheta(\ep; s).
\end{align}
Similarly, using \cref{eq:mean-mun1-bnd}, we get
\begin{align}
\emu_s(\Ea_{t / \ep^2}) &= \sum_{n=0}^\infty (2\ep \|\ho\|_1)^n \frac{(s / \ep^2)^n}{n!}
    \mean{\mu_n(S_1 / \ep^2, \ldots, S_n / \ep^2; \Ea_{t / \ep^2})} \notag \\
&\leq \sum_{n=0}^\infty \frac{1}{n!} \p{\frac{2s}{\ep} \|\ho\|_1}^n
    \sum_{\substack{G \in \cG \\ |G| = n + 1}} \vartheta_G
    2^n e^{-\eta |t - s| / \ep^2} \p{\frac{2 \ep^2}{s \eta}}^{|G| - c(G)} \notag \\
&= e^{-\eta |t - s| / \ep^2} \sum_{G \in \cG} \vartheta_G \frac{1}{(|G| - 1)!}
    \p{\frac{4 s}{\ep} \|\ho\|_1}^{|G| - 1} \p{2\frac{\ep^2}{s \eta}}^{|G| - c(G)} \notag \\
&= e^{-\eta |t - s| / \ep^2} \sum_{G \in \cG} \vartheta_G \frac{1}{(|G| - 1)!}
    2^{2|G| - 2 + |G| - c(G)} \|\ho\|_1^{|G| - 1} \,\, \eta^{-(|G| - c(G))}
    \cdot s^{|G| - 1 - (|G| - c(G))} \ep^{-(|G| - 1) + 2|G| - 2c(G)} \notag \\
\label{eq:vartheta-form1}
&= \ep e^{-\eta |t - s| / \ep^2} \sum_{G \in \cG} \vartheta_G \frac{1}{(|G| - 1)!}
    2^{3|G| - c(G) - 2} \|\ho\|_1^{|G| - 1} \,\, \eta^{c(G) - |G|}
    \cdot s^{c(G) - 1} \ep^{|G| - 2c(G)}
= \ep e^{-\eta |t - s| / \ep^2} \vartheta(\ep; s).
\end{align}
Analogously, from \cref{eq:mean-mun2-bnd}, one has
\[
\emu_s\p{\Ea_{t / \ep^2} \Eb_{\tau / \ep^2} - \mean{\Ea_{t / \ep^2} \Eb_{\tau / \ep^2}}}
\leq \ep e^{-\eta |\tau - t|} \cdot e^{-\eta |t - s|} \vartheta(\ep; s)
= \ep e^{-\eta |\tau - s|} \vartheta(\ep; s).
\]
Lastly, from the definition of $ \etmu_\bt $, for any $ A \in \cE $,
one has $ \etmu_s(A) = \etmu_s(A \otimes 1) \leq \emu_s(A) \cdot \nrM{1} = \emu_s(A) $.
\end{proof}

\begin{coro} \label{coro:egxi-etmu-lip}
For any times $ 0 \leq s \leq t $, parameter $ \ep > 0 $,
and $ f \in \cM $, we have $ \etmu_s(f) \leq 2s\|\ho\|_1 \vartheta(\ep; s) \cdot \|f\| $ and
\[
\etmu_s(f - \egxi_{s \gets t}(f))
\leq 2(t - s) \vartheta(\ep; t) \|\ho\|_1 \cdot \|f\|.
\]
\end{coro}
\begin{proof}  
First, from \cref{lem:etmu-bimod} and \cref{prop:etmu-vartheta-bnd}, we obtain
\[ \etmu_s(f) \leq \etmu_s(1) \cdot \|f\| \leq 2s \|\ho\|_1 \vartheta(\ep; s) \cdot \|f\|. \]
Next, using \cref{lem:etmu-bimod}, \cref{lem:etmu-ctl-egxi}, and \cref{prop:etmu-vartheta-bnd},
we may write
\begin{align*}
\etmu_s(f - \egxi_{s \gets t}(f))
&= \etmu_s\p{\int_s^t \dr[\tau] \egxi_{s \gets \tau}(\egdf_\tau f)} \\
&\leq \frac{1}{\ep} \int_s^t \dr[\tau] \sum_{\alpha=1}^\infty
    \etmu_s(\egxi_{s \gets \tau}(
        [\Ea_{\tau / \ep^2} \Re(e^{i\tau\wa / \ep^2} \ghoa_\tau), f]
    )) \\
&\leq \frac{1}{\ep} \int_s^t \dr[\tau] \sum_{\alpha=1}^\infty
    2\etmu_\tau(\Ea_{\tau / \ep^2}) \cdot \|\hoa\| \,\, \|f\|
\leq \frac{2}{\ep} \|f\| \int_s^t \dr[\tau] \sum_{\alpha=1}^\infty
    \ep \vartheta(\ep; \tau) \cdot \|\hoa\| \\
&\leq 2 (t - s) \|f\| \vartheta(\ep; t) \sum_{\alpha=1}^\infty \|\hoa\|
\leq 2(t - s) \vartheta(\ep; t) \|\ho\|_1 \cdot \|f\|.
\end{align*}
\end{proof}

\subsection{Stochastic Limit}
\label{subsect:stoch-limit}

To prove the time-centered weak convergence of $ \eX $,
we will first prove the convergence of the generators of $ \egX $
with respect to $ \etmu_\bt $.
Though, it must be kept in mind that the processes $ \egX $ are not themselves Markov.
Next, we will use the convergence of generators to demonstrate that $ \egX $
asymptotically fulfills the martingale problem for $ L $.
Then, we may prove the convergence of (time-centered) $ n $-point functions of $ \eX $
to the $ n $-point functions of a process generated by $ L $.
\medskip

\subsubsection{Generator Convergence}

To clarify the following computations,
we use a form of big-$ O $ notation in this section.
More precisely, suppose $ V $ is a $ \bC $-vector space
with a seminorm $ q : V \to \ebRp $.
We fix $ \bar{\ep} > 0 $ and $ \bar{t} \geq 0 $.
Then, for any real-valued expression $ r $, we will use $ O_q(r) $
to indicate an element $ g \in V $ such that $ q(g) \leq rC $
where $ C > 0 $ is a constant
which may depend on $ \bar{\ep} $, $ \bar{t} $, and $ \eta $ from \cref{cond:graph-ctl}
as well as the collections $ \{\Ea\} $ and $ \{\hoa\} $
but not on $ \ep $, $ t $, $ \alpha $, $ \beta $, or other introduced variables.
Abusing notation, we use $ O_\cM(\cdot) $ to refer to $ O_{\|\cdot\|_\cM}(\cdot) $.
When $ q = |\cdot| $ and $ V = \bR $, we simply say $ O(\cdot) $.
\medskip

\begin{lem} \label{lem:cab-integ-approx}
For any $ \alpha, \beta \in \bN $
and times $ 0 \leq s \leq t $, one has
\begin{align}
\label{eq:re-cab-integ-approx}
\int_s^t \dr[\tau] \mean{\{\Ea_{s / \ep^2}, \Eb_{\tau / \ep^2}\}} \Re(e^{i\tau\wb / \ep^2} \ghob_\tau)
&= 2\ep^2 \Re(\cab^R(\wb)^* \,\, e^{is\wb / \ep^2} \ghob_s)
    + O_\cM\p{\frac{\ep^4}{\eta^2} \|\fks \hob\| + \frac{\ep^2}{\eta} \|\hob\| e^{-\eta |t - s| / \ep^2}} \\
\label{eq:im-cab-integ-approx}
\int_s^t \dr[\tau] \mean{[\Ea_{s / \ep^2}, \Eb_{\tau / \ep^2}]} \Re(e^{i\tau\wb / \ep^2} \ghob_\tau)
&= 2i\ep^2 \Re(\cab^I(\wb)^* \,\, e^{is\wb / \ep^2} \ghob_s)
    + O_\cM\p{\frac{\ep^4}{\eta^2} \|\fks \hob\| + \frac{\ep^2}{\eta} \|\hob\| e^{-\eta |t - s| / \ep^2}}.
\end{align}
\end{lem}
\begin{proof}  
First, we decompose \cref{eq:im-cab-integ-approx}
using $ 2\Re(e^{i\tau\wb / \ep^2} \ghob_\tau) = e^{i\tau\wb / \ep^2} \ghob_\tau + e^{-i\tau\wb / \ep^2} (\ghob_\tau)^* $.
From \cref{cond:Ea-cov-stat}, we know $ \{\Ea\} $ are covariance stationary.
So, by substituting $ \tau = s + \ep^2 \dl $ so $ \dl = (\tau - s) / \ep^2 $, we have
\begin{align*}
\int_s^t \dr[\tau]
  \mean{[\Ea_{s / \ep^2}, \Eb_{\tau / \ep^2}]} e^{i\tau\wb / \ep^2} \ghob_\tau
&= \ep^2 \int_0^{\frac{t - s}{\ep^2}} \dr[\dl]
  \mean{[\Ea_{s / \ep^2}, \Eb_{\dl + s / \ep^2}]}
  e^{i\dl\wb} e^{is\wb / \ep^2} \ghob_{s + \ep^2 \dl} \\
&= \ep^2 e^{is\wb / \ep^2} \int_0^{\frac{t - s}{\ep^2}} \dr[\dl]
  e^{i\dl\wb} \mean{[\Ea_0, \Eb_\dl]}
  \ghob_{s + \ep^2 \dl}.
\end{align*}
We can write the conjugate term similarly:
\[
\int_s^t \dr[\tau]
  \mean{[\Ea_{s / \ep^2}, \Eb_{t / \ep^2}]} e^{-i\tau\wb / \ep^2} (\ghob_\tau)^*
= \ep^2 e^{-is\wb / \ep^2} \int_0^{\frac{t - s}{\ep^2}} \dr[\dl]
  e^{-i\dl\wb} \mean{[\Ea_0, \Eb_\dl]}
  (\ghob_{s + \ep^2 \dl})^*.
\]
From \cref{cond:ha-eigen}, we know $ \hob \in \dom(\fks) $
so for any $ \dl > 0 $,
\[
\nrM{\ghob_{s + \ep^2 \dl} - \ghob_s}
\leq \int_s^{s + \ep^2 \dl} \dr[\tau] \|\kps_\tau(\fks \hob)\|
= \ep^2 \dl \|\fks \hob\|
\]
so $ \ghob_{s + \ep^2 \dl} \in \ghob_s + O_\cM(\ep^2 \dl \|\fks \hob\|) $.
From \cref{cond:graph-ctl}, we know $ \ab{\mean{[\Ea_0, \Eb_\dl]}} \leq e^{-\eta \dl} $.
Thus, using \cref{eq:cab-im-defn}
and the fact that $ \cab^I $ is a Hermitian function, we get
\begin{align}
\ep^2 e^{is\wb / \ep^2} &\int_0^{\frac{t - s}{\ep^2}} \dr[\dl] e^{i\dl\wb} \mean{[\Ea_0, \Eb_\dl]} \ghob_{s + \ep^2 \dl} \\
&= \ep^2 e^{is\wb / \ep^2} \int_0^\infty \dr[\dl] e^{i\dl\wb} \mean{[\Ea_0, \Eb_\dl]} \ghob_{s + \ep^2 \dl}
    - \ep^2 e^{is\wb / \ep^2} \int_{\frac{t - s}{\ep^2}}^\infty \dr[\dl] e^{i\dl\wb} \mean{[\Ea_0, \Eb_\dl]} \ghob_{s + \ep^2 \dl}
\notag \\
&= \ep^2 e^{is\wb / \ep^2} \int_0^\infty \dr[\dl] e^{i\dl\wb} \mean{[\Ea_0, \Eb_\dl]} (\ghob_s + O_\cM(\ep^2 \dl \|\fks \hob\|))
    + \ep^2 \int_{\frac{t - s}{\ep^2}}^\infty \dr[\dl] e^{-\eta \dl} \,\, O_\cM(\|\hob\|) \notag \\
&= \ep^2 e^{is\wb / \ep^2} \int_0^\infty \dr[\dl] e^{i\dl\wb} \mean{[\Ea_0, \Eb_\dl]} \ghob_s
    + \ep^2 \int_0^\infty \dr[\dl] e^{-\eta \dl} \dl \cdot O_\cM(\ep^2 \|\fks \hob\|)
    + \frac{\ep^2}{\eta} e^{-\eta |t - s| / \ep^2} O_\cM(\|\hob\|) \notag \\
&= \ep^2 e^{is\wb / \ep^2} \cdot 2i \cab^I(-\wb) \ghob_s
    + \frac{1}{\eta^2} O_\cM(\ep^4 \|\fks \hob\|)
    + e^{-\eta |t - s| / \ep^2} \frac{\ep^2}{\eta} O_\cM(\|\hob\|) \notag \\
\label{eq:re-cab-std}
&= 2i\ep^2 \cab^I(\wb)^* \,\, e^{is\wb / \ep^2} \ghob_s
    + O_\cM\p{\frac{\ep^4}{\eta^2} \|\fks \hob\| + \frac{\ep^2}{\eta} \|\hob\| e^{-\eta |t - s| / \ep^2}}.
\end{align}
Similarly, the conjugate term can be approximated by
\begin{equation} \label{eq:re-cab-conj} \begin{split}
\ep^2 e^{-is\wb / \ep^2} \int_0^{\frac{t - s}{\ep^2}} \dr[\dl] e^{-i\dl\wb} \mean{[\Ea_0, \Eb_\dl]} (\ghob_{s + \ep^2 \dl})^*
&= 2i\ep^2 \cab^I(\wb) \,\, e^{-is\wb / \ep^2} (\ghob_s)^* \\
&\quad + O_\cM\p{\frac{\ep^4}{\eta^2} \|\fks \hob\| + \frac{\ep^2}{\eta} \|\hob\| e^{-\eta |t - s| / \ep^2}}.
\end{split} \end{equation}
Then, taking the mean of \cref{eq:re-cab-std} and \cref{eq:re-cab-conj}
gives \cref{eq:im-cab-integ-approx}.
By a similar argument, one obtains \cref{eq:re-cab-integ-approx}.
\end{proof}

The following expression will be used in the statements and proofs
of \cref{lem:egxi-kps-lip} and \cref{prop:sreso-approx}.
For any $ h_1, h_2 \in \dom(\fks) \subseteq \cM $ and $ t \geq 0 $, we define
\[
B(h_1, h_2) := \|h_1\| \,\, \|h_2\| +
    \|\fks h_1\| \,\, \|h_2\| + \|h_1\| \,\, \|\fks h_2\|.
\]
\smallskip

\begin{lem} \label{lem:egxi-kps-lip}
For any $ \ep > 0 $, times $ 0 \leq a \leq s \leq \tau $,
and $ h_1, h_2, f \in \cM $ with $ h_1, h_2 \in \dom(\fks) $,
\begin{equation} \label{eq:egxi-kps-lip} \begin{split}
&\etmu_a\p{\egxi_{a \gets s}(\kps_s(h_1 h_2) f) - \egxi_{a \gets \tau}(\kps_\tau(h_1 h_2) f)},
\etmu_a\p{\egxi_{a \gets s}(\kps_s(h_1) f \kps_s(h_2)) - \egxi_{a \gets \tau}(\kps_\tau(h_1) f \kps_\tau(h_2))} \\
&\qquad \leq 2 (s + 1) (\tau - s) \|\ho\|_1 \vartheta(\ep; \tau) \cdot \|f\| B(h_1, h_2).
\end{split} \end{equation}
Moreover, for any $ t \geq a $,
the functions $ s \mapsto \egxi_{a \gets s}(\kps_s(h_1 h_2) f) $
and $ s \mapsto \egxi_{a \gets s}(\kps_s(h_1) f \kps_s(h_2)) $
are Lipschitz for $ s \in [a, t] $ with respect to $ \etmu_a $
with constant $ 2 (t + 1) \|\ho\|_1 \vartheta(\ep; t) \cdot \|f\| B(h_1, h_2) $.
\end{lem}
\begin{proof}  
First, we show the inequality for the $ \kps_s(h_1 h_2) f $ term.
Using \cref{lem:etmu-ctl-egxi}, we can rewrite the expression as
\begin{align*}
\etmu_a\p{\egxi_{a \gets s}(\kps_s(h_1 h_2) f) - \egxi_{a \gets \tau}(\kps_\tau(h_1 h_2) f)}
&= \etmu_a\p{\egxi_{a \gets s}\p{\kps_s(h_1 h_2) f - \egxi_{s \gets \tau}(\kps_\tau(h_1 h_2) f)}} \\
&\leq \etmu_s\p{\kps_s(h_1 h_2) f - \egxi_{s \gets \tau}(\kps_\tau(h_1 h_2) f)} \\
&\leq \etmu_s\p{\kps_s(h_1 h_2) f - \kps_\tau(h_1 h_2) f}
    + \etmu_s\p{\kps_\tau(h_1 h_2) f - \egxi_{s \gets \tau}(\kps_\tau(h_1 h_2) f)}.
\end{align*}
Now, observe that
\[
\nr{\kps_s(h_1 h_2) - \kps_\tau(h_1 h_2)}
\leq \int_s^\tau \dr[r] \|\kps_r(\fks h_1)\|
= (\tau - s) \|\fks h_1\|.
\]
Thus, from \cref{coro:egxi-etmu-lip}, we have
\begin{align}
\etmu_s\p{\kps_s(h_1 h_2) f - \kps_\tau(h_1 h_2) f}
&\leq 2s \|\ho\|_1 \vartheta(\ep; s) \cdot \|\kps_s(h_1 h_2) f - \kps_\tau(h_1 h_2) f\| \notag \\
&\leq 2s \|\ho\|_1 \vartheta(\ep; s) \cdot (\tau - s) \|\fks (h_1 h_2)\| \,\, \|f\| \notag \\
&\leq 2s (\tau - s) \|\ho\|_1 \vartheta(\ep; s) \cdot \|f\| \p{\big.
    \|\fks h_1\| \,\, \|h_2\| + \|h_1\| \,\, \|\fks h_2\|
} \notag \\
\label{eq:kps-change-bnd}
&\leq 2s (\tau - s) \|\ho\|_1 \vartheta(\ep; s) \cdot \|f\| B(h_1, h_2).
\end{align}
Also, using \cref{coro:egxi-etmu-lip}, we obtain the bound
\begin{align}
\etmu_s\p{\kps_\tau(h_1 h_2) f - \egxi_{s \gets \tau}(\kps_\tau(h_1 h_2) f)}
&\leq 2(\tau - s) \vartheta(\ep; \tau) \|\ho\|_1 \cdot \|\kps_\tau(h_1 h_2) f\| \notag \\
&\leq 2(\tau - s) \|\ho\|_1 \vartheta(\ep; \tau) \cdot \|h_1\| \,\, \|h_2\| \,\, \|f\| \notag \\
\label{eq:egxi-change-bnd}
&\leq 2(\tau - s) \|\ho\|_1 \vartheta(\ep; \tau) \cdot \|f\| B(h_1, h_2).
\end{align}
From \cref{lem:vartheta-conv}, we know $ \vartheta(\ep; s) \leq \vartheta(\ep; \tau) $.
Then, combining \cref{eq:kps-change-bnd} and \cref{eq:egxi-change-bnd}, we get
\[
\etmu_s\p{\kps_s(h_1 h_2) f - \kps_\tau(h_1 h_2) f}
\leq 2s (\tau - s) \|\ho\|_1 \vartheta(\ep; s) \cdot \|f\| B(h_1, h_2)
    + 2(\tau - s) \|\ho\|_1 \vartheta(\ep; \tau) \cdot \|f\| B(h_1, h_2).
\]
yielding the first bound in \cref{eq:egxi-kps-lip}.
\smallskip

Next, we consider the $ \kps_s(h_1) f \kps_s(h_2) $ term.
Note that
\begin{align*}
\nr{\kps_s(h_1) f \kps_s(h_2) - \kps_\tau(h_1) f \kps_\tau(h_2)}
&\leq \nr{\kps_s(h_1) f \kps_s(h_2) - \kps_\tau(h_1) f \kps_s(h_2)}
    + \nr{\kps_\tau(h_1) f \kps_s(h_2) - \kps_\tau(h_1) f \kps_\tau(h_2)} \\
&\leq (\tau - s) \|\fks h_1\| \,\, \|f\| \,\, \|\kps_s(h_2)\|
    + (\tau - s) \|\kps_\tau(h_1)\| \,\, \|f\| \,\, \|\fks h_2\| \\
&\leq (\tau - s) \|f\| (\|\fks h_1\| \,\, \|h_2\| + \|h_1\| \,\, \|\fks h_2\|) \\
&\leq (\tau - s) \|f\| B(h_1, h_2).
\end{align*}
Then, using \cref{coro:egxi-etmu-lip},
\begin{align*}
&\etmu_a\p{\egxi_{a \gets s}(\kps_s(h_1) f \kps_s(h_2)) - \egxi_{a \gets \tau}(\kps_\tau(h_1) f \kps_\tau)} \\
&\qquad \leq \etmu_s\p{\kps_s(h_1) f \kps_s(h_2) - \kps_\tau(h_1) f \kps_\tau(h_2)} \\
    &\qquad\qquad + \etmu_s\p{\kps_\tau(h_1) f \kps_\tau(h_2) - \egxi_{s \gets \tau}(\kps_\tau(h_1) f \kps_\tau(h_2))} \\
&\qquad \leq 2s\|\ho\|_1 \vartheta(\ep; s) \cdot (\tau - s) \|f\| B(h_1, h_2)
    + 2(\tau - s) \|\ho\|_1 \vartheta(\ep; \tau) \cdot \|f\| B(h_1, h_2) \\
&\qquad \leq 2 (s + 1) (\tau - s) \|\ho\|_1 \vartheta(\ep; \tau) \cdot \|f\| B(h_1, h_2; s).
\end{align*}
yielding the second bound in \cref{eq:egxi-kps-lip}.
Lastly, for any $ t \geq a $ and all $ s \in [a, t] $,
we know $ s + 1 \leq t + 1 $ and $ \vartheta(\ep; s) \leq \vartheta(\ep; t) $
so the Lipschitz condition follows from \cref{eq:egxi-kps-lip}.
\end{proof}

\begin{lem} \label{lem:fast-osc-vanish}
Suppose $ V $ is a $ \bC $-vector space with seminorm $ q : V \to \ebRp $
and $ t \geq a \geq 0 $.
Let $ C_1, C_2 > 0 $ be real and $ \gamma : [a, t] \to V $
be Lipschitz with respect to $ q $ with constant $ C_1 $
with $ q(\gamma(s)) \leq C_2 $ for all $ s \in [a, t] $.
Then, for any $ \w \in \bR $ with $ \w \neq 0 $,
\[
q\p{\int_a^t \dr[s] e^{i\w s} \gamma(s)}
\leq \frac{\pi}{|\w|} ((t - a) C_1 + C_2).
\]
\end{lem}
\begin{proof}  
We use $ P := \frac{2\pi}{|\w|} $ to refer to the period of $ s \mapsto e^{i\w s} $.
Then, because $ \gamma $ is $ C_1 $-Lipschitz, we have
\begin{align}
q\p{\gamma(s) - \frac{1}{P} \int_s^{s + P} \dr[\tau] \gamma(\tau)}
&\leq \frac{1}{P} \int_s^{s + P} \dr[\tau] q(\gamma(s) - \gamma(\tau)) \notag \\
&\leq \frac{1}{P} \int_s^{s + P} \dr[\tau] C_1 (\tau - s) \notag \\
\label{eq:osc-lip-bnd}
&= \frac{C_1}{P} \int_0^P \dl \dr[\dl]
= \frac{C_1}{P} \cdot \frac{P^2}{2} = \frac{P C_1}{2}.
\end{align}
Because $ P $ is the period of $ e^{i\w s} $, we have
\[
\int_{\tau - P}^\tau \dr[s] e^{i\w s} = 0
\quad \text{for all} \,\, \tau \geq 0.
\]
Then, we can bound the integral of $ \frac{1}{P} \int_s^{s + P} \dr[\tau] \gamma(\tau) $ by
\begin{align}
&q\p{\int_a^t \dr[s] e^{i\w s} \cdot \frac{1}{P} \int_s^{s + P} \dr[\tau] \gamma(\tau)} \notag
= \frac{1}{P} q\p{\int_a^{t + P} \dr[\tau] \gamma(\tau) \cdot \int_{a \lor (\tau - P)}^\tau \dr[s] e^{i\w s}} \\
&\qquad \leq \frac{1}{P} q\p{\int_{a + P}^{t + P} \dr[\tau] \gamma(\tau) \cdot \int_{\tau - P}^\tau \dr[s] e^{i\w s}}
    + \frac{1}{P} q\p{\int_a^{a + P} \dr[\tau] \gamma(\tau) \cdot \int_a^\tau \dr[s] e^{i\w s}} \notag \\
&\qquad \leq \frac{1}{P} q\p{\int_{a + P}^{t + P} \dr[\tau] \gamma(\tau) \cdot 0}
    + \frac{1}{P} \int_a^{a + P} \dr[\tau] C_2 \cdot \int_a^\tau \dr[s] |e^{i\w s}| \notag \\
\label{eq:osc-period-bnd}
&\qquad = \frac{C_2}{P} \int_a^{a + P} \dr[\tau] (\tau - a)
= \frac{C_2}{P} \int_0^P \dl \dr[\dl]
= \frac{C_2}{P} \cdot \frac{P^2}{2} = \frac{P C_2}{2}.
\end{align}
Now, we integrate \cref{eq:osc-lip-bnd} and combine it with \cref{eq:osc-period-bnd} to get
\begin{align*}
q\p{\int_a^t \dr[s] e^{i\w s} \gamma(s)}
&\leq q\p{\int_a^t \dr[s] e^{i\w s} \p{\gamma(s) - \frac{1}{P} \int_s^{s + P} \dr[\tau] \gamma(\tau)}}
    + q\p{\int_a^t \dr[s] e^{i\w s} \cdot \frac{1}{P} \int_s^{s + P} \dr[\tau] \gamma(\tau)} \\
&\leq \int_a^t \dr[s] |e^{i\w s}| \cdot \frac{P C_1}{2}
    + \frac{P C_2}{2}
= \frac{P}{2} ((t - a) C_1 + C_2)
= \frac{\pi}{|\w|} ((t - a) C_1 + C_2).
\end{align*}
\end{proof}

The property demonstrated by \cref{prop:sreso-approx}
is referred to as the \textsl{stochastic resonance principle}
by Accardi, Lu, and Volovich (see section 11.8 of \cite{ALV02}).
As $ \ep \dto 0 $, this principle forces $ \hoa $ and $ \hob $
to have the same frequency with respect to the fast dynamics
(i.e.\ $ \wa = \wb $) in order to interact.
\medskip

\begin{prop} \label{prop:sreso-approx}
Suppose $ \ep \in (0, \bar{\ep}) $ and $ a, t \in [0, \bar{t}] $ with $ a \leq t $.
For any $ f \in \cM $ with $ f^* = f $ and $ \alpha, \beta \in \bN $,
\begin{align}
\int_a^t \dr[s] & \egxi_{a \gets s}\p{\big.
    [\Re(e^{is\wa / \ep^2} \ghoa_s), [\Re(\cab^R(\wb)^* \,\, e^{is\wb / \ep^2} \ghob_s), f]]
} \notag \\
\label{eq:sreso-approx-re}
&= \bbb{\wa = \wb} \int_a^t \dr[s] \Re\p{
    \cab^R(\wb)^* \egxi_{a \gets s}([(\ghoa_s)^*, [\ghob_s, f]])
} + \frac{\ep^2 \nr{f}}{\Delta\wa \lor \Delta\wb} O_{\etmu_a}(B(\hoa, \hob)) \\
\int_a^t \dr[s] & \egxi_{a \gets s}\p{\big.
    \{\Re(e^{is\wa / \ep^2} \ghoa_s), [i\Re(\cab^I(\wb)^* \,\, e^{is\wb / \ep^2} \ghob_s), f]\}
} \notag \\
\label{eq:sreso-approx-im}
&= - \bbb{\wa = \wb} \int_a^t \dr[s] \Im\p{
    \cab^I(\wb)^* \egxi_{a \gets s}(\{(\ghoa_s)^*, [\ghob_s, f]\})
} + \frac{\ep^2 \|f\|}{\Delta\wa \lor \Delta\wb} O_{\etmu_a}(B(\hoa, \hob))
\end{align}
with $ \Delta\wa $ and $ \Delta\wb $ defined as in \cref{cond:ha-freq-sep}.
\end{prop}
\begin{proof}  
First, we consider \cref{eq:sreso-approx-re}.
We can write
\begin{align}
&\egxi_{a \gets s}\p{ \big.
    [\Re(e^{is\wa / \ep^2} \ghoa_s), [\Re(\cab^R(\wb)^* \;\; e^{is\wb / \ep^2} \ghob_s), f]]
} \notag \\
&\qquad = \frac{1}{4} e^{is(\wb + \wa) / \ep^2} \cab^R(\wb)^* \;\; \egxi_{a \gets s}([\ghoa_s, [\ghob_s, f]])
  + \frac{1}{4} e^{-is(\wb + \wa) / \ep^2} \cab^R(\wb) \;\; \egxi_{a \gets s}([(\ghoa_s)^*, [(\ghob_s)^*, f]]) \notag \\
&\qquad\qquad + \frac{1}{4} e^{is(\wb - \wa) / \ep^2} \cab^R(\wb)^* \;\; \egxi_{a \gets s}([(\ghoa_s)^*, [\ghob_s, f]])
  + \frac{1}{4} e^{-is(\wb - \wa) / \ep^2} \cab^R(\wb) \;\; \egxi_{a \gets s}([\ghoa_s, [(\ghob_s)^*, f]]) \notag \\
\label{eq:sreso-approx-pieces-re}
&\qquad = \frac{1}{2} \Re\p{e^{is(\wb + \wa) / \ep^2} \cab^R(\wb)^* \;\; \egxi_{a \gets s}([\ghoa_s, [\ghob_s, f]])}
+ \frac{1}{2} \Re\p{e^{is(\wb - \wa) / \ep^2} \cab^R(\wb)^* \;\; \egxi_{a \gets s}([(\ghoa_s)^*, [\ghob_s, f]])}
\end{align}
using the fact that $ [\ghoa_s, [\ghob_s, f]]^* = [(\ghoa_s)^*, [(\ghob_s)^*, f]] $.
We now treat each term separately.
From \cref{cond:graph-ctl}, we know
\begin{align*}
\ab{\mean{\Ea_0 \Eb_\dl}} &\leq e^{-\eta \dl} \quad \text{for all} \;\; \dl \geq 0 \\
\text{so for any} \,\, \w, \quad
|\cab^R(\w)|, |\cab^I(\w)| &\leq \int_0^\infty \dr[\dl] e^{-\eta \dl} = \frac{1}{\eta}.
\end{align*}
From \cref{lem:egxi-kps-lip}, we know
the functions $ s \mapsto \egxi_{a \gets s}([\ghoa_s, [\ghob_s, f]]) $
and $ s \mapsto \egxi_{a \gets s}([((\ghoa_s)^*, [\ghob_s, f])]) $
are Lipschitz for $ s \in [a, t] $ with respect to $ \etmu_a $ with constant
\[ 8 (t + 1) \|\ho\|_1 \vartheta(\ep; t) \cdot \|f\| B(\hoa, \hob). \]
Additionally, this function is uniformly bounded
by $ 4 \|\hoa\| \,\, \|\hob\| \,\, \|f\| \leq 4 \|f\| B(\hoa, \hob) $.
So, from \cref{lem:fast-osc-vanish},
since $ |\wa - \wb| \geq \Delta\wa \lor \Delta\wb $, we get
\begin{align}
&\etmu_a\p{\int_a^t \dr[s] e^{is(\wb + \wa) / \ep^2} \cab^R(\wb)^* \;\; \egxi_{a \gets s}([\ghoa_s, [\ghob_s, f]])} \notag \\
&\qquad \leq \frac{1}{\eta} \cdot \frac{\pi \ep^2}{|\wa + \wb|} \p{\big.
    (t - a) \cdot 8 (t + 1) \|\ho\|_1 \vartheta(\ep; t) \cdot \|f\| B(\hoa, \hob) + 4\|f\| B(\hoa, \hob)
} \notag \\
&\qquad \leq \frac{1}{\eta} \cdot \frac{\pi \ep^2}{|\wa - \wb|} \cdot \|f\| B(\hoa, \hob) \p{
    \bar{t} \cdot 8 (\bar{t} + 1) \|\ho\|_1 \vartheta(\bar{\ep}; \bar{t}) + 4
} \notag \\
&\qquad = \frac{\ep^2 \|f\|}{|\wa - \wb|} O(B(\hoa, \hob)) \notag \\
&\qquad = \frac{\ep^2 \|f\|}{\Delta\wa \lor \Delta\wb} O(B(\hoa, \hob)) \notag \\
\label{eq:sreso-oppos-piece}
\text{and so} &\qquad
\int_a^t \dr[s] \Re(e^{is(\wb + \wa) / \ep^2} \cab^R(\wb)^* \;\; \egxi_{a \gets s}([\ghoa_s, [\ghob_s, f]]))
= \frac{\ep^2 \|f\|}{\Delta\wa \lor \Delta\wb} O_{\etmu_a}(B(\hoa, \hob)).
\end{align}
When $ \wa \neq \wb $, from \cref{lem:egxi-kps-lip}, we know
the function $ s \mapsto \egxi_{} $
is Lipschitz for $ s \in [a, t] $ with respect to $ \etmu_a $ with constant
\[ 8 (t + 1) \|\ho\|_1 \vartheta(\ep; t) \cdot \|f\| B(\hoa, \hob) \]
and uniformly bounded by $ 4 \|f\| B(\hoa, \hob) $.
Again, from \cref{lem:fast-osc-vanish}, we get
\begin{align*}
&\etmu_a\p{\int_a^t \dr[s] e^{is(\wb - \wa) / \ep^2} \cab^R(\wb)^* \;\; \xi_{a \gets s}([(\ghoa_s)^*, [\ghob_s, f]])} \\
&\qquad \leq \frac{1}{\eta} \cdot \frac{\pi \ep^2}{|\wa - \wb|} \p{\big.
    (t - a) \cdot 8 (t + 1) \|\ho\|_1 \vartheta(\ep; t) \cdot \|f\| B(\hoa, \hob) + 4\|f\| B(\hoa, \hob)
} \\
&\qquad = \frac{\ep^2 \|f\|}{\Delta\wa \lor \Delta\wb} O(B(\hoa, \hob)).
\end{align*}
Otherwise, $ \wb - \wa = 0 $ and
\[
\int_a^t \dr[s] e^{is(\wb - \wa) / \ep^2} \cab^R(\wb)^* \;\; \egxi_{a \gets s}([(\ghoa_s)^*, [\ghob_s, f]])
= \int_a^t \dr[s] \cab^R(\wb)^* \;\; \egxi_{a \gets s}([(\ghoa_s)^*, [\ghob_s, f]]).
\]
Thus, in either case,
\begin{align}
\int_a^t \dr[s] e^{is(\wb - \wa) / \ep^2} \cab^R(\wb)^* \;\; \egxi_{a \gets s}([(\ghoa_s)^*, [\ghob_s, f]])
&= \bbb{\wa = \wb} \int_0^t \dr[s] \cab^R(\wb)^* \;\; \egxi_{a \gets s}([(\ghoa_s)^*, [\ghob_s, f]]) \notag \\
    &\qquad + \frac{\ep^2 \|f\|}{\Delta\wa \lor \Delta\wb} O_{\etmu_a}(B(\hoa, \hob)) \notag \\
\label{eq:sreso-align-piece}
\text{and so} \quad
\int_a^t \dr[s] \Re(e^{is(\wb - \wa) / \ep^2} \cab^R(\wb)^* \;\; \egxi_{a \gets s}([(\ghoa_s)^*, [\ghob_s, f]]))
&= \bbb{\wa = \wb} \int_0^t \dr[s] \Re(\cab^R(\wb)^* \;\; \egxi_{a \gets s}([(\ghoa_s)^*, [\ghob_s, f]])) \\
    &\qquad + \frac{\ep^2 \|f\|}{\Delta\wa \lor \Delta\wb} O_{\etmu_a}(B(\hoa, \hob)). \notag
\end{align}
Combining the bounds from \cref{eq:sreso-oppos-piece}
and \cref{eq:sreso-align-piece},
we can approximate \cref{eq:sreso-approx-pieces-re} by
\begin{align*}
\int_0^t \dr[s] & [\Re(e^{is\wa / \ep^2} \ghoa_s), [\Re(\cab^R(\wb)^* e^{is\wb / \ep^2} \ghob_s), f]] \\
&= \bbb{\wa = \wb} \int_0^t \dr[s] \Re\p{
        \cab^R(\wb)^* \,\, \egxi_{a \gets s}([(\ghoa_s)^*, [\ghob_s, f]])
    } + \frac{\ep^2 \|f\|}{\Delta\wa \lor \Delta\wb} O(B(\hoa, \hob))
\end{align*}
proving \cref{eq:sreso-approx-re}.
\medskip

Similarly, for \cref{eq:sreso-approx-im}, we can write
\begin{align}
&\egxi_{a \gets s}\p{
    \{\Re(e^{is\wa / \ep^2} \ghoa_s), [\Re(\cab^I(\wb)^* \;\; e^{is\wb / \ep^2} \ghob_s), f]\}
} \notag \\
&\qquad = \frac{1}{4} e^{is(\wb + \wa) / \ep^2} \cab^I(\wb)^* \;\; \egxi_{a \gets s}(\{\ghoa_s, [\ghob_s, f]\})
  + \frac{1}{4} e^{-is(\wb + \wa) / \ep^2} \cab^I(\wb) \;\; \egxi_{a \gets s}(\{(\ghoa_s)^*, [(\ghob_s)^*, f]\}) \notag \\
&\qquad \quad + \frac{1}{4} e^{is(\wb - \wa) / \ep^2} \cab^I(\wb)^* \;\; \egxi_{a \gets s}(\{(\ghoa_s)^*, [\ghob_s, f]\})
  + \frac{1}{4} e^{-is(\wb - \wa) / \ep^2} \cab^I(\wb) \;\; \egxi_{a \gets s}(\{\ghoa_s, [(\ghob_s)^*, f]\}) \notag \\
\label{eq:sreso-approx-pieces-im}
&\qquad = \frac{1}{2} i\Im\p{e^{is(\wb + \wa) / \ep^2} \cab^I(\wb)^* \;\; \egxi_{a \gets s}(\{\ghoa_s, [\ghob_s, f]\})}
+ \frac{1}{2} i\Im\p{e^{is(\wb - \wa) / \ep^2} \cab^I(\wb)^* \;\; \egxi_{a \gets s}(\{(\ghoa_s)^*, [\ghob_s, f]\})}.
\end{align}
using the fact that $ \{\ghoa_s, [\ghob_s, f]\}^* = -\{(\ghoa_s)^*, [(\ghob_s)^*, f]\} $.
Then, we have the bounds analogous to \cref{eq:sreso-oppos-piece}
and \cref{eq:sreso-align-piece} for \cref{eq:sreso-approx-pieces-im} and so
\begin{align*}
\int_a^t \dr[s] & \egxi_{a \gets s}\p{
    \{\Re(e^{is\wa / \ep^2} \ghoa_s), [i\Re(\cab^I(\wb)^* \;\; e^{is\wb / \ep^2} \ghob_s), f]\}
} \\
&= -\bbb{\wa = \wb} \int_0^t \dr[s] \Im\p{
        \cab^I(\wb)^* \;\; \egxi_{a \gets s}(\{(\ghoa_s)^*, [\ghob_s, f]\})
    } + \frac{\ep^2 \|f\|}{\Delta\wa \lor \Delta\wb} O(B(\hoa, \hob))
\end{align*}
proving \cref{eq:sreso-approx-im}.
\end{proof}

\begin{lem} \label{lem:egdf-mean-approx}
For any $ \ep \in (0, \bar{\ep}) $ and times $ s, t \in [0, \bar{t}] $ and $ t \geq s $,
\[
\int_s^t \dr[\tau] \egdf_s \egdf_\tau f
= O_{\etmu_s}(\ep \|f\|) + \int_s^t \dr[\tau] \meanM{\egdf_s \egdf_\tau f}.
\]
\end{lem}
\begin{proof}  
Expanding and applying \cref{lem:dbl-drv-parts} to $ \egdf_s \egdf_\tau f $, we get
\begin{align*}
\egdf_s \egdf_\tau f
&= -\frac{1}{\ep^2} \sum_{\alpha,\beta=1}^\infty
    [\Ea_{s / \ep^2} \Re(e^{is\wa / \ep^2} \ghoa_s),
        [\Eb_{\tau / \ep^2} \Re(e^{i\tau\wb / \ep^2} \ghob_\tau), f]
    ] \\
&= -\frac{1}{4 \ep^2} \sum_{\alpha,\beta=1}^\infty \Big(
    [\Ea_{s / \ep^2}, \Eb_{\tau / \ep^2}]
    \{\Re(e^{is\wa / \ep^2} \ghoa_s),
        [\Re(e^{i\tau\wb / \ep^2} \ghob_\tau), f]
    \} \\
&\quad  + \{\Ea_{s / \ep^2}, \Eb_{\tau / \ep^2}\}
    [\Re(e^{is\wa / \ep^2} \ghoa_s),
        [\Re(e^{i\tau\wb / \ep^2} \ghob_\tau), f]
    ]
\Big).
\end{align*}
For each $ \alpha, \beta \in \bN $,
using \cref{lem:etmu-bimod} and \cref{prop:etmu-vartheta-bnd},
\begin{align*}
&\etmu_s\p{\int_s^t \dr[\tau] 
    \p{[\Ea_{s / \ep^2}, \Eb_{\tau / \ep^2}] - \mean{[\Ea_{s / \ep^2}, \Eb_{\tau / \ep^2}]}}
    \{\Re(e^{is\wa / \ep^2} \ghoa_s), [\Re(e^{i\tau\wb / \ep^2} \ghob_\tau), f]\}} \\
&\qquad \leq \int_s^t \dr[\tau]
    \etmu_s\p{[\Ea_{s / \ep^2}, \Eb_{\tau / \ep^2}] - \mean{[\Ea_{s / \ep^2}, \Eb_{\tau / \ep^2}]}}
    \cdot \|\hoa\| \,\, \|\hob\| \,\, \|f\| \\
&\qquad \leq \int_s^t \dr[\tau] 2 \ep e^{-\eta |\tau - s| / \ep^2} \vartheta(\ep; s)
    \cdot \|\hoa\| \,\, \|\hob\| \,\, \|f\| \\
&\qquad \leq \|\hoa\| \,\, \|\hob\| \,\, \|f\| \int_0^\infty \ep^2 \dr[\dl] 2\ep e^{-\eta \dl} \vartheta(\ep; s)
= 2\ep^3 \vartheta(\ep; s) \frac{1}{\eta} \cdot \|\hoa\| \,\, \|\hob\| \,\, \|f\|.
\end{align*}
By an analogous argument, the same bound holds for the $ \{\Ea_{s / \ep^2}, \Eb_{\tau / \ep^2}\} $ term.
Then, by summing over $ \alpha $ and $ \beta $, one obtains
\begin{align*}
\etmu_s\p{\int_s^t \dr[\tau] \p{\big. \egdf_s \egdf_\tau f - \meanM{\egdf_s \egdf_\tau f}}}
&\leq \frac{1}{4 \ep^2} \sum_{\alpha,\beta=1}^\infty
    \frac{4\ep^3}{\eta} \vartheta(\ep; s) \cdot \|\hoa\| \,\, \|\hob\| \,\, \|f\| \\
&= \frac{\ep}{\eta} \vartheta(\ep; s) \|f\|
    \p{\sum_{\alpha=1}^\infty \|\hoa\|}
    \p{\sum_{\beta=1}^\infty \|\hob\|} \\
&\leq \frac{\ep \|\ho\|_1^2}{\eta} \vartheta(\ep; s) \|f\|
\leq \frac{\ep \|\ho\|_1^2}{\eta} \vartheta(\bar{\ep}; \bar{t}) \|f\|
\end{align*}
and $ \frac{\ep \|\ho\|_1^2}{\eta} \vartheta(\bar{\ep}; \bar{t}) \|f\| = O(\ep \|f\|) $.
\end{proof}

\begin{prop} \label{prop:gener-limit}
For any $ \bar{t} \geq 0 $ and $ \bar{\ep} > 0 $,
there exists a real $ C_{\bar{\ep}, \bar{t}} > 0 $ such that
for any times $ a, t \in [0, \bar{t}] $ with $ a \leq t $
and any $ f \in \cM $,
\begin{equation} \label{eq:gener-limit}
\int_a^t \dr[\tau] \int_a^\tau \dr[s] \egxi_{a \gets s}(\egdf_s \egdf_\tau (1 \otimes f))
= O_{\etmu_a}(\ep \|f\|) + \int_a^t \dr[s] \egxi_{a \gets s}(\gL_s(f)).
\end{equation}
\end{prop}
\begin{proof}  
Because $ \gL_s $ and $ \egdf_s \egdf_\tau $ are $ \bC $-linear,
it suffices to consider self-adjoint $ f $.
First, since $ a \leq s \leq \tau \leq t $, we can rewrite the integral as
\[
\int_a^t \dr[\tau] \int_a^\tau \dr[s] \egxi_{a \gets s}(\egdf_s \egdf_\tau (1 \otimes f))
= \int_a^t \dr[s] \egxi_{a \gets s}\p{\int_s^t \dr[\tau] \egdf_s \egdf_\tau f}.
\]
Then, by \cref{lem:egdf-mean-approx},
\begin{align*}
\int_a^t \dr[s] \egxi_{a \gets s}\p{\int_s^t \dr[\tau] \egdf_s \egdf_\tau f}
&= \int_a^t \dr[s] \egxi_{a \gets s}\p{O_{\etmu_s}(\ep \|f\|) + \int_s^t \dr[\tau] \meanM{\egdf_s \egdf_\tau f}} \\
&= (t - a) O_{\etmu_a}(\ep \|f\|) + \int_a^t \dr[s] \egxi_{a \gets s}\p{\int_s^t \dr[\tau] \meanM{\egdf_s \egdf_\tau f}}
\end{align*}
where $ t - a \leq \bar{t} $ so $ (t - a) O_{\etmu_a}(\ep \|f\|) \leq O_{\etmu_a}(\ep \|f\|) $.
Hence, it only remains to show that
\begin{equation} \label{eq:mean-to-gener}
\int_a^t \dr[s] \int_s^t \dr[\tau] \egxi_{a \gets s}\p{\meanM{\egdf_s \egdf_\tau f}}
= \int_a^t \dr[s] \egxi_{a \gets s}(\gL_s(f)) + O_{\etmu_a}(\ep \|f\|).
\end{equation}
Applying \cref{lem:dbl-drv-parts} to $ \egdf_s \egdf_\tau f $, the double integral becomes
\begin{align}
&\int_a^t \dr[s] \int_s^t \dr[\tau] \egxi_{a \gets s}\p{\meanM{\egdf_s \egdf_\tau f}} \\
&\qquad = -\frac{1}{\ep^2} \int_a^t \dr[s] \int_s^t \dr[\tau] \sum_{\alpha,\beta=1}^\infty
    \egxi_{a \gets s}\p{\meanM{
        [\Ea_{s / \ep^2} \Re(e^{is\wa / \ep^2} \ghoa_s),
            [\Eb_{\tau / \ep^2} \Re(e^{i\tau\wb / \ep^2} \ghob_\tau), f]
        ]
    }} \notag \\
&\qquad = -\frac{1}{4 \ep^2} \int_a^t \dr[s] \int_s^t \dr[\tau] \sum_{\alpha,\beta=1}^\infty \Bigg(
    \mean{\{\Ea_{s / \ep^2}, \Eb_{\tau / \ep^2}\}}
    \egxi_{a \gets s}([\Re(e^{is\wa / \ep^2} \ghoa_s),
        [\Re(e^{i\tau\wb / \ep^2} \ghob_\tau), f]
    ]) \notag \\
&\qquad\quad + \mean{[\Ea_{s / \ep^2}, \Eb_{\tau / \ep^2}]}
    \egxi_{a \gets s}(\{\Re(e^{is\wa / \ep^2} \ghoa_s),
        [\Re(e^{i\tau\wb / \ep^2} \ghob_\tau), f]
    \})
\Bigg) \notag \\
\label{eq:integ-EaEb-inside} \begin{split}
&\qquad= -\frac{1}{4 \ep^2} \int_a^t \dr[s] \sum_{\alpha,\beta=1}^\infty \Bigg(
    \egxi_{a \gets s}\p{\left[\Re(e^{is\wa / \ep^2} \ghoa_s), \left[
        \int_s^t \dr[\tau] \mean{\{\Ea_{s / \ep^2}, \Eb_{\tau / \ep^2}\}}
        \Re(e^{i\tau\wb / \ep^2} \ghob_\tau),
    f\right] \right]} \Bigg) \\
&\qquad\quad + \egxi_{a \gets s}\p{\left\{\Re(e^{is\wa / \ep^2} \ghoa_s), \left[
        \int_s^t \dr[\tau] \mean{[\Ea(s / \ep^2, \Eb_{\tau / \ep^2})]}
        \Re(e^{i\tau\wb / \ep^2} \ghob_\tau),
    f\right] \right\}}
\end{split} &
\end{align}
Then, applying \cref{lem:cab-integ-approx}, \cref{eq:integ-EaEb-inside} becomes
\begin{align}
&\int_a^t \dr[s] \int_s^t \dr[\tau] \egxi_{a \gets s}\p{\meanM{\egdf_s \egdf_\tau f}} \notag \\
&\qquad = -\frac{1}{4 \ep^2} \sum_{\alpha,\beta=1}^\infty \Bigg(
\label{eq:cab-integ-re-sum}
    2\ep^2 \int_a^t \dr[s] \egxi_{a \gets s}\p{
        [\Re(e^{is\wa / \ep^2} \ghoa_s), [\Re(\cab^R(\wb)^* \,\, e^{is\wb / \ep^2} \ghob_s), f]]
    } \\
\label{eq:cab-integ-im-sum}
&\qquad \qquad + 2\ep^2 \int_a^t \dr[s] \egxi_{a \gets s}\p{
        \{\Re(e^{is\wa / \ep^2} \ghoa_s), [i\Re(\cab^I(\wb)^* \,\, e^{is\wb / \ep^2} \ghob_s), f]\}
    }\\
\label{eq:cab-integ-crc-sum}
&\qquad \qquad + 4\|f\| \,\, \|\hoa\| \int_a^t \dr[s] \egxi_{a \gets s}\p{O_\cM\p{
    \frac{\ep^4}{\eta^2} \|\fks \hob\| + \frac{\ep^2}{\eta} \|\hob\| e^{-\eta |t - s| / \ep^2}
}}\Bigg)
\end{align}
Now, using the substitution $ s = t - \ep^2 u $ so $ u = \frac{t - s}{\ep^2} $,
the integral in \cref{eq:cab-integ-crc-sum} can be written as
\[
\int_a^t \dr[s] e^{-\eta |t - s| / \ep^2}
= \int_0^{\frac{t - a}{\ep^2}} \ep^2 \dr[u] e^{-\eta u}
\leq \frac{\ep^2}{\eta}.
\]
Thus, we write the bound in \cref{eq:cab-integ-crc-sum} as
\begin{align*}
\frac{1}{4\ep^2} \sum_{\alpha,\beta=1}^\infty & 4\|f\| \,\, \|\hoa\| \p{
    \int_a^t \dr[s] \frac{\ep^4}{\eta^2} \|\fks \hob\|
    + \frac{\ep^2}{\eta} \|\hob\| \int_a^t \dr[s] e^{-\eta |t - s| / \ep^2}
} \\
&\leq \frac{1}{4 \ep^2} \sum_{\alpha,\beta=1}^\infty 4\|f\| \,\, \|\hoa\| \p{
    (t - a) \frac{\ep^4}{\eta^2} \|\fks \hob\|
    + \frac{\ep^4}{\eta^2} \|\hob\| 
} \\
&= \frac{\ep^2}{\eta^2} \|f\| \p{\sum_{\alpha=1}^\infty \|\hoa\|}
    \sum_{\beta=1}^\infty \p{(t - a) \|\fks \hob\| + \|\hob\|} \\
&\leq \frac{\ep^2}{\eta^2} \|f\| \,\, \|\ho\|_1
    \cdot \p{\big. (t - a) \|\fks \ho\|_1 + \|\ho\|_1} \\
&\leq \frac{\ep^2}{\eta^2} \|f\| \,\, \|\ho\|_1
    \cdot \p{\big. \bar{t} \|\fks \ho\|_1 + \|\ho\|_1}
= O(\ep^2 \|f\|).
\end{align*}
From \cref{lem:etmu-bimod} and \cref{coro:egxi-etmu-lip},
since $ s \leq \bar{t} $ and $ \vartheta(\ep; s) \leq \vartheta(\bar{\ep}; \bar{s}) $ by \cref{lem:vartheta-conv},
we know $ O_\cM(\ep^2 \|f\|) = O_{\etmu_s}(\ep^2 \|f\|) $.
Then, from \cref{lem:etmu-ctl-egxi},
we get $ \egxi_{a \gets s}(O_{\etmu_s}(\ep^2 \|f\|)) = O_{\etmu_a}(\ep^2 \|f\|) $.
Thus, \cref{eq:cab-integ-crc-sum} becomes
\begin{align*}
\int_a^t \dr[s] \egxi_{a \gets s}(O_\cM(\ep^2 \|f\|))
&= \int_a^t \dr[s] \egxi_{a \gets s}(O_{\etmu_s}(\ep^2 \|f\|))
= \int_a^t \dr[s] O_{\etmu_a}(\ep^2 \|f\|) \\
&= (t - a) O_{\etmu_a}(\ep^2 \|f\|)
= \bar{t} O_{\etmu_a}(\ep^2 \|f\|) = O_{\etmu_a}(\ep^2 \|f\|).
\end{align*}
Furthermore, by applying \cref{prop:sreso-approx},
\cref{eq:cab-integ-re-sum} and \cref{eq:cab-integ-im-sum} become
\begin{align}
&-\frac{1}{4\ep^2} \sum_{\alpha,\beta=1}^\infty \Bigg(
    2\ep^2 \int_a^t \dr[s] \egxi_{a \gets s}\p{
        [\Re(e^{is\wa / \ep^2} \ghoa_s), [\Re(\cab^R(\wb)^* \,\, e^{is\wb / \ep^2} \ghob_s), f]]
    } \notag \\
    &\qquad + 2\ep^2 \int_a^t \dr[s] \egxi_{a \gets s}\p{
        \{\Re(e^{is\wa / \ep^2} \ghoa_s), [i\Re(\cab^I(\wb)^* \,\, e^{is\wb / \ep^2} \ghob_s), f]\}
    }\Bigg) \notag \\
\label{eq:sreso-re-im-sum}
&= -\frac{1}{2} \sum_{\alpha,\beta=1}^\infty \bbb{\wa = \wb} \int_a^t \dr[s] \p{
    \Re(\cab^R(\wb)^* \,\, \egxi_{a \gets s}([(\ghoa_s)^*, [\ghob_s, f]]))
    - \Im(\cab^I(\wb)^* \,\, \egxi_{a \gets s}(\{(\ghoa_s)^*, [\ghob_s, f]\}))
} \\
\label{eq:sreso-crc-sum}
&\qquad + \frac{1}{2} \sum_{\alpha,\beta=1}^\infty
    2\frac{\ep^2 \|f\|}{\Delta\wa \lor \Delta\wb} O_{\etmu_a}(B(\hoa, \hob)).
\end{align}
Then, the bound in \cref{eq:sreso-crc-sum} can be written as
\begin{align*}
&\sum_{\alpha,\beta=1}^\infty
    \frac{\ep^2 \|f\|}{\Delta\wa \lor \Delta\wb} B(\hoa, \hob) \\
&\qquad = \ep^2 \|f\| \sum_{\alpha,\beta=1}^\infty \p{
    \frac{\|\fks \hoa\| \,\, \|\hob\|}{\Delta\wa \lor \Delta\wb}
    + \frac{\|\hoa\| \,\, \|\fks \hob\|}{\Delta\wa \lor \Delta\wb}
    + \frac{\|\hoa\| \,\, \|\hob\|}{\Delta\wa \lor \Delta\wb}
} \\
&\qquad \leq \ep^2 \|f\| \p{\sum_{\alpha=1}^\infty \|\fks \hoa\|} \p{\sum_{\beta=1}^\infty \frac{\|\hob\|}{\Delta\wb}}
    + \ep^2 \|f\| \p{\sum_{\alpha=1}^\infty \frac{\|\hoa\|}{\Delta\wa}} \p{\sum_{\beta=1}^\infty \|\fks \hob\|}
    + \ep^2 \|f\| \p{\sum_{\alpha=1}^\infty \frac{\|\hoa\|}{\Delta\wa}} \p{\sum_{\beta=1}^\infty \|\hob\|} \\
&\qquad \leq \ep^2 \|f\| \cdot \|\ho\|_{\Delta,1} \p{\big. 2 \|\fks \ho\|_1 + \|\ho\|_1}
\end{align*}
with $ \|\ho\|_{\Delta,1} $ defined as in \cref{cond:ha-freq-sep}.
Thus, \cref{eq:sreso-crc-sum} is $ O_{\etmu_a}(\ep^2 \|f\|) $.
Next, using \cref{prop:sum-dbl-drv-to-gksl}, we write \cref{eq:sreso-re-im-sum} as
\begin{align}
-\frac{1}{2} \int_a^t \dr[s] &\sum_{\alpha,\beta=1}^\infty \bbb{\wa = \wb} \p{
    \Re(\cab^R(\wb)^* \,\, \egxi_{a \gets s}([(\ghoa_s)^*, [\ghob_s, f]]))
    - \Im(\cab^I(\wb)^* \,\, \egxi_{a \gets s}(\{(\ghoa_s)^*, [\ghob_s, f]\}))
} \notag \\
&= -\frac{1}{2} \int_a^t \dr[s] \sum_{\alpha,\beta=1}^\infty \bbb{\wa = \wb} \Bigg(
    \p{-2\cab^{R+}(\wa) \,\, \egxi_{a \gets s}(\gksl^+_{\ghoa_s,\ghob_s}(f))
    + i\cab^{R-}(\wa) \,\, \egxi_{a \gets s}([[(\ghoa_s)^*, \ghob_s], f])} \\
    &\qquad - \p{-2\cab^{I-}(\wa) \,\, \egxi_{a \gets s}(\gksl^-_{\ghoa_s,\ghob_s}(f))
    - i\cab^{I+}(\wa) \,\, \egxi_{a \gets s}([\{(\ghoa_s)^*, \ghob_s\}, f])}
\Bigg) \notag \\
&= \int_a^t \dr[s] \egxi_{a \gets s}\Bigg(
    \frac{1}{2} \sum_{\substack{\alpha, \beta \in \bN \\ \wa = \wb}} \Big(
        2\cab^{R+}(\wa) \gksl^+_{\ghoa_s,\ghob_s}(f)
        - i\cab^{R-}(\wa) [[(\ghoa_s)^*, \ghob_s], f] \\
        &\qquad - 2\cab^{I-}(\wa) \gksl^-_{\ghoa_s,\ghob_s}(f)
        - i\cab^{I+}(\wa) [\{(\ghoa_s)^*, \ghob_s\}, f]
\Big)\Bigg) \notag \\
\label{eq:terms-to-gener}
&= \int_a^t \dr[s] \egxi_{a \gets s}(\gL_s(f))
\end{align}
where we use the expression for $ \gL_s $
from \cref{eq:int-gener-alt-form} in \cref{lem:gener-alt-form}.
Combining \cref{eq:terms-to-gener} with the corrections
from \cref{eq:cab-integ-crc-sum} and \cref{eq:sreso-crc-sum}, we obtain
\[
\int_a^t \dr[\tau] \int_a^\tau \dr[s] \egxi_{a \gets s}\p{\meanM{\egdf_s \egdf_\tau f}}
= O_{\etmu_a}(\ep^2 \|f\|) + \int_a^t \dr[s] \egxi_{a \gets s}(\gL_s(f))
\]
proving \cref{eq:mean-to-gener}.
\end{proof}

\subsubsection{Asymptotically Martingale}
\label{subsubsect:asymp-martin}
We now have the tools necessary to prove
that $ \etX $ and $ \egtX $ asymptotically behave as martingales
and thus the time-centered weak convergence.
\smallskip

\begin{lem} \label{lem:gener-diff-bnd}
For all $ \ep \in (0, \bar{\ep}) $, $ a, t \in [0, \bar{t}] $ with $ a \leq t $,
and any $ f \in \cM $,
\begin{equation} \label{eq:gener-diff-bnd}
\egxi_{a \gets t}(f) = O_{\etmu_a}(\ep \|f\|) + f + \int_a^b \dr[s] \egxi_{a \gets s}(\gL_s f).
\end{equation}
\end{lem}
\begin{proof}  
We can expand $ \egxi_{a \gets t}(f) $ twice to obtain
\begin{equation} \label{eq:egxi-dbl-expand} \begin{split}
\egxi_{a \gets t}(f) &= f + \int_a^t \dr[s]
  \egxi_{a \gets \tau}(\egdf_\tau f) \\
&= f + \int_a^t \dr[\tau] \egdf_\tau f
  + \int_a^t \dr[\tau] \int_a^\tau \dr[s] \egxi_{a \gets s}
    \p{\egdf_s \egdf_\tau f}.
\end{split} \end{equation}
Now, we consider the first order and second order terms separately.
For the first order term, from \cref{prop:etmu-vartheta-bnd}, we know
\[
\etmu_a(\Ea_{\tau / \ep^2})
\leq \ep e^{-\eta |\tau - a| / \ep^2} \vartheta(\ep; a)
\leq \ep e^{-\eta |\tau - a| / \ep^2} \vartheta(\bar{\ep}; \bar{t}).
\]
Therefore, using \cref{lem:etmu-bimod}, we get
\begin{align*}
\etmu_a\p{\int_a^t \dr[\tau] \egdf_\tau f}
&\leq \int_a^t \dr[\tau] \frac{1}{\ep} \sum_{\alpha=1}^\infty
  \etmu_a\p{[\Ea_{\tau / \ep^2} \Re(e^{i\tau\wa / \ep^2} \ghoa_\tau), f]} \\
&\leq \int_a^t \dr[\tau] \frac{1}{\ep} \sum_{\alpha=1}^\infty
  2 \etmu_a\p{\Ea_{\tau / \ep^2}} \cdot \|\Re(e^{i\tau\wa / \ep^2} \ghoa_\tau)\| \,\, \|f\| \\
&\leq \int_a^t \dr[\tau] \frac{1}{\ep} \sum_{\alpha=1}^\infty
  2 \ep e^{-\eta |\tau - a| / \ep^2} \vartheta(\bar{\ep}; \bar{t}) \cdot \|\hoa\| \,\, \|f\| \\
&\leq 2 \p{\sum_{\alpha=1}^\infty \|\hoa\|} \|f\| \vartheta(\bar{\ep}; \bar{t})
    \int_0^{\frac{t - a}{\ep^2}} \ep^2 \dr[\dl] e^{-\eta \dl} \\
&\leq 2 \ep^2 \|\ho\|_1 \,\, \|f\| \vartheta(\bar{\ep}; \bar{t}) \frac{1}{\eta} \\
\text{and so} \quad
\int_a^t \dr[\tau] \egdf_\tau f &= O_{\etmu_a}(\ep^2 \|f\|).
\end{align*}
Next, from \cref{prop:gener-limit}, we have
\[
\int_a^t \dr[\tau] \int_a^\tau \dr[s] \egxi_{a \gets s}\p{\egdf_s \egdf_\tau f}
= O(\ep \|f\|) + \int_a^t \dr[s] \egxi_{a \gets s}(\gL_s f)
\]
proving \cref{eq:gener-diff-bnd}.
\end{proof}

\begin{prop} \label{prop:approx-martingale}
For any $ \ep \in (0, \bar{\ep}) $ and times $ s, t \in [0, \bar{t}] $ with $ s \leq t $,
centered times $ s_1, \ldots, s_n \in [0, s] $,
and $ g_1, \ldots, g_n, f \in \cM $,
\begin{equation} \label{eq:gen-approx-martingale} \begin{split}
&\nrM{\meanM{g_1(\egtX_{s_1}) \cdots g_k(\egtX_{s_k}) \p{
    f(\egtX_t) - f(\egtX_s) - \int_s^t \dr[\tau] (\gL_\tau f)(\egtX_\tau)
  } g_{k+1}(\egtX_{s_{k+1}}) \cdots g_n(\egtX_{s_n})
}} \\
&\qquad \leq O\p{\ep \|f\| \prod_{j=1}^n \|g_j\|}.
\end{split} \end{equation}
Moreover, for any initial condition $ X_0 = X_\init $
that is independent of $ \cE $, one has
\begin{equation} \label{eq:part-approx-martingale} \begin{split}
&\ab{\mean{g_1(\egX_{s_1}) \cdots g_k(\egX_{s_k}) \p{
    f(\egX_t) - f(\egX_s) - \int_s^t \dr[\tau] (\gL_\tau f)(\egX_\tau)
  } g_{k+1}(\egX_{s_{k+1}}) \cdots g_n(\egX_{s_n})
}} \\
&\qquad \leq O\p{\ep \|f\| \prod_{j=1}^n \|g_j\|}.
\end{split} \end{equation}
\end{prop}
\begin{proof}  
If $ 1 \leq j \leq k $ and $ s_j \notin \{s_{k+1}, \ldots, s_n\} $
then we can introduce $ s_j $ into the sequence $ s_{k+1} \geq \ldots \geq s_n $
by inserting the term $ 1(\egtX_{s_j}) = 1 $ at the appropriate location.
Similarly, if $ k + 1 \leq j \leq n $ and $ s_j \notin \{s_1, \ldots, s_k\} $
then we can introduce it.
So, without loss of generality, we may assume $ 2k = n $
with $ s_j = s_{n + 1 - j} $ for all $ j \in [1, k] $.
Hence, the product becomes
\begin{align*}
&\meanM{g_1(\egtX_{s_1}) \cdots g_k(\egtX_{s_k}) \p{
    f(\egtX_t) - f(\egtX_s) - \int_s^t \dr[\tau] (\gL_\tau f)(\egtX_\tau)
  } g_{k+1}(\egtX_{s_k}) \cdots g_n(\egtX_{s_1})
} \\
&\qquad = \meanM{ \egxi_{0 \gets s_1}\p{
  g_1 \exi_{s_1 \gets s_2}\p{g_2 \cdots
    \egxi_{s_j \gets s}\p{
      \egxi_{s \gets t}(f) - f - \int_s^t \dr[\tau] \egxi_{s \gets \tau}(\gL_\tau f)
    }
  \cdots g_{2k - 1} }
g_{2k} }}.
\end{align*}
Therefore, applying \cref{lem:mean-bnd-etmu0}
then repeatedly using \cref{lem:etmu-bimod} and \cref{lem:etmu-ctl-egxi},
we obtain the bound
\begin{align*}
&\nr{\meanM{\egxi_{0 \gets s_1}\p{\cdots
  \egxi_{s_j \gets s}\p{
    \egxi_{s \gets t}(f) - f - \int_s^t \dr{\tau} \egxi_{s \gets \tau}(\gL_\tau f)
  }
\cdots }}} \\
&\qquad \leq \etmu_0\p{\egxi_{0 \gets s_1}\p{\cdots
  \egxi_{s_j \gets s}\p{
    \egxi_{s \gets t}(f) - f - \int_s^t \dr{\tau} \egxi_{s \gets \tau}(\gL_\tau f)
  }
\cdots }} \\
&\qquad \leq \p{ \prod_{j=1}^k \|g_{s_j}\| \;\; \|g_{s_{2k + 1 - j}}\| }
  \cdot \etmu_s\p{
    \exi_{s \gets t}(f) - f - \int_s^t \dr{\tau} \exi_{s \gets \tau}(\gL_\tau f)
  }.
\end{align*}
Lastly, we apply the bound from \cref{lem:gener-diff-bnd} to obtain \cref{eq:gen-approx-martingale}.
Recall that $ f(\egX_t) = \wbar{X}_\init(f(\egtX_t)) $
where $ \wbar{X}_\init : \tcA \to \cA $ is defined
as $ \wbar{X}_\init(A \otimes f) := A f(X_0) $
for any $ A \otimes f \in \cE \mxtens \cM $
and extended by linearity and closure.
Then, for any $ A \in \cE $ and $ f \in \cM $,
because $ X_0 $ is independent of $ \cE $,
\begin{align*}
\mean{\wbar{X}_\init(A \otimes f)} &= \mean{A f(X_0)}
= \mean{A} \mean{f(X_0)} \\
&= \mean{\mean{A} f(X_0)}
= \mean{\wbar{X}_\init(\mean{A} f)}
= \mean{\wbar{X}_\init(\meanM{A \otimes f})}
\end{align*}
Thus, the absolute value of the expected value is bounded by
\[
\ab{\mean{\wbar{X}_\init(E \otimes f)}}
= \ab{\mean{\wbar{X}_\init(\meanM{E \otimes f})}}
\leq \nrM{\meanM{E \otimes f}}
\]
and, since this holds for all $ E $ and $ f $,
we have $ \ab{\mean{\wbar{X}_\init(U)}} \leq \nrM{\meanM{U}} $
for all $ U \in \tcA $.
Therefore, \cref{eq:part-approx-martingale}
follows from \cref{eq:gen-approx-martingale}.
\end{proof}

\subsubsection{Time-Centered Weak Convergence}
\label{subsubsect:tc-weak-conv-proof}

\begin{proof}[Proof of \cref{thm:int-weak-conv}]  
First, let $ s, t \in [0, \bar{t}] $ with $ s \leq t $.
Then, from \cref{eq:Y-markov-gener}, we get
\begin{align*}
\meanWhen[B]{f(\gX_t)}{\cF_s}
&= \meanWhen[B]{f(\gX_s)}{\cF_s} + \int_s^t \dr[\tau] \frac{\dr}{\dr[\tau]}
  \meanWhen[B]{f(\gX_\tau)}{\cF_s} \\
&= f(\gX_s) + \int_s^t \dr[\tau] (\gL_\tau f)(\gX_\tau) \\
\text{so} \quad
0 &= \meanWhen[B]{f(\gX_t) - f(\gX_s) - \int_s^t \dr[\tau] (\gL_\tau f)(\gX_\tau)}{\cF_s}.
\end{align*}
Then, we observe that for any $ s_1, \ldots, s_n \in [0, s] $,
integer $ k \in [1, n] $,
and $ g_1, \ldots, g_n, f \in \cM $,
\begin{equation} \label{eq:Y-martingale} \begin{split}
&\mean[B]{g_1(\gX_{s_1}) \cdots g_k(\gX_{s_k}) \p{
  f(\gX_t) - f(\gX_s) - \int_s^t \dr{\tau} (\gL_\tau f)(\gX_\tau)
} g_{k+1}(\gX_{s_{k+1}}) \cdots g_n(\gX_{s_n})} \\
&\qquad = \mean[B]{g_1(\gX_{s_1}) \cdots g_k(\gX_{s_k}) \meanWhen[B]{
  f(\gX_t) - f(\gX_s) - \int_s^t \dr{\tau} (\gL_\tau f)(\gX_\tau)
}{\cF_s} g_{k+1}(\gX_{s_{k+1}}) \cdots g_n(\gX_{s_n})}
= 0.
\end{split} \end{equation}
\smallskip

Let $ \ep \in (0, \bar{\ep}) $ and $ \dl > 0 $.
From \cref{eq:limit-int-gener-defn}, one observes that
the operator norm $ \|\gL_t\| $ of $ \gL_t $ is uniformly bounded in $ t $.
We write $ \|\gL\| < \infty $ to denote such a uniform bound on $ \|\gL_t\| $.
So, we define $ K_\ep \in \ebRp $ as
\[ \begin{split}
K_\ep &:= \sup_{t \in [0, \bar{t}]} e^{-2t \nr{\gL}} \sup\Bigg\{ \ab{
    \mean{f_1\p{\egX_{t_1}} \cdots f_n\p{\egX_{t_n}}}
    - \mean[B]{f_1(\gX_{t_1}) \cdots f_n(\gX_{t_n})}
  } \\
  &\qquad \,\Bigg|\,
    f_1, \ldots, f_n \in \cM,
    \, \|f_1\| = \ldots = \|f_n\| = 1, \,
    0 = t_1, \ldots, t_n \in [0, t]
    \, \text{are centered and} \,
    \forall i \neq j, |t_i - t_j| \geq \dl
\Bigg\}.
\end{split} \]
Now, let $ t \in [0, \bar{t}] $ and $ 0 = t_1, \ldots, t_n \in [0, t] $ be centered
with $ |t_i - t_j| \geq \dl $ for $ i \neq j $.
We also let $ f_1, \ldots, f_n \in \cM $ with $ \nr{f_1} = \ldots = \nr{f_n} = 1 $.
First, we have
\begin{align*}
&\ab{ \mean{f_1(\egX_{t_1}) \cdots f_n(\egX_{t_n})}
  - \mean[B]{f_1(\gX_{t_1}) \cdots f_n(\gX_{t_n})} } \\
&\qquad \leq \|f_1(\egX_{t_1})\| \cdots \|f_n(\egX_{t_n})\|
  + \|f_1(\gX_{t_1})\| \cdots \|f_n(\gX_{t_n})\| \\
&\qquad \leq 2 \nrM{f_1} \cdots \nrM{f_n} \leq 2.
\end{align*}
Because this holds for all $ t \in [0, \bar{t}] $,
we see that $ K_t \leq \sup_{t \in [0, \bar{t}]} 2 e^{-2t\nr{\gL}} \leq 2 < \infty $.
\medskip

Now, either $ n = 1 $ or $ n > 1 $.
If $ n = 1 $ then $ t_1 = 0 $ and so
\[
\mean{f_1\p{\egX_{t_1}}} = \mean{f_1\p{\egX_0}}
= \mean[B]{f_1(\gX_0)} = \mean[B]{f_1(\gX_{t_1})}
\]
implying $ \ab{\mean{f_1(\egX_{t_1})} - \mean[B]{f_1(\gX_{t_1})}} = 0 $.
\smallskip

Otherwise, $ n > 1 $ and because $ t_1, \ldots, t_n $ are centered,
there exists $ k \in [1, n] $
such that $ t_1, \ldots, t_{k-1}, t_{k+1}, \ldots, t_n \leq t_k $.
Without loss of generality, we can assume $ t_{k-1} \geq t_{k+1} $.
Then, from \cref{prop:approx-martingale},
there exists $ C > 0 $ depending on $ \bar{\ep} $, $ \bar{t} $, and $ \eta $
as well as the collections $ \{\Ea\} $ and $ \{\hoa\} $ such that
\begin{equation} \label{eq:X-mart} \begin{split}
&\ab{\mean{f_1(\egX_{t_1}) \cdots f_n(\egX_{t_n})}
  - \mean{f_1(\egX_{t_1}) \cdots \p{
    f_k(\egX_{t_k}) + \int_{t_{k-1}}^{t_k} \dr[s] (\gL_s f_k)(\egX_s)
  } \cdots f_n(\egX_{t_n})}} \\
&= \ab{\mean{f_1(\egX_{t_1}) \cdots \p{
    f_k(\egX_{t_k}) - f_k(\egX_{t_{k-1}}) - \int_{t_{k-1}}^{t_k} \dr[s] (\gL_s f)(\egX_s)
  } \cdots f_n(\egX_{t_n})}}
\leq \ep C.
\end{split} \end{equation}
From \cref{eq:Y-martingale}, we also get
\begin{equation} \label{eq:Y-mart}
\ab{\mean[B]{f_1(\gX_{t_1}) \cdots f_n(\gX_{t_n})}
  - \mean[B]{f_1(\gX_{t_1}) \cdots \p{
    f_k(\gX_{t_k}) + \int_{t_{k-1}}^{t_k} \dr[s] (\gL_s f)(\gX_s)
  } \cdots f_n(\gX_{t_n})}
} = 0.
\end{equation}
By the definition of $ K_\ep $, for $ s \in [t_{k-1}, t_k] $, we have
\begin{align}
\Big|\mean{f_1\p{\egX_{t_1}} \cdots f_k\p{\egX_{t_{k-1}}} \cdots f_n\p{\egX_{t_n}}}
  &- \mean[B]{f_1(\gX_{t_1}) \cdots f_k(\gX_{t_{k-1}}) \cdots f_n(\gX_{t_n})} \Big| \notag \\
\label{eq:init-diff-mart}
&\leq K_\ep e^{2t_{k-1} \|\gL\|} \\
\Big|\mean{f_1\p{\egX_{t_1}} \cdots (\gL_s f_k)\p{\egX_s} \cdots f_n\p{\egX_{t_n}}}
  &- \mean[B]{f_1(\gX_{t_1}) \cdots (\gL_s f_k)(\gX_\tau) \cdots f_n(\gX_{t_n})} \Big| \notag \\
\label{eq:gen-diff-mart}
&\leq K_\ep e^{2s \|\gL\|} \nr{\gL_s f_k}
\leq K_\ep \|\gL\| e^{2s \|\gL\|}
\end{align}
Combining the bounds in \cref{eq:X-mart}, \cref{eq:Y-mart},
\cref{eq:init-diff-mart}, and \cref{eq:gen-diff-mart}, we get
\begin{align}
&\ab{ \mean{f_1\p{\egX_{t_1}} \cdots f_n\p{\egX_{t_n}}}
  - \mean[B]{f_1(\gX_{t_1}) \cdots f_n(\gX_{t_n})} } \notag \\
&\qquad \leq \ep C + 0 + K_\ep e^{2t_{k-1} \|\gL\|}
  + \int_{t_{k-1}}^{t_k} \dr[s] K_\ep \|\gL\| e^{2s \|\gL\|} \notag \\
&\qquad = \ep C + K_\ep e^{2t_{k-1} \|\gL\|}
  + \frac{1}{2} K_\ep \p{e^{2t_k \|\gL\|} - e^{2t_{k-1} \|\gL\|}} \notag \\
\label{eq:comb-mart-bnd}
&\qquad = \ep C + \frac{1}{2} K_\ep e^{2t_{k-1} \|\gL\|}
+ \frac{1}{2} K_\ep e^{2t_k \|\gL\|}
\end{align}
Since $ |t_{k-1} - t_k| > \dl $,
we have $ t_{k-1} < t_k - \dl $
and so $ e^{2t_{k-1} \|\gL\|} \leq e^{-2\dl \|\gL\|} e^{2t_k \|\gL\|} $.
Hence, \cref{eq:comb-mart-bnd} becomes
\begin{align*}
\ep C + \frac{1}{2} K_\ep e^{2t_{k-1} \|\gL\|} + \frac{1}{2} K_\ep e^{2t_k \|\gL\|}
&\leq \ep C + \frac{1}{2} K_\ep e^{2t_k \|\gL\|} \p{1 + e^{-2\dl \|\gL\|}} \\
&\leq \ep C + e^{2t \|\gL\|} K_\ep \frac{1 + e^{-2\dl \|\gL\|}}{2}
\end{align*}
Because this holds for all $ t \in [0, \bar{t}] $ and all centered $ 0 = t_1, \ldots, t_n \in [0, t] $,
one obtains
\begin{align*}
K_\ep \leq \ep C e^{-0 \cdot \|\gL\|}
  &+ K_\ep \frac{1 + e^{-2\dl \|\gL\|}}{2} \\
K_\ep \frac{1 - e^{-2\dl \|\gL\|}}{2} \leq \ep C
\quad &\text{and so} \quad
K_\ep \leq \frac{2 \ep C}{1 - e^{-2 \dl \|\gL\|}}.
\end{align*}
\medskip

Now, let $ 0 = t_1, \ldots, t_n \in [0, \bar{t}] $ be distinct centered times.
We take $ \dl := \min_{i \neq j} |t_i - t_j| $
and let $ \epsilon > 0 $.
Observe that for any $ f_1, \ldots, f_n \in \cM $
with $ \nr{f_1} = \ldots = \nr{f_n} = 1 $,
\[ \begin{split}
&\ab{\mean{f_1\p{\egX_{t_1}} \cdots f_n\p{\egX_{t_n}}}
  - \mean[B]{f_1(\gX_{t_1}) \cdots f_n(\gX_{t_n})}} \\
&\qquad \leq K_\ep e^{-2b \|\gL\|}
\leq \frac{2 \ep C}{1 - e^{-2 \dl \|\gL\|}} \to 0
\end{split} \]
as $ \ep \dto 0 $
meaning that $ \{\egX_t\}_{t \in [0, \bar{t}]} \to \{\gX_t\}_{t \in [0, \bar{t}]} $ time-centered weakly.
\end{proof}

\begin{proof}[Proof of \cref{coro:weak-conv}]  
For each $ t \geq 0 $ and $ f \in \cM $,
we define $ f(\gX_t) = (\kps_{-t} f)(X_t) $.
The semigroup for $ X $ is
\[ (S^X_{t - s} f)(X_s) := \meanWhen{f(X_t)}{\cF_s}. \]
Let $ s \in [0, t] $ and $ f \in \dom(\fk) $ then
\begin{align*}
\frac{\dr}{\dr[t]} \meanWhen[B]{f(\gX_t)}{\cF_s}
&= \frac{\dr}{\dr[t]} \meanWhen[B]{(\kps_{-t} f)(X_t)}{\cF_s} \\
&= \frac{\dr}{\dr[t]} S^X_{t - s}(\kps_{-t}(f))(X_s) \\
&= - S^X_{t - s}(\fks \kps_{-t}(f))(X_s)
  + S^X_{t - s}(L \kps_{-t}(f))(X_s) \\
&= S^X_{t - s}((L - \fks) \kps_{-t}(f))(X_s).
\end{align*}
Thus, at $ t = s $, one has
\begin{align*}
\frac{\dr}{\dr[t]} \meanWhen[B]{f(\gX_t)}{\cF_s} \Big|_{t=s}
&= S^X_{s - s}((L - \fks) \kps_{-s}(f))(X_s) \\
&= \p{\big. (L - \fks) \kps_{-s}(f)}(X_s) \\
&= \p{\big. \kps_{-s}(\gL_s f)}(X_s)
= (\gL_s f)(\gX_s)
\end{align*}
implying that $ \gL_s $ is the generator of $ \gX $ at time $ s $.
Moreover, $ f(\gX_0) = (\kps_0 f)(X_0) = (\kps_0 f)(\eX_0) = f(\egX_0) $
for all $ \ep > 0 $. 
Thus, by \cref{thm:int-weak-conv}, $ \{\egX_t\}_{t \in [0, \bar{t}]} \to \{\gX\}_{t \in [0, \bar{t}]} $
time-centered weakly as $ \ep \dto 0 $.
Then, for any $ f_1, \ldots, f_n \in \cM $
with $ \nr{f_1} = \ldots = \nr{f_n} = 1 $
and centered $ t_1, \ldots, t_n \in [0, \bar{t}] $,
\begin{align*}
&\ab{\mean{f_1(\eX_{t_1}) \cdots f_n(\eX_{t_n})}
  - \mean[B]{\Big. f_1(X_{t_1}) \cdots f_n(X_{t_n})}} \\
&\qquad = \ab{\mean{(\kps_{t_1} f_1)(\egX_{t_1}) \cdots (\kps_{t_n} f_n)(\egX_{t_n})}
  - \mean[B]{(\kps_{t_1} f_1)(\gX_{t_1}) \cdots (\kps_{t_n} f_n)(\gX_{t_n})}}
\to 0
\end{align*}
uniformly in $ f_1, \ldots, f_n $ so $ \eX \to X $ time-centered weakly.
\end{proof}

\subsection{Decorrelation for Gaussian Systems}
\label{subsect:gauss-sys-proofs}

\begin{proof}[Proof of \cref{thm:gauss-graph-ctl}]  
For each integer $ n \geq 0 $,
if $ n $ is even then we define $ \cP_n $ to be
the collection of forests on $ [1, n] $
such that each constituent tree has exactly two vertices.
Such forests are in bijection with the pairings on $ [1, n] $
so $ \# \cP_n = n!! = n \cdot (n - 2) \cdots 4 \cdot 2 $.
\smallskip

Otherwise, $ n $ is odd and we define $ \cP_n $ to be
the collection of forests on $ [1, n] $
such that exactly one tree has three vertices
while all of the rest have two.
If we define the set
\[ \wtld{\cP}_n := \{
  (P, v) \,:\, P \in \cP_n, v \in [1, n], \,
  \text{$ v $ is a leaf of the three-vertex tree in $ P $}
\} \]
then $ \# \wtld{\cP}_n = 2 \# \cP_n $.
Also, for each $ (P, v) \in \wtld{\cP}_n $,
the forest $ P' = P \setminus v $ is a pairing on $ [1, n] \setminus \{k\} $.
Moreover, for each pairing $ P' $ on $ [1, n] \setminus \{k\} $,
there are $ n - 1 $ vertices to connect $ k $ to.
Hence, $ \# \wtld{\cP}_n = (n - 1) \cdot \# \cP_{n-1} = (n-1) \cdot (n-1)!! $.
Therefore, $ \# \cP_n = \frac{n - 1}{2} \cdot (n - 1)!! \leq n!! $.
\smallskip

Now, we take $ \cG := \bigcup_{n=0}^\infty \cP_n $
and $ \vartheta_G := C^{\flr{|G| / 2}} $ for all $ G \in \cG $.
Then, for any $ \alpha_1, \ldots, \alpha_n \in \bN $
and $ t_1, \ldots, t_n \geq 0 $,
by \cref{eq:isserlis-form}, when $ n $ is even,
\[
\ab{\mean{E^{\alpha_1}_{t_1} \cdots E^{\alpha_n}_{t_n}}}
= \ab{\sum_{P \in \cP_n}
  \prod_{\substack{x, y \in [1, n] \\ x \sim_P y}}
  \mean{E^{\alpha_x}_{t_x} E^{\alpha_y}_{t_y}}}
\leq \sum_{P \in \cP_n} \prod_{x \sim_P y}
  \ab{\mean{E^{\alpha_x}_{t_x} E^{\alpha_y}_{t_y}}}
\leq \sum_{P \in \cP_n} \prod_{x \sim_P y}
  C e^{-\eta |t_x - t_y|}
= \Theta_\cG(t_1, \ldots, t_n)
\]
and, when $ n $ is odd, the expected value of the product is zero
demonstrating \cref{eq:graph-mean-ctl}.
\smallskip

Next, let $ j \in [1, n) $ and $ t_k \leq t_j \land t_{j+1} $
for all $ k \in [1, n] \setminus \{j, j+1\} $.
Without loss of generality, by reversing the ordering by conjugation,
we assume $ t_j \leq t_{j+1} $.
We use $ \cP_{[1, j) \cup (j+1, n]} $ for the pairings on $ [1, j) \cup (j+1, n] $.
We define the set
\[ \cP^j_n := \{P \sqcup (j \sim (j+1)) \,:\, \cP_{[1, j) \cup (j+1, n]}\} \subseteq \cP_n \]
where $ P \sqcup (j \sim (j+1)) $ is the disjoint union
of $ P $ and the graph consisting of vertices $ j $ and $ j + 1 $
with a single edge between them.
Again using \cref{eq:isserlis-form}, we obtain
\begin{align}
&\mean{E^{\alpha_1}_{t_1} \cdots E^{\alpha_i}_{t_i} \p{
    E^{\alpha_{n+1}}_{t_{n+1}} E^{\alpha_{n+2}}_{t_{n+2}}
    - \mean{E^{\alpha_{n+1}}_{t_{n+1}} E^{\alpha_{n+2}}_{t_{n+2}}}
  } E^{\alpha_{i+1}}_{t_{i+1}} \cdots E^{\alpha_n}_{t_n}} \notag \\
&\qquad = \sum_{P \in \cP_n} \prod_{\substack{x, y \in [1, n] \\ x \sim_P y}}
  \mean{E^{\alpha_x}_{t_x} E^{\alpha_y}_{t_y}}
- \mean{E^{\alpha_{n+1}}_{t_{n+1}} E^{\alpha_{n+2}}_{t_{n+2}}}
  \sum_{P \in \cP_{[1, j) \cup (j+1, n]}} \,\, \prod_{\substack{x, y \in [1, j) \cup (j+1, n] \\ x \sim_P y}}
  \mean{E^{\alpha_x}_{t_x} E^{\alpha_y}_{t_y}} \notag \\
&\qquad = \sum_{P \in \cP_n} \prod_{x \sim_P y}
  \mean{E^{\alpha_x}_{t_x} E^{\alpha_y}_{t_y}}
- \sum_{P \in \cP^j_n} \prod_{x \sim_P y}
  \mean{E^{\alpha_x}_{t_x} E^{\alpha_y}_{t_y}} \notag \\
\label{eq:odd-gauss-bnd}
&\qquad = \sum_{P \in \cP_n \setminus \cP^j_n}
  \prod_{x \sim_P y} \mean{E^{\alpha_x}_{t_x} E^{\alpha_y}_{t_y}}.
\end{align}
Now, let $ P \in \cP_n \setminus \cP^j_n $.
If $ j \sim_P (j+1) $ then, because $ P $ is a pairing,
we would have $ P \in \cP_n \sqcup (j \sim (j+1)) $.
So, $ j \nsim_P (j+1) $ meaning there exists $ v \in [1, n] $
such that $ v \sim_P (j+1) $.
Then, $ t_v \leq t_j \leq t_{j+1} $
so $ |t_{j+1} - t_v| = |t_{j+1} - t_j| + |t_j - t_v| $
and thus
\[
\ab{\mean{E^{\alpha_{j+1}}_{t_{j+1}} E^{\alpha_v}_{t_v}}}
\leq C e^{-\eta |t_{j+1} - t_v|}
= e^{-\eta |t_{j+1} - t_j|} \cdot C e^{-\eta |t_j - t_v|}.
\]
Therefore, we can bound the product by
\begin{align*}
\ab{\prod_{x \sim_P y} \mean{E^{\alpha_x}_{t_x} E^{\alpha_y}_{t_y}}}
&= \ab{\mean{E^{\alpha_{j+1}}_{t_{j+1}} E^{\alpha_v}_{t_v}} \cdot
  \prod_{\substack{x \neq j+1 \neq y \\ x \sim_P y}}
  \mean{E^{\alpha_x}_{t_x} E^{\alpha_y}_{t_y}}} \\
&\leq e^{-\eta |t_{j+1} - t_j|} \cdot C e^{-\eta |t_j - t_v|}
  \prod_{\substack{x \neq j+1 \neq y \\ x \sim_P y}}
   C e^{-\eta |t_x - t_y|} \\
&= e^{-\eta |t_{j+1} - t_j|} \cdot
  \prod_{\substack{x, y \in [1, n] \setminus \{j+1\} \\ x \sim_{\wtld{P}} y}}
  C e^{-\eta |t_x - t_y|}
\end{align*}
where $ \wtld{P} = P / \{j, j+1\} $ is the forest obtained
by contracting $ j+1 $ into $ j $.
Since the contractions of each forest are unique
and $ \wtld{P} \in \cP_{n-1} $,
we can bound \cref{eq:odd-gauss-bnd} by
\begin{align*}
\ab{\sum_{P \in \cP_n \setminus \cP^j_n}
  \prod_{x \sim_P y} \mean{E^{\alpha_x}_{t_x} E^{\alpha_y}_{t_y}}
}
&\leq e^{-\eta |t_{j+1} - t_j|} \cdot
  \sum_{P \in \cP_{[1, n] \setminus \{j+1\}}}
  \prod_{x \sim_P y} C e^{-\eta |t_x - t_y|} \\
&= e^{-\eta |t_{j+1} - t_j|} \cdot \Theta_\cG(t_1, \ldots, t_j, t_{j+2}, \ldots, t_n)
\end{align*}
proving \cref{eq:graph-cov-ctl}.
\medskip

Lastly, we show \cref{eq:graph-inf-rad-conv}.
For each $ n \geq 0 $, we know $ \# \{G \in \cG \,:\, |G| = n\} = \#\cP_n \leq n!! $ so
\[
\sum_{\substack{G \in \cG \\ |G| = n}} \vartheta_G
= C^{\flr{n/2}} \cdot \# \cP_n
\leq C^n \cdot n!!
\leq C^n \cdot \sqrt{(n+1)!}.
\]
Therefore,
\[
\lim_{n \to \infty} \frac{1}{n}
  \sqrt[n]{\sum_{\substack{G \in \cG \\ |G| = n}} \vartheta_G}
= \lim_{n \to \infty} \frac{1}{n} \sqrt[n]{C^n \cdot \sqrt{(n+1)!}}
= \lim_{n \to \infty} \frac{C}{n} \cdot \sqrt{\frac{n}{e}} = 0
\]
showing \cref{eq:graph-inf-rad-conv} holds.
\end{proof}


\subsection{Differential Weyl Operators on Bosonic Fock Space}
\label{subsect:diff-weyl-ops-proofs}

Ultimately, we will prove \cref{thm:gauge-graph-ctl}.
First, however, we prove an Isserlis-like formula for the differential Weyl operators.
The following result may be found in proposition 20.1 from \cite{Part92}.

\begin{lem} \label{lem:weyl-comb}
For any $ x, y \in \cH $ and $ S, T \in \UNI(\cH) $,
\[ W(x, S) W(y, T) = e^{-i\Im\ang{x, S y}} W(x + S y, S T). \]
\end{lem}
\begin{proof}  
Let $ v \in \cH $ then, using \cref{eq:weyl-coh-st}
and the fact that $ S $ is unitary,
\begin{align}
W(x, S) W(y, T) \vp(v)
&= W(x, S) e^{-i\Im\ang{y, T v}} \vp(y + T v) \\
&= e^{-i\Im\ang{y, T v}} e^{-i\Im\ang{x, S (y + T v)}} \vp(x + S (y + T v)) \\
&= e^{-i\Im\ang{Sy, S T v}} e^{-i\Im\ang{x, S T v}} e^{-i\Im\ang{x, S y}} \vp(x + S y + S T v) \\
&= e^{-i\Im\ang{x, S y}} e^{-i\Im\ang{x + S y, S T v}} \vp(x + S y + S T v) \\
&= e^{-i\Im\ang{x, S y}} W(x + S y, S T) \vp(v).
\end{align}
Because this holds for all $ v \in \cH $,
by density of the linear hull of coherent states,
we obtain the identity for all of $ \BosFock(\cH) $.
\end{proof}
\bigskip

To facilitate the following calculations,
we introduce some notation.
\medskip

\begin{defn} \label{defn:prod-seq}
Let $ \{T_j\}_{j=1}^n $ be a sequence of operators.
Then, for any subset $ J := \{j_1, \ldots, j_{|J|}\} \subseteq [1, n] $
where $ j_1 < \ldots < j_{|J|} $, we say
\[ T_J := T_{j_1} \cdots T_{j_{|J|}} \]
and, by convention, we take $ T_\emptyset := 1 $.
\end{defn}

So, for example, $ T_{[a, b]} = T_a T_{a+1} \cdots T_{b-1} T_b $
for any $ 1 \leq a \leq b \leq n $.
Note that for any $ a, b, c \in [1, n] $ with $ a \leq b \leq c $,
\begin{equation} \label{eq:prod-seq-cancel} \begin{split}
(T_{[a, b)})^* T_{[a, c)}
&= (T_a \cdots T_{b-1})^* (T_a \cdots T_{b-1}) T_b \cdots T_{c-1} \\
&= T_b \cdots T_{c-1} = T_{[b, c)}.
\end{split} \end{equation}
It will also be necessary to consider inner products involving $ T_J $.

\begin{defn} \label{defn:capped-prod-seq}
Let $ \{x_j\}_{j=1}^n $ and $ \{y_j\}_{j=1}^n $ be sequences of elements in $ \cH $
and $ \{T_j\}_{j=1}^n $ be a sequence of bounded operators in $ \cH $.
Then, let $ J := \{j_1, \ldots, j_{|J|}\} \subseteq [1, n] $ be a subset
with $ j_1 < \ldots < j_{|J|} $.
When $ |J| \geq 3 $, we say
\[ \ang{x | T | y}_J := \ang{ x_{j_1}, T_{j_2} \cdots T_{j_{|J| - 1}} y_{j_{|J|}} }. \]
If $ J = \{j_1, j_2\} $ then we say $ \ang{x | T | y}_J = \ang{x_{j_1}, y_{j_2}} $.
Otherwise, $ J = \emptyset $ or $ J = \{j_1\} $ and we say $ \ang{x | T | y}_J = 0 $.
\end{defn}

Also, all products $ \prod $ are interpreted as running from left to right so that
\[ \prod_{j=1}^n T_j = T_1 \cdots T_n. \]
\bigskip

\subsubsection{Isserlis-like Formula for Differential Weyl Operators}
\label{subsubsect:gauge-isserlis}  

In order to prove \cref{cond:graph-ctl} for the differential Weyl operators,
we prove the following formula similar to Isserlis' formula for Gaussian systems.
\medskip

\begin{thm} \label{thm:gauge-isserlis}
For any $ v_1, \ldots, v_n \in \cH $
and $ A_1, \ldots, A_n \in i\uni(n) $,
\[
\ang{\vp(0), \dW(v_1, A_1) \cdots \dW(v_n, A_n) \vp(0)}
= \sum_{\bigcup P = [1, n]} (-1)^{|P|} \prod_{K \in P} \ang{v | A | v}_K
\]
where $ P $ ranges over partitions of $ [1, n] $.
\end{thm}
\vspace{1em}

\begin{lem} \label{lem:Wprod-form}
For any $ x_1, \ldots, x_n \in \cH $
and $ T_1, \ldots, T_n \in \UNI(\cH) $,
using the notation from \cref{defn:prod-seq},
\[
W(x_1, T_1) \cdots W(x_n, T_n)
= \exp\p{-i\Im\p{\sum_{j < k} \ang{x_j, T_{[j, k)} x_k} }}
    W\p{\sum_j T_{[1, j)} x_j, T_{[1, n]}}
\]
where $ j, k \in [1, n] $.
\end{lem}
\begin{proof} 
We prove the result by induction on $ n \geq 1 $.
When $ n = 1 $, $ j < k $ is impossible so one has
\[ W(x_1, T_1) = W(T_{[1, 1)} x_1, T_{[1, 1]}) \]
fulfilling the identity.
Otherwise, $ n \geq 2 $ and, by the induction hypothesis,
\begin{equation} \label{eq:Wprod-induc-hypo}
W(x_1, T_1) \cdots W(x_{n-1}, T_{n-1})
= \exp\p{-i\Im\p{\sum_{j=1}^{n-1} \sum_{k=j+1}^{n-1} \ang{x_j, T_{[j, k)} x_k}}}
    W\p{\sum_{j=1}^{n-1} T_{[1, j)} x_j, T_{[1, n-1]}}.
\end{equation}
Applying $ W(x_n, T_n) $ to the right, using \cref{lem:weyl-comb}
and \cref{eq:prod-seq-cancel}, we obtain
\begin{align}
&W\p{\sum_{j=1}^{n-1} T_{[1, j)} x_j, T_{[1, n-1]}} W(x_n, T_n) \notag \\
&\qquad = \exp\p{-i\Im\ang{\sum_{j=1}^{n-1} T_{[1, j)} x_j, T_{[1, n-1]} x_n}}
    W\p{\sum_{j=1}^{n-1} T_{[1, j)} x_j + T_{[1, n-1]} x_n, T_{[1, n-1]} T_n} \notag \\
&\qquad = \exp\p{-i\Im\p{\sum_{j=1}^{n-1} \ang{x_j, (T_{[1, j)})^* T_{[1, n)} x_n}}}
    W\p{\sum_{j=1}^{n-1} T_{[1, j)} x_j + T_{[1, n)} x_n, T_{[1, n]}} \notag \\
\label{eq:Wprod-sans-exp}
&\qquad = \exp\p{-i\Im\p{\sum_{j=1}^{n-1} \ang{x_j, T_{[j, n)} x_n}}}
    W\p{\sum_{j=1}^n T_{[1, j)} x_j, T_{[1, n]}}
\end{align}
Then, the product of the $ \exp(\ldots) $ terms
from \cref{eq:Wprod-induc-hypo} and \cref{eq:Wprod-sans-exp} is 
\begin{align}
&\exp\p{-i\Im\p{\sum_{j=1}^{n-1} \sum_{k=j+1}^{n-1} \ang{x_j, T_{[j, k)} x_k}}}
\exp\p{-i\Im\p{\sum_{j=1}^{n-1} \ang{x_j, T_{[j, n)} x_n}}} \notag \\
&\qquad = \exp\p{-i\Im\p{\sum_{j=1}^{n-1} \p{ \ang{x_j, T_{[j, n)} x_n} +
    \sum_{k=j+1}^{n-1} \ang{x_j, T_{[j, k)} x_k}
}}} \notag \\
\label{eq:Wprod-exp-term}
&\qquad = \exp\p{-i\Im\p{\sum_{j=1}^{n-1} \sum_{k=j+1}^n \ang{x_j, T_{[j, k)} x_k}}}
= \exp\p{-i\Im\p{\sum_{j=1}^n \sum_{k=j+1}^n \ang{x_j, T_{[j, k)} x_k}}}
\end{align}
since $ \sum_{k=j+1}^n \ang{x_j, T_{[j, k)} x_k} = 0 $ when $ j = n $.
Combining \cref{eq:Wprod-exp-term} with
the $ W(\ldots) $ term in \cref{eq:Wprod-sans-exp}, we get
\[
W(x_1, T_1) \cdots W(x_{n-1}, T_{n-1}) W(x_n, T_n)
= \exp\p{-i\Im\p{\sum_{j=1}^n \sum_{k=j+1}^n \ang{x_j, T_{[j, k)}, x_k}}}
    W\p{\sum_{j=1}^n T_{[1, j)} x_j, T_{[1, n]}}
\]
completing the induction.
\end{proof}

\begin{lem} \label{lem:weyl-vacuum}
For any $ (x, T) \in \ISO(\cH) $,
one has $ \ang{\vp(0), W(x, T) \vp(0)} = e^{-\frac{1}{2} \|x\|^2} $.
\end{lem}
\begin{proof}  
From \cref{eq:weyl-coh-st}, we get
\begin{align*}
\ang{\vp(0), W(x, T) \vp(0)} &= \ang{\vp(0), e^{-i\Im\ang{x, T0}} \vp(x + T 0)}
= \ang{\vp(0), \vp(x)} \\
&= e^{-\frac{1}{2} \|x\|^2} \ang{e(0), e(x)}
= e^{-\frac{1}{2} \|x\|^2} e^{\ang{0, x}} = e^{-\frac{1}{2} \|x\|^2}.
\end{align*}
\end{proof}

\begin{lem} \label{lem:Wprod-vac-st}
For any $ x_1, \ldots, x_n \in \cH $
and $ T_1, \ldots, T_n \in \UNI(\cH) $,
using the notation from \cref{defn:capped-prod-seq},
\[
\ang{\vp(0), W(x_1, T_1) \cdots W(x_n, T_n) \vp(0)}
= \exp\p{-\frac{1}{2} \sum_{j,k}
    \ang{T^* x | T | x}_{[j \land k, j \lor k]}
}
\]
where $ j, k \in [1, n] $
and $ \{(T^* x)_j\}_{j=1}^n $ is a sequence in $ \cH $
with $ (T^* x)_j = T_j^* x_j $.
\end{lem}
\begin{proof}  
Using \cref{lem:Wprod-form} and \cref{lem:weyl-vacuum},
\begin{align}
\ang{\vp(0), W(x_1, T_1) \cdots W(x_n, T_n) \vp(0)}
&= \exp\p{-i\Im\p{\sum_{j < k} \ang{x_j, T_{[j, k)} x_k}}}
    \ang{\vp(0), W\p{\sum_{j=1}^n T_{[1, j)} x_j, T_{[1, n]}} \vp(0)} \notag \\
\label{eq:Wprod-vac-expand}
&= \exp\p{-i\Im\p{\sum_{j < k} \ang{x_j, T_{[j, k)} x_k}}}
    \exp\p{-\frac{1}{2} \nr{\sum_{j=1}^n T_{[1, j)} x_j}^2}.
\end{align}
Next, using \cref{eq:prod-seq-cancel}, we may calculate the squared norm to be
\begin{align}
\nr{\sum_{j=1}^n T_{[1, j)} x_j}^2
&= \ang{\sum_{j=1}^n T_{[1, j)} x_j, \sum_{k=1}^n T_{[1, k)} x_k}
= \sum_{j=1}^n \sum_{k=1}^n \ang{T_{[1, j)} x_j, T_{[1, k)} x_k} \notag \\
&= \sum_{j = k} \ang{T_{[1, j)} x_j, T_{[1, j)} x_j}
    + \sum_{j < k} \ang{x_j, (T_{[1, j)})^* T_{[1, k)} x_k}
    + \sum_{j > k} \ang{(T_{[1, k)})^* T_{[1, j)} x_j, x_k} \notag \\
&= \sum_{j=1}^n \ang{x_j, x_j}
    + \sum_{j < k} \ang{x_j, T_{[j, k)} x_k}
    + \sum_{j > k} \ang{T_{[k, j)} x_j, x_k} \notag \\
&= \sum_{j=1}^n \ang{x_j, x_j}
    + \sum_{j < k} \ang{x_j, T_{[j, k)} x_k}
    + \sum_{k > j} \ang{T_{[j, k)} x_k, x_j} \notag \\
&= \sum_{j=1}^n \ang{x_j, x_j} + \sum_{j < k} \p{
    \ang{x_j, T_{[j, k)} x_k} + \ang{x_j, T_{[j, k)} x_k}^*
} \notag \\
\label{eq:Wprod-vac-sqnr}
&= \sum_{j=1}^n \ang{x_j, T_{[j, j)} x_j} + 2\Re\p{\sum_{j < k} \ang{x_j, T_{[j, k)} x_k}}.
\end{align}
Substituting \cref{eq:Wprod-vac-sqnr} into \cref{eq:Wprod-vac-expand}, one obtains
\begin{align}
&\ang{\vp(0), W(x_1, T_1) \cdots W(x_n, T_n) \vp(0)} \notag \\
&\qquad = \exp\p{-i\Im\p{\sum_{j < k} \ang{x_j, T_{[j, k)} x_k}}}
    \exp\p{-\frac{1}{2} \sum_j \ang{x_j, T_{[j, j)} x_j}
        - \Re\p{\sum_{j < k} \ang{x_j, T_{[j, k)} x_k}}} \notag \\
&\qquad = \exp\p{-\frac{1}{2} \p{\sum_j \ang{x_j, T_{[j, j)} x_j}
    + 2\sum_{j < k} \ang{x_j, T_{[j, k)} x_k}}} \notag \\
&\qquad = \exp\p{-\frac{1}{2} \p{\sum_{j=k} \ang{x_j, T_{[j, k)} x_k}
    + \sum_{j \neq k} \ang{x_{j \land k}, T_{[j \land k, j \lor k)} x_{j \lor k}} }} \notag \\
\label{eq:Wprod-vac-unsimple}
&\qquad = \exp\p{-\frac{1}{2} \sum_{j,k} \ang{x_{j \land k}, T_{[j \land k, j \lor k)} x_{j \lor k}}}.
\end{align}
For any $ j, k \in [1, n] $ with $ j \leq k $, we can write
\[
\ang{x_j, T_{[j, k)} x_k}
= \ang{x_j, T_j T_{j+1} \cdots T_{k-1} x_k}
= \ang{T_j^* x_j, T_{(j, k)} x_k}
= \ang{T^* x | T | x}_{[j, k]}.
\]
Applying this to \cref{eq:Wprod-vac-unsimple},
we obtain the desired identity.
\end{proof}

\begin{lem} \label{lem:iso-expgrp-form}
Let $ v \in \cH $ and $ A $ be a (bounded) Hermitian operator on $ \cH $
then $ (v, -iA) \in \iso(\cH) $ and
\begin{equation} \label{eq:iso-expgrp-form}
e^{t (v, -iA)}
= \p{i\frac{e^{-itA} - 1}{A} v, e^{-itA}} \in \ISO(\cH)
\end{equation}
for all $ t \in \bR $ where
\[
\frac{e^{-itA} - 1}{A}
= \sum_{n=0}^\infty \frac{(-it)^{n+1}}{(n + 1)!} A^n
\]
is well-defined for any bounded operator $ A $ and all $ t \in \bR $.
\end{lem}
\begin{proof}  
First, we prove that the expression in \cref{eq:iso-expgrp-form}
produces a group homomorphism from $ \bR $ into $ \ISO(\cH) $.
Let $ s, t \in \bR $ then
\begin{align*}
\p{i\frac{e^{-isA} - 1}{A} v, e^{-isA}} \p{i\frac{e^{-itA} - 1}{A} v, e^{-itA}}
&= \p{i\frac{e^{-isA} - 1}{A} v + i e^{-isA} \frac{e^{-itA} - 1}{A} v, e^{-isA} e^{-itA}} \\
&= \p{i\frac{e^{-isA} - 1 + e^{-i(s + t)A} - e^{-isA}}{A} v, e^{-i(s + t)A}} \\
&= \p{i\frac{e^{-i(s + t)A} - 1}{A} v, e^{-i(s + t)A}}.
\end{align*}
\medskip

Second, we show that the derivative \cref{eq:iso-expgrp-form}
at $ t = 0 $ is $ (v, -iA) $:
\[
\at{\frac{\dr}{\dr[t]} \p{i\frac{e^{-itA} - 1}{A} v, e^{-itA}} }_{t=0}
= \at{ \p{i\frac{-iA e^{-itA}}{A} v, -iA e^{-itA}} }_{t=0}
= (v, -iA).
\]
\end{proof}

\begin{lem} \label{lem:diff-K-capd-pseq}
Let $ v_1, \ldots, v_n \in \cH $, $ A_1, \ldots, A_n \in i\uni(\cH) $,
and $ t_1, \ldots, t_n \in \bR $.
For each $ j \in [1, n] $,
we take $ (x_j(t_j), T_j(t_j)) := e^{t_j (v_j, -iA_j)} \in \ISO(\cH) $
such that $ x(t) = \{x_j(t_j)\}_{j=1}^n $
and $ T(t) = \{T_j(t_j)\}_{j=1}^n $.
For each $ k \in [1, n] $, we write $ \df_k $ to mean $ \frac{\df}{\df t_k} $.
Then, for any $ J, K \subseteq [1, n] $,
\begin{equation} \label{eq:drv-pseq}
\at{ \df_K \ang{T^*(t) x(t) \,|\, T(t) \,|\, x(t)}_J }_{t_1 = \ldots = t_n = 0}
= -(-i)^{|K|} \bbb{j_1, j_{|J|} \in K \subseteq J} \ang{v | A | v}_K
\end{equation}
where $ \bbb{\cdot} $ is an indicator function.
\end{lem}
\begin{proof}  
For any $ j \in [1, n] $,
using \cref{lem:iso-expgrp-form}, we note that
\begin{align*}
x_j(t_j) &= i\frac{e^{-it_j A_j} - 1}{A_j} v_j, \quad
T_j(t_j) = e^{-it_j A_j}, \\
\text{and} \quad
T_j^*(t_j) x_j(t_j) &= e^{it_j A_j} \cdot i\frac{e^{-it_j A_j} - 1}{A_j} v_j
= i\frac{1 - e^{it_j A_j}}{A_j} v_j
\end{align*}
so $ x_j(0) = 0 $, $ T_j(0) = 1 $, and $ T_j^*(0) x_j(0) = 0 $.
Then, the derivatives at zero are
\begin{align*}
(\df_j x_j)(0) = i\frac{-iA_j e^{-i0 \cdot A_j}}{A_j} v_j &= v_j, \qquad
(\df_j T_j)(0) = -iA_j e^{-i0 \cdot A_j} = -iA_j, \\
\text{and} \quad
(\df_j T_j^* x_j)(0) &= i \frac{- iA_j e^{i0 \cdot A_j}}{A_j} v_j = v_j.
\end{align*}
So, for any $ s \in \{0, 1\} $, we have
\begin{equation} \label{eq:cond-drvs}
(\df_j^s x_j)(0) = s v_j, \qquad
(\df_j^s T_j)(0) = (-i A_j)^s,
\quad \text{and}  \quad
(\df_j^s T_j^* x_j)(0) = s v_j
\end{equation}
where, by convention, $ A^0 = 1 $ for any operator $ A $.
\medskip

Now, if $ K \nsubseteq J $ then there exists $ \wtld{k} \in K \nsubseteq J $ such that
\[
\ang{T^*(t) x(t) | T(t) | x(t)}_J = \ang{T^*_{j_1}(t_{j_1}) x(t_{j_1}),
    T_{j_2}(t_{j_2}) \cdots T_{j_{|J| - 1}}(t_{j_{|J| - 1}}) x_{j_{|J|}}(t_{|J|})}
\]
is constant with respect to $ t_{\wtld{k}} $
so applying $ \frac{\df}{\df t_{\wtld{k}}} $ produces zero.
Otherwise, $ K \subseteq J $.
For each $ k \in K $ and $ j \in J $, if $ j \neq k $ then
\[
\df_k x_j(t_j) = \df_{t_k} T_j(t_j)
= \df_{t_k} T_j^*(t_j) x_j(t_j) = 0.
\]
Because of this, using \cref{eq:cond-drvs},
we can write the left-hand side of \cref{eq:drv-pseq} as
\begin{align*}
&\at{\df_{k_1} \cdots \df_{k_{|K|}}
    \ang{T_{j_1}^*(t_{j_1}) x_{j_1}(t_{j_1}),
        \p{\prod_{\ell=2}^{|J| - 1} T_{j_\ell}(t_{j_\ell})}  x_{j_{|J|}}(t_{j_{|J|}})
    }
}_{t_1 = \ldots = t_n = 0} \\
&\qquad = \ang{
    \big( \df_{t_{j_1}}^{\bbb{j_1 \in K}} T_{j_1}^* x_{j_1} \big)(0),
    \p{\prod_{\ell=2}^{|J| - 1} \big( \df_{j_\ell}^{\bbb{j_\ell \in K}} T_{j_\ell} \big)(0)}
    \cdot \Big( \df_{t_{j_{|J|}}}^{\bbb{j_{|J|} \in K}} x_{j_{|J|}} \Big)(0)
} \\
&\qquad = \ang{\bbb{j_1 \in K} v_{j_1},
    \p{\prod_{\ell=2}^{|J| - 1} (-i A_{j_\ell})^{\bbb{j_\ell \in K}}}
    \bbb{j_{|J|} \in K} v_{j_{|J|}}
} \\
&\qquad = \bbb{j_1, j_{|J|} \in K} (-i)^{|K| - 2}
    \ang{v_{j_1}, A_{K \setminus \{j_1, j_{|J|}\}} v_{j_{|J|}}} \\
&\qquad = -\bbb{j_1, j_{|J|} \in K} (-i)^{|K|} \ang{v | A | v}_K.
\end{align*}
\end{proof}

\begin{lem} \label{lem:diff-set-exp}
Suppose $ t_1, \ldots, t_n \in \bR $ and $ f : \bR^n \to \bC $ is first-differentiable.
Then, using $ \df_k $ to mean $ \frac{\df}{\df t_k} $,
for any set $ S \subseteq [1, n] $,
\[ \df_S e^{f(t)} = e^{f(t)} \sum_{\bigcup P = S} \;\; \prod_{J \in P} (\df_J f)(t) \]
where $ P $ ranges over partitions of $ S $
(i.e. disjoint families of nonempty subsets of $ S $ which cover $ S $).
\end{lem}
\begin{proof}  
We prove by induction on $ |S| $.
For $ |S| = 0 $, the only partition on $ S = \emptyset $ is $ P = \emptyset $
and the product over $ P $ is one so
\[ \df_\emptyset e^{f(t)} = e^{f(t)} = e^{f(t)} \prod_{J \in \emptyset} (\df_J f)(t). \]
Next, when $ |S| \geq 1 $, let $ j = \min S $ and, by the induction hypothesis,
\[ \df_{S \setminus \{j\}} e^{f(t)} = e^{f(t)} \sum_{\bigcup P = S \setminus \{j\}} \prod_{J \in P} (\df_J f)(t). \]
Applying $ \df_j $ to both sides, we get
\begin{align}
\df_S e^{f(t)} &= \df_j \df_{S \setminus \{j\}} e^{f(t)} \notag \\
&= (\df_j e^{f(t)}) \sum_{\bigcup P = S \setminus \{j\}} \prod_{J \in P} (\df_J f)(t)
    + e^{f(t)} \sum_{\bigcup P = S \setminus \{j\}} \sum_{J' \in P} (\df_j \df_{J'} f)(t)
        \prod_{J \in P \setminus \{J'\}} (\df_J f)(t) \notag \\
&= e^{f(t)} (\df_j f)(t) \sum_{\bigcup P = S \setminus \{j\}} \prod_{J \in P} (\df_J f)(t)
    + e^{f(t)} \sum_{\bigcup P = S \setminus \{j\}} \sum_{J' \in P} (\df_{J \cup \{j'\}} f)(t)
        \prod_{J \in P \setminus \{J'\}} (\df_J f)(t) \notag \\
&= e^{f(t)} \sum_{\bigcup P = S \setminus \{j\}} \p{\prod_{J \in P \cup \{\{j\}\}} (\df_J f)(t)}
    + e^{f(t)} \sum_{\bigcup P = S \setminus \{j\}} \sum_{J' \in P}
        \p{\prod_{J \in P \setminus \{J'\} \cup \{J' \cup \{j\}\}} (\df_J f)(t)} \notag \\
\label{eq:part-sum-minus-j}
&= e^{f(t)} \sum_{\bigcup P = S \setminus \{j\}} \sum_{J' \in P \cup \{\emptyset\}}
    \p{\prod_{J \in P \setminus \{J'\} \cup \{J' \cup \{j\}\}} (\df_J f)(t)}.
\end{align}
Every partition of $ S $ is obtained
from a partition $ P $ of $ S \setminus \{j\} $
by adjoining the singleton set $ \{j\} $
or by adjoining $ j $ to one of the sets in $ P $.
Thus, the summation in \cref{eq:part-sum-minus-j} becomes
\[
\df_S e^{f(t)} = e^{f(t)} \sum_{\bigcup P = S \setminus \{j\}} \sum_{J' \in P \cup \{\emptyset\}}
    \p{\prod_{J \in P \setminus \{J'\} \cup \{J' \cup \{j\}\}} (\df_J f)(t)}
= e^{f(t)} \sum_{\bigcup P = S} \;\; \prod_{J \in P} (\df_J f)(t)
\]
completing the induction.
\end{proof}

\begin{proof}[Proof of \cref{thm:gauge-isserlis}]  
Let $ t_1, \ldots, t_n \in \bR $.
Then, for each $ j \in [1, n] $,
we define $ (x_j(t_j), T_j(t_j)) := e^{t_j (v_j, -iA_j)} \in \ISO(\cH) $.
From \cref{lem:Wprod-vac-st}, we have
\begin{equation} \label{eq:Wprod-vac-st-exp}
\ang{\vp(0), W(x_1(t_1), T_1(t_1)) \cdots W(x_n(t_n), T_n(t_n)) \vp(0)}
= \exp\p{-\frac{1}{2} \sum_{a,b}
    \ang{T^*(t) x(t) \,|\, T(t) \,|\, x(t)}_{[a \land b, a \lor b]}
} =: e^{f(t)}.
\end{equation}
where $ a, b \in [1, n] $.
Let $ K = \{k_1, \ldots, k_{|K|}\} \subseteq [1, n] $
with $ k_1 < \ldots < k_{|K|} $.
Then, either $ |K| \leq 1 $ or $ |K| \geq 2 $.
If $ |K| \leq 1 $ then, from \cref{defn:capped-prod-seq}, $ \ang{v | A | v}_K = 0 $
so, by \cref{lem:diff-K-capd-pseq},
\begin{align}
\label{eq:diff-K-expo}
(\df_K f)(0)
&= \frac{1}{2} \sum_{a,b} (-i)^{|K|} \bbb{a, b \in K \subseteq [a \land b, a \lor b]} \ang{v | A | v}_K \\
&= 0 = -(-i)^{|K|} \ang{v | A | v}_K.  \notag
\end{align}
Otherwise, $ |K| \geq 2 $ and $ k_1 \neq k_{|K|} $
so we must have $ (a, b) = (k_1, k_{|K|}) $ or $ (b, a) = (k_1, k_{|K|}) $
in order for $ \bbb{a, b \in K \subseteq [a \land b, a \lor b]} \neq 0 $.
Thus, there are exactly two terms in the summation in \cref{eq:diff-K-expo} that are non-zero so
\[
(\df_K f)(0)
= -\frac{1}{2} \sum_{a,b} (-i)^{|K|} \bbb{a, b \in K \subseteq [a \land b, a \lor b]} \ang{v | A | v}_K
= -(-i)^{|K|} \ang{v | A | v}_K.
\]
Hence, in either case, $ (\df_K f)(0) = -(-i)^{|K|} \ang{v | A | v}_K $.
Applying $ \df_{[1, n]} $ to \cref{eq:Wprod-vac-st-exp} and evaluating at $ t_1 = \ldots = t_n = 0 $,
using \cref{lem:diff-set-exp} and \cref{lem:diff-K-capd-pseq}, we get
\begin{align*}
\at{\df_{[1, n]} \ang{\vp(0), \prod_{j=1}^n W(x_j(t_j), T_j(t_j)) \vp(0)}}_{t = 0}
&= \sum_{\bigcup P = [1, n]} \prod_{K \in P} (\df_K f)(0) \\
&= \sum_{\bigcup P = [1, n]} \prod_{K \in P} (-1) (-i)^{|K|} \ang{v | A | v}_K \\
&= (-i)^n \sum_{\bigcup P = [1, n]} (-1)^{|P|} \prod_{K \in P} \ang{v | A | v}_K.
\end{align*}
Lastly, we know $ W(x_j(t_j), T_j(t_j)) = e^{-it \dW(v_j, A_j)} $
for all $ j \in [1, n] $ and thus
\begin{align*}
\at{\df_{[1, n]} \ang{\vp(0), \prod_{j=1}^n W(x_j(t_j), T_j(t_j)) \vp(0)} }_{t = 0}
&= \ang{\vp(0), \p{ \prod_{j=1}^n (-i) \dW(v_j, A_j) W(x_j(0), T_j(0)) } \vp(0)} \\
&= (-i)^n \ang{\vp(0), \dW(v_1, A_1) \cdots \dW(v_n, A_n) \vp(0)}
\end{align*}
proving the desired result.
\end{proof}

\subsection{Control of Correlations for Differential Second Quantization}

We recall from \cref{subsubsect:diff-weyl-ops} that $ \cH $ is a Hilbert space
and $ \{\cF_t\}_{t \geq 0} $ is a collection of closed subspaces of $ \cH $.
We assume there exists $ \eta > 0 $
such that for all $ s, t \geq 0 $, $ u \in \cF_s $, and $ v \in \cF_t $,
\begin{equation} \label{eq:Ft-angle-bnd}
\ab{\ang{u, v}} \leq e^{-\eta |t - s|} \|u\| \,\, \|v\|.
\end{equation}
For each $ t \geq 0 $, we also have vectors $ \{v^\alpha_t\}_{\alpha \in \bN} \subseteq \cF_t $
and bounded self-adjoint operators $ \{A^\alpha_t\}_{\alpha \in \bN} \in i\uni(\cF_t) $.
Moreover, $ \{v^\alpha_t\} $ and $ \{A^\alpha_t\} $ are uniformly bounded
so there exists $ C > 0 $
such that $ \|v^\alpha_t\|, \|A^\alpha_t\| \leq C $ for all $ \alpha $ and $ t $.
\medskip

\begin{lem} \label{lem:capd-pseq-bnd}
Let $ \alpha_1, \ldots, \alpha_n \in \bN $
and $ t_1, \ldots, t_n \geq 0 $.
We say $ v_j := v^{\alpha_j}_{t_j} $
and $ A_j := A^{\alpha_j}_{t_j} $ for every $ j \in [1, n] $.
Then, for any subset $ K \subseteq [1, n] $
with $ k_1 < \ldots < k_{|K|} $,
\[
\ab{\ang{v | A | v}_K}
= \ab{\ang{v^{\alpha_{k_1}}_{t_{k_1}} \,|\,
    A^{\alpha_{k_2}}_{t_{k_2}} \cdots A^{\alpha_{k_{|K| - 1}}}_{t_{k_{|K| - 1}}}
    \,|\, v^{\alpha_{k_{|K|}}}_{t_{k_{|K|}}}
}}
\leq C^{|K|} \prod_{\ell=1}^{|K| - 1} e^{-\eta |t_{k_{\ell + 1}} - t_{k_\ell}|}.
\]
\end{lem}
\begin{proof}  
For each $ t \geq 0 $, we use $ P_t : \cH \to \cH $
to refer to the projection onto $ \cF_t $.
Then, by \cref{eq:Ft-angle-bnd}, for any $ s, t \geq 0 $,
the operator norm of the product of projections is
\[ \|P_t P_s\| = \|P_s P_t\| \leq e^{-\eta |t - s|}. \]
Moreover, for each $ j \in [1, n] $,
we have $ v_j = v^{\alpha_j}_{t_j} \in \cF_{t_j} $
and $ A_j = A^{\alpha_j}_{t_j} \in i\uni(\cF_{t_j}) $
so we can write $ v_j = P_{t_j} v_j $ and $ A_j = P_{t_j} A_j P_{t_j} $.
Thus, for any subset $ K = \{k_1, \ldots, k_{|K|}\} \subseteq [1, n] $ with $ k_1 < \ldots < k_{|K|} $,
$ \ab{\ang{v | A | v}_K} $ can be written as
\begin{align*}
\ab{\ang{v | A | v}_K}
&= \ab{\ang{P_{t_{k_1}} v_{k_1},
    \p{\prod_{\ell=2}^{|K| - 1} P_{t_{k_\ell}} A_{k_\ell} P_{t_{k_\ell}}}
    P_{t_{k_{|K|}}} v_{k_{|K|}}
}} \\
&\leq \|v_{k_1}\| \,\, \|P_{t_{k_2}} P_{t_{k_1}}\| \cdot
    \p{\prod_{\ell=2}^{|K| - 1} \|A_{k_\ell}\| \,\, \|P_{t_{k_{\ell + 1}}} P_{t_{k_\ell}}\|}
    \cdot \|v_{k_{|K|}}\| \\
&\leq C e^{-\eta |t_{k_2} - t_{k_1}|} \cdot
    \p{\prod_{\ell=2}^{|K| - 1} C e^{-\eta |t_{k_{\ell + 1}} - t_{k_\ell}|}} \cdot C
= C^{|K|} \prod_{\ell=1}^{|K| - 1} e^{-\eta |t_{k_{\ell + 1}} - t_{k_\ell}|}.
\end{align*}
\end{proof}

\begin{proof}[Proof of \cref{thm:gauge-graph-ctl}]  
We use $ \cP_n $ to refer to the set of partitions on $ [1, n] $.
For any $ P \in \cP_n $, we define $ G_P $ to be the graph on vertices $ [1, n] $
such that $ x \sim_{G_P} y $ if and only if there exists $ J = \{j_1, \ldots, j_{|J|}\} \in P $
with $ x = j_\ell $ and $ y = j_{\ell \pm 1} $.
Hence, each set in $ P $ corresponds to a connected component of $ G_P $
each of which is a path graph in which the vertices are connected consecutively.
Next, we define
\[
\cG_\prt := \{G_P \,:\, n \in \bN, \ P \in \cP_n,
    \ \forall K \in P, \, |K| \geq 2
\}
\]
and since every set in $ P $ has at least two elements,
every vertex of $ G_P $ has an edge.
Next, for each $ j \in [1, n) $, we say $ \wtld{\cP}^j_n \subseteq \cP_n $
is the subset of partitions $ P $ with $ \{j, j+1\} \in P $.
Then, we define
\[
\cG_\prt' := \{G_P / \{j, j+1\} \,:\, n \in \bN, j \in [1, n),
    \ P \in \cP_n \setminus \wtld{\cP}^j_n
\}
\]
where $ G / \{x, y\} $ is the graph obtained by identifying $ x $ and $ y $.
Let $ G_P / \{j, j+1\} \in \cG_\prt' $ then $ P \in \cP \setminus \wtld{\cP}^j_n $.
Now, let $ K, K' \in P $ such that $ j \in K $ and $ j+1 \in K' $.
If $ K = K' $ then, since $ P \notin \wtld{\cP}^j_n $, we know $ K \neq \{j, j+1\} $
so $ |K| \geq 3 $ and there exists $ x \in K \setminus \{j, j+1\} $
with $ x \sim_{G_P} j $ or $ x \sim_{G_P} j + 1 $.
Otherwise, $ K \neq K' $ so $ K \cap K' = \emptyset $ and there exists $ x \in K \setminus \{j\} $
with $ x \neq j + 1 $ and $ x \sim_{G_P} j $.
Thus, in either case, there exists $ x \in [1, n] \setminus \{j, j+1\} $
with $ x \sim_{G_P} j $ or $ x \sim_{G_P} j + 1 $
implying that the vertex $ \{j, j+1\} $ in $ G_P / \{j, j+1\} $ has an edge.
\medskip

Then, since every vertex in every graph of $ \cG_\prt $ and $ \cG_\prt' $ have edges,
we can take $ \cG := \cG_\prt \cup \cG_\prt' $.
We also take $ \vartheta_G := C^{|G|} $ for all $ G \in \cG $.
\medskip

Now, for any $ \alpha_1, \ldots, \alpha_n \in \bN $ and $ t_1, \ldots, t_n \geq 0 $,
we define $ v_j := v^{\alpha_j}_{t_j} $ and $ A_j := A^{\alpha_j}_{t_j} $ for all $ j \in [1, n] $.
Then, by \cref{thm:gauge-isserlis},
\begin{align}
\mean{E^{\alpha_1}_{t_1} \cdots E^{\alpha_n}_{t_n}}
&=\ang{\vp(0), \dW(v_1, A_1) \cdots \dW(v_n, A_n) \vp(0)} \notag \\
&= \sum_{\bigcup P = [1, n]} \,\, \prod_{K \in P} \ang{v | A | v}_K
\label{eq:gauge-n-point-form}
= \sum_{P \in \cP_n} \,\, \prod_{K \in P} \ang{v | A | v}_K.
\end{align}
Then, applying \cref{lem:capd-pseq-bnd} to \cref{eq:gauge-n-point-form}, one gets
\begin{align*}
\ab{\mean{E^{\alpha_1}_{t_1} \cdots E^{\alpha_n}_{t_n}}}
&\leq \sum_{P \in \cP_n} \prod_{K \in P} \ab{\ang{v | A | v}_K}
\leq \sum_{P \in \cP_n} \prod_{K \in P} C^{|K|}
    \prod_{\ell=1}^{|K| - 1} e^{-\eta |t_{k_{\ell + 1}} - t_{k_\ell}|}
= \sum_{P \in \cP_n} C^n \Theta_{G_P}(t_1, \ldots, t_n) \\
&\leq \sum_{\substack{G \in \cG_\prt \\ |G| = n}} \vartheta_G \Theta_G(t_1, \ldots, t_n)
\leq \Theta_{\cG_\prt}(t_1, \ldots, t_n) \leq \Theta_\cG(t_1, \ldots, t_n)
\end{align*}
fulfilling \cref{eq:graph-mean-ctl}.
\medskip

Next, let $ j \in [1, n) $ and assume that $ t_k \leq t_j \land t_{j+1} $
for all $ k \in [1, n] \setminus \{j, j+1\} $.
Without loss of generality, by reversing the ordering by conjugation,
we assume $ t_j \leq t_{j+1} $.
Applying \cref{thm:gauge-isserlis}, we obtain
\begin{align}
&\mean{E^{\alpha_j}_{t_j} E^{\alpha_{j+1}}_{t_{j+1}}}
\mean{E^{\alpha_1}_{t_1} \cdots E^{\alpha_{j-1}}_{t_{j-1}}
    E^{\alpha_{j+2}}_{t_{j+2}} \cdots E^{\alpha_n}_{t_n}} \notag \\
&\qquad = \ang{v_j, v_{j+1}} \sum_{\bigcup P = [1, j) \cup (j + 1, n]} \,\, \prod_{K \in P} \ang{v | A | v}_K \notag \\
&\qquad = \sum_{\bigcup P = [1, j) \cup (j+1, n]} \,\, \prod_{K \in P \cup \{j, j+1\}} \ang{v | A | v}_K
\label{eq:split-n-point-form}
= \sum_{P \in \wtld{\cP}^j_n} \,\, \prod_{K \in P} \ang{v | A | v}_K.
\end{align}
Subtracting \cref{eq:split-n-point-form} from \cref{eq:gauge-n-point-form}, one gets
\begin{align}
&\ab{\mean{E^{\alpha_1}_{t_1} \cdots E^{\alpha_{j-1}}_{t_{j-1}} \p{
        E^{\alpha_j}_{t_j} E^{\alpha_{j+1}}_{t_{j+1}} - \mean{E^{\alpha_j}_{t_j} E^{\alpha_{j+1}}_{t_{j+1}}}
    } E^{\alpha_{j+2}}_{t_{j+2}} \cdots E^{\alpha_n}_{t_n}}} \notag \\
&\qquad = \ab{\sum_{P \in \cP_n \setminus \wtld{\cP}^j_n} \,\, \prod_{K \in P} \ang{v | A | v}_K}
\leq \sum_{P \in \cP_n \setminus \wtld{\cP}^j_n}  \,\, \prod_{K \in P} \ab{\ang{v | A | v}_K} \notag \\
&\qquad \leq \sum_{P \in \cP_n \setminus \wtld{\cP}^j_n}  \,\, \prod_{K \in P} C^{|K|}
    \prod_{\ell=1}^{|K| - 1} e^{-\eta |t_{k_{\ell + 1}} - t_{k_\ell}|}
\label{eq:2crc-n-point}
= \sum_{P \in \cP_n \setminus \wtld{\cP}^j_n} C^n \Theta_{G_P}(t_1, \ldots, t_n).
\end{align}
Now, let $ P \in \cP_n \setminus \wtld{\cP}^j_n $.
since $ \{j, j+1\} \notin P $
For any $ x \in [1, n] $ with $ x \sim_{G_P} (j+1) $,
since $ t_x \leq t_j \leq t_{j+1} $, one has $ |t_{j+1} - t_x| = |t_{j+1} - t_j| + |t_j - t_x| $
and so $ e^{-\eta |t_{j+1} - t_x|} = e^{-\eta |t_{j+1} - t_j|} e^{-\eta |t_j - t_x|} $.
Thus, by contracting $ j + 1 $ into $ j $, we have
\[ \Theta_{G_P}(t_1, \ldots, t_n) = e^{-\eta |t_{j+1} - t_j|} \Theta_{G_P / \{j, j+1\}}(t_1, \ldots, t_j, t_{j+2}, \ldots, t_n) \]
and thus \cref{eq:2crc-n-point} becomes
\begin{align*}
\sum_{P \in \cP_n \setminus \wtld{\cP}^j_n} \vartheta_G \Theta_{G_P}(t_1, \ldots, t_n)
&= e^{-\eta |t_{j+1} - t_j|} \Theta_{\cG_\prt'}(t_1, \ldots, t_j, t_{j+2}, \ldots, t_n) \\
&\leq e^{-\eta |t_{j+1} - t_j|} \Theta_\cG(t_1, \ldots, t_j, t_{j+2}, \ldots, t_n)
\end{align*}
proving \cref{eq:graph-cov-ctl}.
\medskip

Lastly, we prove \cref{eq:graph-inf-rad-conv}.
Using $ B_n $ for the $ n $-th Bell number,
observe that $ \# \{G \in \cG_\prt \,:\, |G| = n\} \leq \# \cP_n = B_n $
and $ \# \{G \in \cG_\prt' \,:\, |G| = n\} \leq \# \cP_{n+1} = B_{n+1} $.
Thus,
\[
\sum_{\substack{G \in \cG \\ |G| = n}} \vartheta_G
\leq C^n (B_n + B_{n+1}).
\]
Recall that the generating function for the Bell numbers is
\[
e^{e^t - 1} = \sum_{n=0}^\infty \frac{B_n}{n!} t^n
\quad \text{and so} \quad
e^t \cdot e^{e^t - 1} = \frac{\dr}{\dr[t]} e^{e^t - 1} = \sum_{n=0}^\infty \frac{B_{n+1}}{n!} t^n.
\]
Thus, by combining these sums,
\[
\sum_{n=0}^\infty \frac{C^n (B_n + B_{n+1})}{n!} t^n
= \sum_{n=0}^\infty \frac{B_n}{n!} (C t)^n + \sum_{n=0}^\infty \frac{B_{n+1}}{n!} (C t)^n
= e^{e^{C t} - 1} + e^{C t} \cdot e^{e^{C t} - 1}
\]
so the power series has infinite radius of convergence.
Hence, by Cauchy's root test, \cref{eq:graph-inf-rad-conv} holds.
\end{proof}

\vspace{1em}

\textbf{AI Statement:}
Large Language Models (LLMs) were \textsl{not} used in any part of
the literature review, research, proof development, writing, editing, or reviewing
for this work.

\vspace{2em}
\pagebreak[2]
\appendix

\section{Operator Semigroupoids}
\label{sect:op-sgprd}

In order to accommodate differential equations driven by non-Markovian processes,
it will be necessary to generalize the theory of operator semigroups
to allow the generator to change with time.
For any vector space $ B $ and seminorm $ p $ on $ B $,
we say $ \dom(p) := \{x \in B \,:\, p(x) < \infty\} $
and use $ B[p] $ to refer to the completion of $ \dom(p) / \ker(p) $ with respect to $ p $.
\smallskip

\begin{defn} \label{defn:op-sgrpd}
Suppose $ p_\bt = \{p_t\}_{t \in [0, T)} $ with $ T \in \ebRp $
is a collection of (potentially unbounded) seminorms on $ B $
such that for all $ s, t \in [0, T) $ with $ s \leq t $, $ p_s \leq p_t $
and $ \dom(p_t) $ is $ p_s $-dense in $ \dom(p_s) $.
We call such a collection of seminorms a \textbf{seminorm process} on $ B $.
Since $ p_s \leq p_t $, there exists the canonical
map $ x \mapsto x + \ker(p_s) : B[p_t] \to B[p_s] $.
\smallskip

We define an \textbf{(operator) semigroupoid} on $ B $
\textbf{controlled} by $ p_\bt $ to be
a collection $ \xi_\bt = \{\xi_{s \gets t} : B[p_t] \to B[p_s]\}_{0 \leq s \leq t < T} $
of linear operators such that
for all $ 0 \leq r \leq s \leq t < T $ and any $ x \in B[p_t] $,
\[
\xi_{t \gets t}(x) = x \, ,
\quad p_s(\xi_{s \gets t}(x)) \leq p_t(x) \, ,
\quad \text{and} \quad
\xi_{r \gets t}(x) = \xi_{r \gets s}(\xi_{s \gets t}(x)).
\]
\end{defn}

Now, suppose $ \df_\bt = \{\df_t\}_{t \in [0, T)} $
is a collection of (potentially unbounded) linear operators on $ B $.
We will call such a collection of operators an \textbf{operator process}.
We say that $ \df_\bt $ \textbf{generates}
the semigroupoid $ \xi_\bt $ controlled by $ p_\bt $
if for all $ t \in [0, T) $ and $ x \in \dom(p_t) $,
we have $ x \in \dom(\df_\tau) $ for all $ \tau \in [0, t] $
and for any $ s \in [0, t] $,
\[ \xi_{s \gets t}(x) = x + \int_s^t \dr[\tau] \xi_{s \gets \tau}(\df_\tau x) \]
where the integral converges in $ p_s $.
\smallskip

In order to show the existence of a semigroupoid generated by $ \df_\bt $,
it will be necessary to enforce some sense of
continuity and boundedness on the operator process.
We say $ p_\bt $ is \textbf{differentiable}
if for every $ t \in [0, T) $ and $ x \in \dom(p_t) $,
the function $ s \mapsto p_s(x) : (0, t) \to \mathbb{R}_+ $ is differentiable
and the derivative with respect to $ s $ (i.e. $ x \mapsto \dot{p}_s(x) $)
is also a seminorm on $ B $.
Next, we say that $ \{p_t\}_{t \in [0, T)} $ \textbf{controls} $ \{\df_t\}_{t \in [0, T)} $
if for all $ t \in [0, T) $ and any $ x \in \dom(p_t) $,
\begin{enumerate}
\item \label{item:strong-pt-cont}
the function $ s \mapsto \df_s x $ exists and is $ p_t $-continuous for $ s \in [0, t] $
\item \label{item:diff-bdd}
$ x \in \dom(\dot{p}_t) $ and $ p_t(\df_t x) \leq \dot{p}_t(x) $.
\end{enumerate}
Note that, given \cref{item:strong-pt-cont},
\cref{item:diff-bdd} is equivalent to the condition
\begin{equation} \label{eq:pt-ctl-integ}
p_s(x) + \int_s^t \dr[\tau] p_\tau(\df_\tau x) \leq p_t(x)
\quad \text{for all} \,\, s, t \in [0, T)
\,\,\text{with}\,\, s \leq t.
\end{equation}
\smallskip

\begin{prop} \label{prop:mu-bnd-sgrpd}
Suppose $ p_\bt $ controls $ \df_\bt $
and $ \{\xi_{s \gets t}\}_{0 \leq s \leq t < T} $ is a semigroupoid on $ B $
controlled by $ p_\bt $ and generated by $ \df_\bt $.
We assume $ \{\mu_t\}_{t \in [0, T)} $ is a seminorm process on $ B $
that controls $ \df_\bt $
and such that for each $ t \in [0, T) $,
there exists $ C_t > 0 $ with $ \mu_t \leq C_t p_t $.
Then, for all $ 0 \leq s \leq t < T $ and $ x \in B[p_t] $,
one has $ \mu_s(\xi_{s \gets t}(x)) \leq \mu_t(x) $.
\end{prop}
\begin{proof}
Fix $ s, t \in [0, T) $ with $ s \leq t $, $ \ep > 0 $,
and $ x \in B[p_t] $ with $ p_s(x) > 0 $.
Then, we define $ b \in [0, \infty] $ as
\[
b := \sup_{\tau \in [s, t]}
  \frac{\ep + \mu_s(\xi_{s \gets \tau}(x))}{\ep e^{\tau - s} + \mu_\tau(x)}.
\]
Notice that for any $ \tau \in [s, t] $,
\[ \begin{split}
\mu_s(\xi_{s \gets \tau}(x)) &\leq C_s p_s(\xi_{s \gets \tau}(x))
\leq C_s p_\tau(x) \leq C_s p_t(x) \\
\text{and} \quad
\ep e^{\tau - s} + \mu_\tau(x) &\geq \ep e^{\tau - s} \geq \ep
\end{split} \]
so $ b \leq \frac{1}{\ep} (\ep + C_s p_t(x)) < \infty $.
Next, we assume for sake of contradiction that $ b > 1 $ so that $ 1 - b < 0 $
Now, because $ \mu_\bt $ controls $ \df_\bt $, for each $ r \in [s, \tau] $,
\[ \begin{split}
\mu_s(\xi_{s \gets \tau}(x))
&\leq \mu_s(x) + \int_s^\tau \dr{r} \mu_s(\xi_{s \gets r}(\df_r x)) \\
&\leq \mu_s(x) + \int_s^\tau \dr{r} b \left(\big. \ep e^{r - s} + \mu_r(\df_r x) \right) \\
&\leq \mu_s(x) + b \int_s^\tau \dr{r} \left(\big. \ep e^{r - s} + \dot{\mu}_r(x)\right) \\
&= \mu_s(x) + b \left(\big. \ep (e^{\tau - s} - 1) + (\mu_\tau(x) - \mu_s(x))\right) \\
&= (1 - b) \mu_s(x) - b \ep + b \left(\big. \ep e^{\tau - s} + \mu_\tau(x)\right)
\end{split} \]
Thus, we may bound the ratio by
\[ \begin{split}
\frac{\ep + \mu_s(\xi_{s \gets \tau}(x))}{\ep e^{\tau - s} + \mu_\tau(x)}
&\leq \frac{(1 - b) \mu_s(x) + (1 - b) \ep}{\ep e^{\tau - s} + \mu_\tau(x)} + b \\
&\leq (1 - b) \frac{\mu_s(x) + \ep}{\ep e^{s - s}} + b
= (1 - b) \left(\frac{1}{\ep} \mu_s(x) + 1\right) + b.
\end{split} \]
Because this holds for all $ \tau \in [s, t] $,
we must have $ b \leq (1 - b) \left(\frac{1}{\ep} \mu_s(x) + 1\right) + b $
so $ 0 \leq (1 - b) \left(\frac{1}{\ep} \mu_s(x) + 1\right) $ where $ \frac{1}{\ep} \mu_s(x) + 1 > 0 $.
Hence, $ 0 \leq 1 - b $ and $ b \leq 1 $ meaning that for $ \tau = t $,
\[
\mu_s(\xi_{s \gets t}(x)) = (\ep + \mu_s(\xi_{s \gets t}(x))) - \ep
\leq \ep (e^{t - s} - 1) + \mu_t(x).
\]
Because this is true for all $ \ep > 0 $,
we see that $ \mu_s(\xi_{s \gets t}(x)) \leq \mu_t(x) $.
\end{proof}

\pagebreak[1]
\providecommand{\bysame}{\leavevmode\hbox to3em{\hrulefill}\thinspace}
\providecommand{\MR}{\relax\ifhmode\unskip\space\fi MR }
\providecommand{\MRhref}[2]{%
  \href{http://www.ams.org/mathscinet-getitem?mr=#1}{#2}
}
\providecommand{\href}[2]{#2}


\begin{thebibliography}{10}

\bibitem{AFL90}
L.~Accardi, A.~Frigerio, and Y.~G. Lu, \emph{The weak coupling limit as a
  quantum functional central limit}, Comm. Math. Phys. \textbf{131} (1990),
  no.~3, 537--570. \MR{1065896}

\bibitem{AC82}
Luigi Accardi and Carlo Cecchini, \emph{Conditional expectations in von
  {Neumann} algebras and a theorem of {Takesaki}}, J. Funct. Anal. \textbf{45}
  (1982), 245--273.

\bibitem{ALV02}
Luigi Accardi, Yun~Gang Lu, and Igor Volovich, \emph{Quantum theory and its
  stochastic limit}, Springer-Verlag, Berlin, 2002. \MR{1925437}

\bibitem{Chan15}
Mou-Hsiung Chang, \emph{Quantum stochastics}, Camb. Ser. Stat. Probab. Math.,
  vol.~37, Cambridge: Cambridge University Press, 2015.

\bibitem{Dav74}
E.~B. Davies, \emph{Markovian master equations}, Comm. Math. Phys. \textbf{39}
  (1974), 91--110. \MR{359633}

\bibitem{DdR07}
Jan Derezi\'nski and Wojciech De~Roeck, \emph{Extended weak coupling limit for
  {F}riedrichs {H}amiltonians}, J. Math. Phys. \textbf{48} (2007), no.~1,
  012103, 19. \MR{2292594}

\bibitem{DdR08}
\bysame, \emph{Extended weak coupling limit for {P}auli-{F}ierz operators},
  Comm. Math. Phys. \textbf{279} (2008), no.~1, 1--30. \MR{2377627}

\bibitem{Dixm77}
Jacques Dixmier, \emph{{$C^*$}-algebras. {T}ranslated by {F}rancis {J}ellett},
  North-Holland Math. Libr., vol.~15, Elsevier (North-Holland), Amsterdam,
  1977.

\bibitem{DB86}
Robert~S. Doran and Victor~A. Belfi, \emph{Characterizations of
  {$C^*$}-algebras}, Monographs and Textbooks in Pure and Applied Mathematics,
  vol. 101, Marcel Dekker, Inc., New York, 1986, The Gelfand-Naimark theorems.
  \MR{831427}

\bibitem{Fell60}
J.~M.~G. Fell, \emph{The dual spaces of {$C\sp{\ast} $}-algebras}, Trans. Amer.
  Math. Soc. \textbf{94} (1960), 365--403. \MR{146681}

\bibitem{Fuji98}
Ichiro Fujimoto, \emph{A {G}elfand-{N}aimark theorem for {$C^*$}-algebras},
  Pacific J. Math. \textbf{184} (1998), no.~1, 95--119. \MR{1626504}

\bibitem{Latr16}
Fr\'ed\'eric Latr\'emoli\`ere, \emph{The quantum {G}romov-{H}ausdorff
  propinquity}, Trans. Amer. Math. Soc. \textbf{368} (2016), no.~1, 365--411.
  \MR{3413867}

\bibitem{Lind76}
G.~Lindblad, \emph{On the generators of quantum dynamical semigroups}, Commun.
  Math. Phys. \textbf{48} (1976), 119--130.

\bibitem{LP25}
Ilya~A. Lopatin and Alexander~N. Pechen, \emph{Weak coupling limit for quantum
  systems with unbounded weakly commuting system operators}, J. Math. Phys.
  \textbf{66} (2025), no.~4, Paper No. 042101, 21. \MR{4890651}

\bibitem{Murp90}
Gerard~J. Murphy, \emph{{$C^*$}-algebras and operator theory}, Academic Press,
  Inc., Boston, MA, 1990. \MR{1074574}

\bibitem{Part92}
K.~R. Parthasarathy, \emph{An introduction to quantum stochastic calculus},
  Modern Birkh\"auser Classics, Birkh\"auser/Springer Basel AG, Basel, 1992,
  [2012 reprint of the 1992 original] [MR1164866]. \MR{3012668}

\bibitem{RW26}
T.J. Reese and J.~Wehr, \emph{Equations driven by fast-oscillating functions of
  an {I}tô diffusion process}, Ann. Henri Poincar{\'e} (2026).

\bibitem{Rief04}
Marc~A. Rieffel, \emph{Compact quantum metric spaces}, Operator algebras,
  quantization, and noncommutative geometry, Contemp. Math., vol. 365, Amer.
  Math. Soc., Providence, RI, 2004, pp.~315--330. \MR{2106826}

\bibitem{Take02}
Masamichi Takesaki, \emph{Theory of operator algebras. {I}}, Springer-Verlag,
  New York-Heidelberg, 1979. \MR{548728}

\bibitem{Umeg54}
Hisaharu Umegaki, \emph{Conditional expectation in an operator algebra}, Tohoku
  Math. J. (2) \textbf{6} (1954), 177--181. \MR{68751}

\bibitem{Hove54}
L.~van Hove, \emph{Quantum-mechanical perturbations giving rise to a
  statistical transport equation}, Physica \textbf{21} (1954), no.~1, 517--540.

\end{thebibliography}
\end{document}